\documentclass[fleqn,usenatbib]{mnras}

\usepackage{newtxtext,newtxmath}

\usepackage[T1]{fontenc}
\usepackage{ae,aecompl}
\usepackage{siunitx}
\usepackage{soul}
\usepackage{subfig}

\usepackage{graphicx}	
\usepackage{amsmath}	
\usepackage{amssymb}	
\usepackage{pdflscape}
\usepackage{booktabs}
\usepackage{threeparttable}
\usepackage{hhline}
\usepackage{longtable}

\usepackage{varwidth}

\usepackage{datetime}

\begin{document}
\label{firstpage}
\pagerange{\pageref{firstpage}--\pageref{lastpage}}
\title[Light travel time effect in MOA eclipsing binaries]{A study of light travel time effect in short-period MOA eclipsing binaries via eclipse timing}

\author[M.~C.~A.~Li et al.]
{\parbox{\textwidth}{M.~C.~A.~Li,$^{1}$\thanks{E-mail: \texttt{mli351@aucklanduni.ac.nz}}
N.~J.~Rattenbury,$^{1}$\thanks{E-mail: \texttt{n.rattenbury@auckland.ac.nz}}
I.~A.~Bond,$^{2}$
T.~Sumi,$^{3}$
D.~P.~Bennett,$^{4}$
N.~Koshimoto,$^{3}$
F.~Abe,$^{5}$
Y.~Asakura,$^{5}$
R.~Barry,$^{6}$
A.~Bhattacharya,$^{4}$
M.~Donachie,$^{1}$
P.~Evans,$^{1}$
A.~Fukui,$^{7}$
Y.~Hirao,$^{3}$
Y.~Itow,$^{5}$
K.~Masuda,$^{5}$
Y.~Matsubara,$^{5}$
Y.~Muraki,$^{5}$
M.~Nagakane,$^{3}$
K.~Ohnishi,$^{8}$
To.~Saito,$^{9}$
A.~Sharan,$^{1}$
D.~J.~Sullivan,$^{10}$
D.~Suzuki,$^{4}$
P.~J.~Tristram$^{11}$ and
A.~Yonehara$^{12}$}\vspace{0.4cm}\\
\parbox{\textwidth}{$^{1}$Department of Physics, University of Auckland, Private Bag 92019, Auckland, New Zealand\\
$^{2}$Institute of Natural and Mathematical Sciences, Massey University, Private Bag 102-904, North Shore Mail Centre, Auckland, New Zealand\\
$^{3}$Department of Earth and Space Science, Graduate School of Science, Osaka University, 1-1 Machikaneyama, Toyonaka, Osaka 560-0043, Japan\\
$^{4}$Department of Physics, University of Notre Dame, Notre Dame, IN 46556, USA\\
$^{5}$Solar-Terrestrial Environment Laboratory, Nagoya University, Nagoya, 464-8601, Japan\\
$^{6}$Astrophysics Science Division, NASA Goddard Space Flight Center, Greenbelt, MD 20771, USA\\
$^{7}$Okayama Astrophysical Observatory, National Astronomical Observatory, 3037-5 Honjo, Kamogata, Asakuchi, Okayama 719-0232, Japan\\
$^{8}$Nagano National College of Technology, Nagano 381-8550, Japan\\
$^{9}$Tokyo Metropolitan College of Industrial Technology, Tokyo 116-8523, Japan\\
$^{10}$School of Chemical and Physical Sciences, Victoria University, Wellington, New Zealand\\
$^{11}$University of Canterbury, Mount John Observatory, P.O. Box 56, Lake Tekapo 8770, New Zealand\\
$^{12}$Department of Physics, Faculty of Science, Kyoto Sangyo University, 603-8555 Kyoto, Japan}}

\date{Accepted XXX. Received YYY; in original form ZZZ}

\pubyear{2018}

\maketitle

\begin{abstract}
A sample of 542 eclipsing binaries (EBs) with periods shorter than $2\,$d were selected from the Microlensing Observations in Astrophysics (MOA) EB catalogue \citep{2017MNRAS.470..539L} for eclipse-time variation analysis. For this sample we were able to obtain the time series from MOA-II that span $9.5\,$yr. We discovered 91 EBs, out of the 542 EBs, with detected light-travel-time effect signals suggesting the presence of tertiary companions of orbiting periods from $250\,$d$-28\,$yr. The frequency of EBs with tertiary companions in our sample increases as the period decreases and reaches a value of $0.65$ for contact binaries with periods shorter than $0.3\,$d. If only the contact binaries of periods $<0.26\,$d are considered, the frequency even goes to the unit. Our results suggest that contact binaries with periods close to the 0.22-d contact binary limit are commonly accompanied by relatively close tertiary companions.
\end{abstract}



\section{Introduction}
Microlensing is a rare astrophysical phenomenon predicted by Einstein's General Relativity \citep{1936Sci....84..506E}. Detection of microlensing events requires the capability of monitoring millions of stars simultaneously, and it had been thought as undetectable until the advent of CCD camera and wide-field observation techniques. Because of the observational strategy, microlensing surveys would result in a large amount of photometric data of variable objects (e.g. \citealt{2016AcA....66..405S,2017AcA....67..297S}). The Microlensing Observations in Astrophysics (MOA-II), for instance, has collected $\sim100\,$TB data of millions of variable objects in the fields towards the Galactic bulge (GB) since the project began in 2006 \citep{2013ApJ...778..150S} and over 8000 eclipsing binaries (EBs) in two MOA fields were identified recently by \citet{2017MNRAS.470..539L}. Amongst them, three contact binaries were further discovered to exhibit light-travel-time effect (LTTE) signals in their observed-minus-calculated (O$-$C) diagrams, indicating the presence of stellar tertiary companions of orbiting periods between 250 and 480 days \citep{2017MNRAS.470..539L}.

The LTTE is an effect associated with the change in orbital motion which appears in an EB with a tertiary companion wherein the EB's centre of mass is no longer stationary but moving around the barycentre of the whole three body system \citep{2016MNRAS.455.4136B}. From an observer's point of view, the movement of the EB's centre of mass might be reflected by the measurement of the times of eclipse minima which occur later or earlier cyclically than expected due to the finite speed of light and varying distances between the conjunction and the observer. Analyzing the eclipse-time variation (ETV) via O$-$C diagrams, otherwise known as the ETV method, has been a traditional technique to detect LTTE in EBs, with or without spectroscopic information \citep{2017MNRAS.469.2952Z, 2016A&A...590A..85Z, 1990BAICz..41..231M}. Nevertheless, before the era of space telescope surveys, the number of EBs with detected LTTE signals was limited and the triple systems found by the ETV method tended to be of outer periods longer than several years or decades because of poor precision in ground-based photometry and insufficient frequency of eclipse timings on the Earth. Majority of the triple systems identified via the ETV method, unsurprisingly, comes form the \textit{Kepler} space mission \citep{2016MNRAS.455.4136B, 2015MNRAS.448..946B, 2012AJ....143..137G}. However, stellar triples with outer periods longer than 4 years are obviously deficient in \textit{Kepler} triple candidates due to the limited duration (i.e. 1470 days) of the mission. Such bias in the population of stellar triples identified via the ETV method may be reduced using the photometric data of long-term ground-based surveys such as MOA-II.

We are interested in searching for and investigating the population of triple systems in the MOA EB catalogue using the time series from MOA-II having a longer time span than the previous work of \citet{2017MNRAS.470..539L}. In this paper, we first review the physics of the LTTE in Section~\ref{sec:ltte} and present the method of eclipse timing in Section~\ref{sec:measure_t_ecl}. The criteria for our sample selection are presented in Section~\ref{sec:sample_sel}. The observation and data reduction processes are described briefly in Section~\ref{sec:obs_dat}. We outlined the analysis works in detail in Section~\ref{sec:analysis} and present the results in Section~\ref{sec:ltte_results}. We finally discuss and conclude in Section~\ref{sec:ltte_conclusion}.

\section{Light Travel Time Effect}
\label{sec:ltte}
Changes in orbital periods were already observed in many EBs a century ago. \citet{1888AJ......7..165C} suggested that the observed period changes in Algol resulted from the LTTE due to the presence of a tertiary object. But this was after \citet{1922BAN.....1...93W} who was able to perform the LTTE calculation so that the LTTE was seriously considered as the plausible explanation. Later, \citet{1959AJ.....64..149I} proposed the analytical model of the LTTE to the O$-$C diagram in terms of stellar masses and orbital parameters. As a simple tool requiring only photometric measurements, O-C diagrams have been traditionally used to detect or study physical phenomena that induce changes in occurrence times of stellar events such as eclipses in EBs and regular pulsations in Cepheid and RR Lyrae variables, etc. For an EB, the O-C diagram represents variations in the times of its eclipse minima, which are determined by the following equation,
\begin{equation}
\Delta=T_{o}(E)-T_{c}(E)=T_{o}(E)-T_{0}-P_{s}E
\label{eq:etv}
\end{equation}
where $T_{o}(E)$ and $T_{c}(E)$ denote the observed and calculated times of the $E$-th eclipse minimum, $T_{0}$ represents the reference epoch and $P_{s}$ denotes the average eclipsing period. The general ETV model involving the LTTE is defined by:
\begin{equation}
	\Delta=c_{0}+c_{1}E+c_{2}E^{2}-\frac{a_{\text{AB}}\sin\\i_{2}}{c}\frac{(1-e_{2}^{2})\sin(\nu_{2}+\omega_{2})}{1+e_2\cos\nu_{2}},
    \label{eq:ltte}
\end{equation}
where the zeroth and first order coefficients, $c_{0}$ and $c_{1}$, in the polynomial of $E$ provide the corrections in $T_{0}$ and $P_{s}$, respectively, while the second order coefficient, $c_{2}$, is equal to half the rate of change in period, regardless of its origin. The parameters in the LTTE term, i.e. the last term in eq.(\ref{eq:ltte}), include eccentricity ($e_{2}$), true anomaly ($\nu_{2}$), argument of periastron ($\omega_{2}$), inclination ($i_{2}$) and the semi-major axis of its absolute orbit, $a_{\text{AB}}$, equal to $(m_{\text{C}}/m_{\text{ABC}})\,a_{2}$. The period ($P_{2}$) and the time of periastron ($\tau_{2}$) of the tertiary object are also needed implicitly when calculating $\nu_{2}$. The LTTE term, therefore, depends on six parameters. Note that $m_{\text{C}}$ is the tertiary object's mass, $m_{\text{ABC}}$ is the total mass of the triple system, and $a_{2}$ is the semi-major axis of the tertiary object's orbit around the EB's centre of mass and $c$ is the speed of light. The amplitude of the LTTE is defined by
\begin{equation}
\mathcal{A}_{\text{LTTE}}=\frac{a_{\text{AB}}\sin i_{2}}{c}\sqrt{1-e_{2}^{2}\cos^{2}\omega_{2}}.
\end{equation}
Unfortunately, the semi-major axis of the absolute orbit, $a_{\text{AB}}$, and the inclination, $i_{2}$, are degenerate in the LTTE model. Yet the mass function, $f(m_{\text{C}})$, defined as
\begin{equation}
f(m_{\text{C}})=\frac{m_{\text{C}}^{3}\sin^{3}i_{2}}{m_{\text{ABC}}^{2}}=\frac{4\pi^{2}a_{\text{AB}}^{3}\sin^{3}i_{2}}{GP_{2}^{2}},
\end{equation}
can be calculated when the LTTE solution is known. Then we can calculate the amplitude of the LTTE via the approximation equation given by
\begin{equation}
\mathcal{A}_{\text{LTTE}}\approx 1.1\times10^{-4}f(m_{\text{C}})^{1/3}P_2^{2/3}\sqrt{1-e_{2}^{2}\cos^{2}\omega_{2}}.
 \label{eq:amp}
\end{equation}
Note that the period and amplitude are in days and the masses are in units of $M_{\sun}$. The minimum mass of the tertiary object can also be estimated by the mass function assuming the inner binary of solar type, i.e., $m_{\text{AB}} = 2\,M_{\sun}$. From eq.(\ref{eq:amp}) we know that the LTTE amplitude decreases as the outer period decreases. Because of this, and owing to insufficient precision in ground-based photometry and difficulty in doing eclipse timing frequently enough on the Earth to satisfactorily cover a short period LTTE cycle, LTTEs associated with triple systems with outer periods shorter than two years were rarely detected by ground-based telescopes, and stellar triples identified on the Earth tends to be those with tertiary periods longer than several years or even decades.

Additional dynamical perturbations may dominate over perturbations due to LTTE, and become observable in an O-C diagram, if the tertiary companion tightly interact with the inner binary in a triple system \citep{2016MNRAS.455.4136B}. In the case of the inner binary being eccentric, the ETV term corresponding to apsidal motion may have to be included as well. The apsidal motion may be simply regarded as linear variation in $\omega_{1}$ as a result of the apsidal line of the inner binary's orbit rotating with a constant angular velocity in the direction of the orbital motion arising from the tidal deformation of the shapes of the binary components or relativistic effects \citep{2015MNRAS.448..946B,1939MNRAS..99..451S,1938MNRAS..98..734C}. Nonetheless, the presence of the tertiary companion may induce the apsidal motion of the inner binary to behave in a complicated manner in which no orbital parameters, except the semimajor axes, would remain constant (\citealt{2015MNRAS.448..946B,2013MNRAS.431.2155N}, and further references therein). Since we restricted our study to short period binaries for which circular orbits should be established, we thus assumed apsidal motions were negligible. 

Although the detection of a LTTE signal with multiple cycles is strong evidence for the existence of a tertiary companion in an EB, several mechanisms such as the mass transfer between the EB's components, magnetic braking and the Applegate effect \citep{1992ApJ...385..621A} can produce quadratic variation in the orbital period that may be confused with the LTTE cycle of period longer than the data time span. Star spots can produce spurious ETVs that mimic LTTE behaviour as well \citep{2013ApJ...774...81T}. In order to justify the plausibility of a LTTE solution to the ETV curve, \citet{1973A&AS...12....1F} suggested four general criteria: (1) the shape of the ETV curve must follow the analytical form of a LTTE solution; (2) the ETVs of the primary and secondary eclipses must be consistent with each other in both phase and amplitude; (3) the estimated mass or lower limit to the mass of the tertiary companion derived from the mass function must be in accord with the photometric measurements or the limit on the third light in the system; (4) the variation in the system's radial velocity must be in accord with the LTTE solution. Obviously, without radial velocity data, criterion (4) could not be satisfied. In addition to these criteria, we also employ the Bayesian information criterion (BIC) as an extra criterion,
\begin{equation}
\text{BIC} = n\,\ln\bigg(\frac{1}{n}\sum_{i}(x_{i}-\hat{x}_{i})^{2}\bigg) + k\ln n,
\label{eq:bic}
\end{equation}
where $x_{i}$ are the measurement values, $\hat{x}_{i}$ are the calculated values from the model fit, $n$ is the number of measurement points and $k$ is the number of variable parameters in the model fit. The goodness of the BIC as a model selection criterion is that it includes the penalty term, i.e., $k\ln n$, to disfavour the case of over-fitting by adding parameters. We shall accept detected LTTE signals as genuine if they satisfy the first and second criteria as well as the fits associated with the LTTE signals have lower values of the BIC compared to the quadratic fits associated with the ETV produced by other mechanisms.

\section{Eclipse Timing Methods}
\label{sec:measure_t_ecl}
The orbital motion of an EB, if it is purely a two-body system, should be exactly described and predicted by Kepler's equation as long as the apsidal motion is not concerned. Thus, EBs can be used as precise clocks in astronomy. Yet accurate eclipse timing is challenging on the Earth. Individual eclipses last usually a few hours. Ground-based observation often fails to obtain the complete coverage of an eclipse because of the poor condition of the night sky. Traditionally, the time of an eclipse minimum would be derived using the Kwee-van Woerden method \citep{1956BAN....12..327K}. Several recent studies to look for circumbinary planets in post-common-envelope binaries applied this method to derive the times of the eclipse minima (e.g., \citealt{2015A&A...577A.146B,2014MNRAS.445.2331L}). The Kwee-van Woerden method, however, cannot work properly if an eclipse is not symmetric about its minimum, or the distribution of the data points over the eclipse is not even (i.e., the observations over the eclipse were not taken in regular cadences), or the number of data points covering the eclipse is too low. The Kwee-van Woerden method also, as mentioned in \citet{2012AN....333..754P}, usually underestimates the uncertainties in the derived times. The eclipse template method, as far as we know, turned out to be an alternative method commonly used nowadays. Various ways to create an eclipse template were proposed and used by different research groups. The high-order polynomial fit (\texttt{polyfit}) was used by the \textit{Kepler} group in order to create an approximate eclipse template. A realistic eclipse template might be derived by fitting the photometric light curve using an EB modeling package such as \texttt{PHOEBE} \citep{2016ApJS..227...29P}.

The template generation methods mentioned so far are, however, either impracticable or unsatisfactory for our study. Although the template of a grazing eclipse can be appropriately generated by a quartic polynomial using the \texttt{polyfit} code, we found that it has trouble producing an appropriate template for a total eclipse. A higher-order polynomial might be adopted to generate templates for total eclipses, but then it often yielded templates with rippling bottoms and the minima did not appropriately correspond to the eclipse minima. We desired a template generation method that was workable for grazing and total eclipses. For these reasons, we finally decided to adopt the phenomenological light curve model of EBs proposed by \cite{2015A&A...584A...8M}. Considering only the portion of an EB's light curve belonging to either the primary or secondary eclipse, the model is reduced to the function of five parameters defined by
\begin{equation}
f(t_{i},\mathbf{\theta})=\alpha_{0}+\alpha_{1}\psi(t_{i},t_{0},d,\Gamma)
\label{eq:miku}
\end{equation}
where \(\alpha_{0}\) is the magnitude zero-point shift (i.e. the relative flux baseline level in our study) and \(\alpha_{1}<0\) is a negative multiplication constant of eclipse profile function, i.e.,
\begin{equation}
\psi(t_{i},t_{0},d,\Gamma)=1-\bigg\{1-\exp\bigg[1-\cosh\bigg(\frac{t_{i}-t_{0}}{d}\bigg)\bigg]\bigg\}^{\Gamma}.
\label{eq:Phi}
\end{equation}
Note that \(t_{0}\) is the time of the minimum of an eclipse, $d>0$ is the minimum width and $\Gamma>0$ is the parameter specifying the pointedness of the minimum such that $\Gamma>1$ corresponds to the flat minimum associated with a total eclipse. The procedures of timing eclipse minima and measuring ETVs are presented in detail in Section \ref{sec:etv_measure}. 

\section{Sample Selection}
\label{sec:sample_sel}
\begin{figure}
\centering
\includegraphics[width=0.45\textwidth]{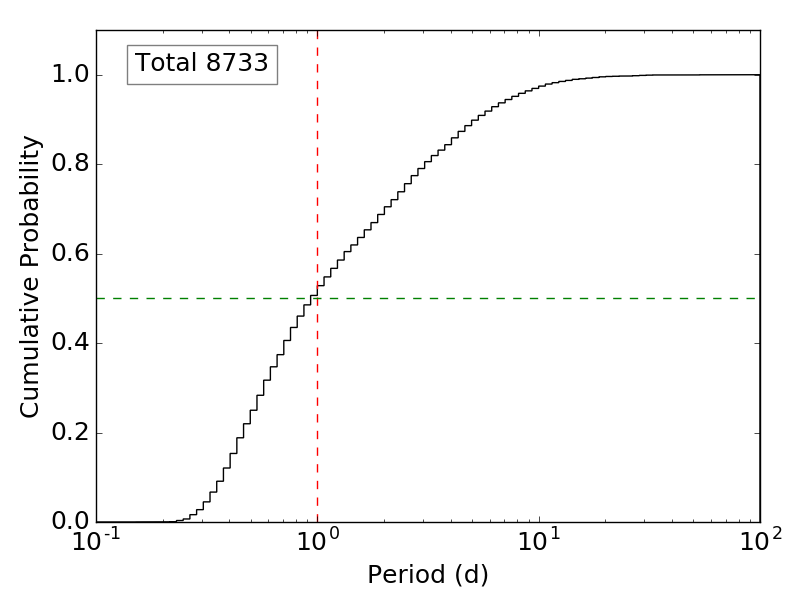}
\caption[Cumulative distribution of period for the
MOA EB candidates]{Cumulative distribution of period for 8733 EB candidates identified in the work of \cite{2017MNRAS.470..539L}. The black curve shows the cumulative distribution of period from $0.1$ day to $100$ days for all MOA EBs identified in the GB9 and GB10 fields. The horizontal dashed line (green) marks the level of the cumulative probability equal to 0.5, while the vertical dashed (red) line marks the period of 1 day.}
\label{fig:cul_eb_pd}
\end{figure}
MOA-II has adopted an observational strategy in which most of its time in the sky is dedicated to routinely monitor the same fields towards the GB every clear night with high cadences. This gives MOA-II an advantage over other microlensing surveys to obtain eclipse time measurements from short period EBs that would be frequent enough to reveal any short period ETV. From the preliminary investigation of \citet{2017MNRAS.470..539L}, we know that there are typically $\sim100$ eclipse time measurement points for the MOA EBs of periods shorter than a day, while this number declines to $\sim19$ for the MOA EBs of periods between 2 and 10 days. This implies that the MOA data should be useful for stellar companion detection and the study of the frequency of tertiary companions in contact or semi-detached binaries. The detection of EBs in the MOA data base was also strongly biased towards EBs of periods $<1$ day. About $50\%$ of the MOA EBs are of periods $<1$ day (see Figure~\ref{fig:cul_eb_pd}). Interestingly, almost all MOA EBs of periods $<1$ day are contact binaries. This implies that the search for tertiary companions in the MOA EBs of periods $<1$ day is equivalent to studying the frequency of contact binaries with tertiary companions. Concerned with the homogeneity of the sample, we focused on studying the MOA EBs of periods $<2$ days and attempted to obtain the full time series of the MOA EBs within this period range. There are over 4000 EBs in the MOA EB catalogue which are of periods shorter than 2 days. However, generating light curves from the full MOA data base is expensive in terms of computational time and data storage space. Therefore, we further restricted our study to two subfields, GB9-9 and GB10-1, from which the full light curves could be generated easily. 542 EBs from the GB9-9 and GB10-1 fields fell into the period range between 0.22 and 2 days and no EB from these two subfields is of period $<0.22$ days.


\section{Observation and Data Reduction}
\label{sec:obs_dat}
The MOA project is a Japan and New Zealand collaboration which began in 1995 and started the second stage of the project in 2006 with a 1.8m telescope located at the University of Canterbury Mount John Observatory, New Zealand. The MOA-II telescope is equipped with the MOA-cam3 wide-field camera which consists of ten \(2\,\text{k}\,\times\,4\,\text{k}\) pixel CCDs with \(15\,\mu\)m pixels and provides a field of view (FOV) of 2.18 deg$^{2}$ given a pixel scale of 0$_{.}^{''}$58 pixel$^{-1}$. The primary mission of the MOA project is always hunting exoplanets via microlensing. For this purpose, it has adopted a special observational strategy that the telescope times are spent mainly for the survey towards 22 fixed fields of the GB. Images of these 22 fields were taken with cadences between 10 minutes to 1 hour through the custom MOA-Red wide-band filter which spans from \(600\,\) to \(900\,\)nm. In each field, there are ten subfields corresponding to ten CCD chips.

The data sets we obtained from the GB9-9 and G10-1 fields span 9.5 years and were collected from February to November every year since 2006. The image reduction was done following the same procedures described in \citet{2017MNRAS.470..539L} using the difference imaging analysis (DIA) method \citep{2001MNRAS.327..868B}. The density of the light curves from the GB9-9 field is approximately uniform, while the density of the light curves from the GB10-1 field is low during the period of the first two years, although big gaps exist as expected due to the off-season periods. The exposure time of 60s was taken for both fields over the entire observational time span. An observation time was recorded in Julian Day and calculated to be the middle between the start and end times of an exposure. 

\begin{figure*}
\centering
\includegraphics[width=0.45\textwidth]{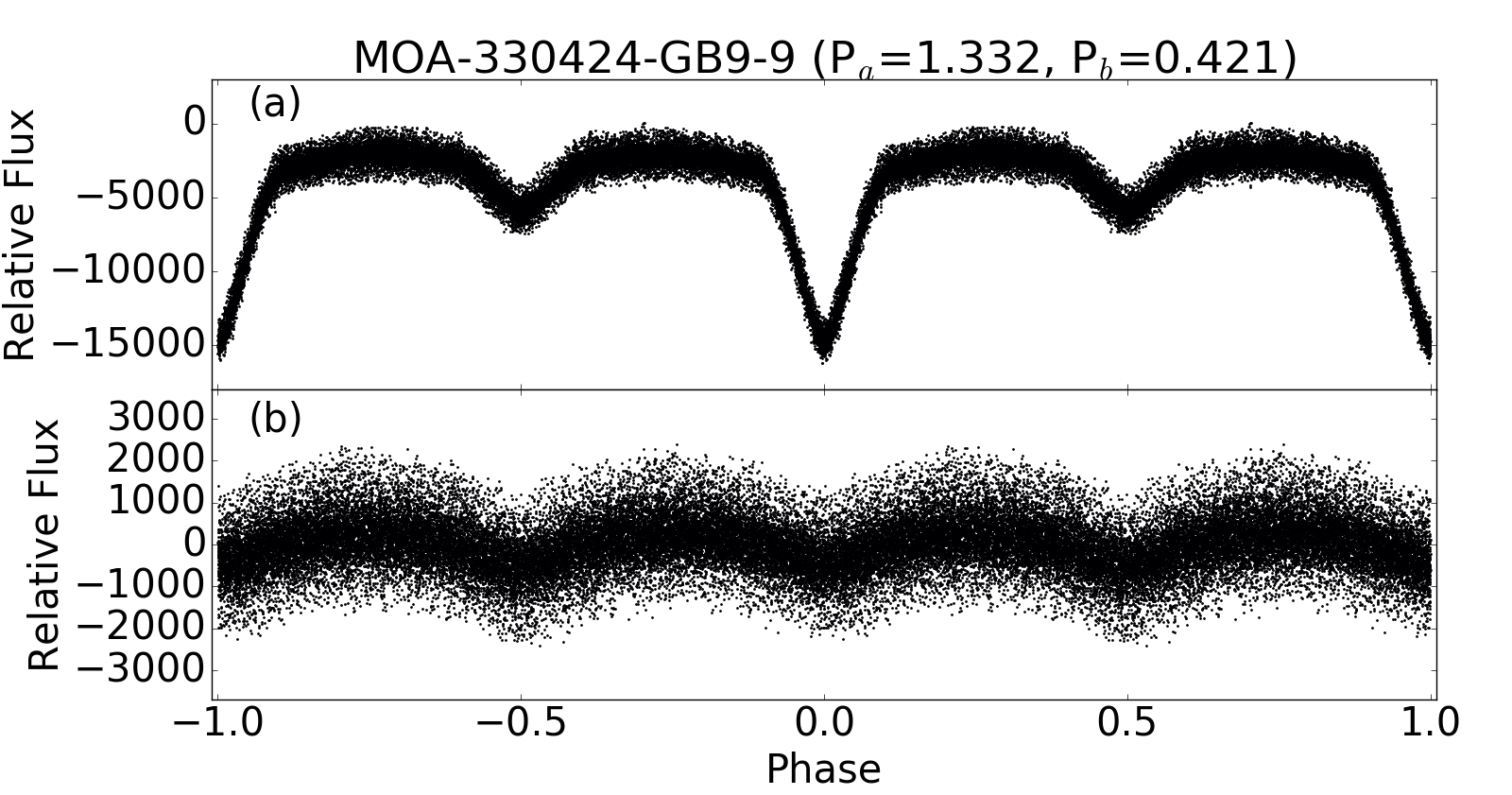}
\includegraphics[width=0.45\textwidth]{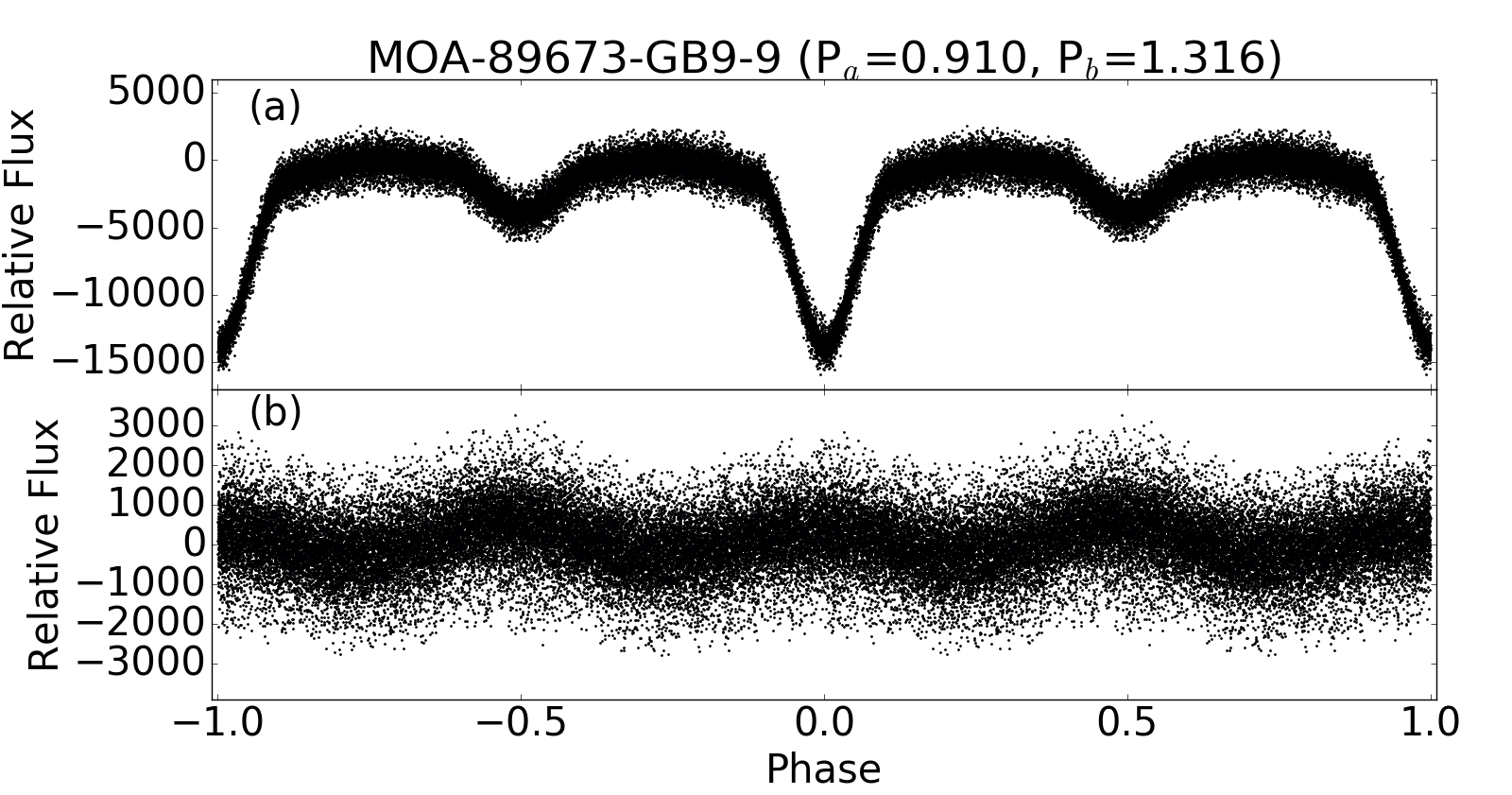}
\includegraphics[width=0.45\textwidth]{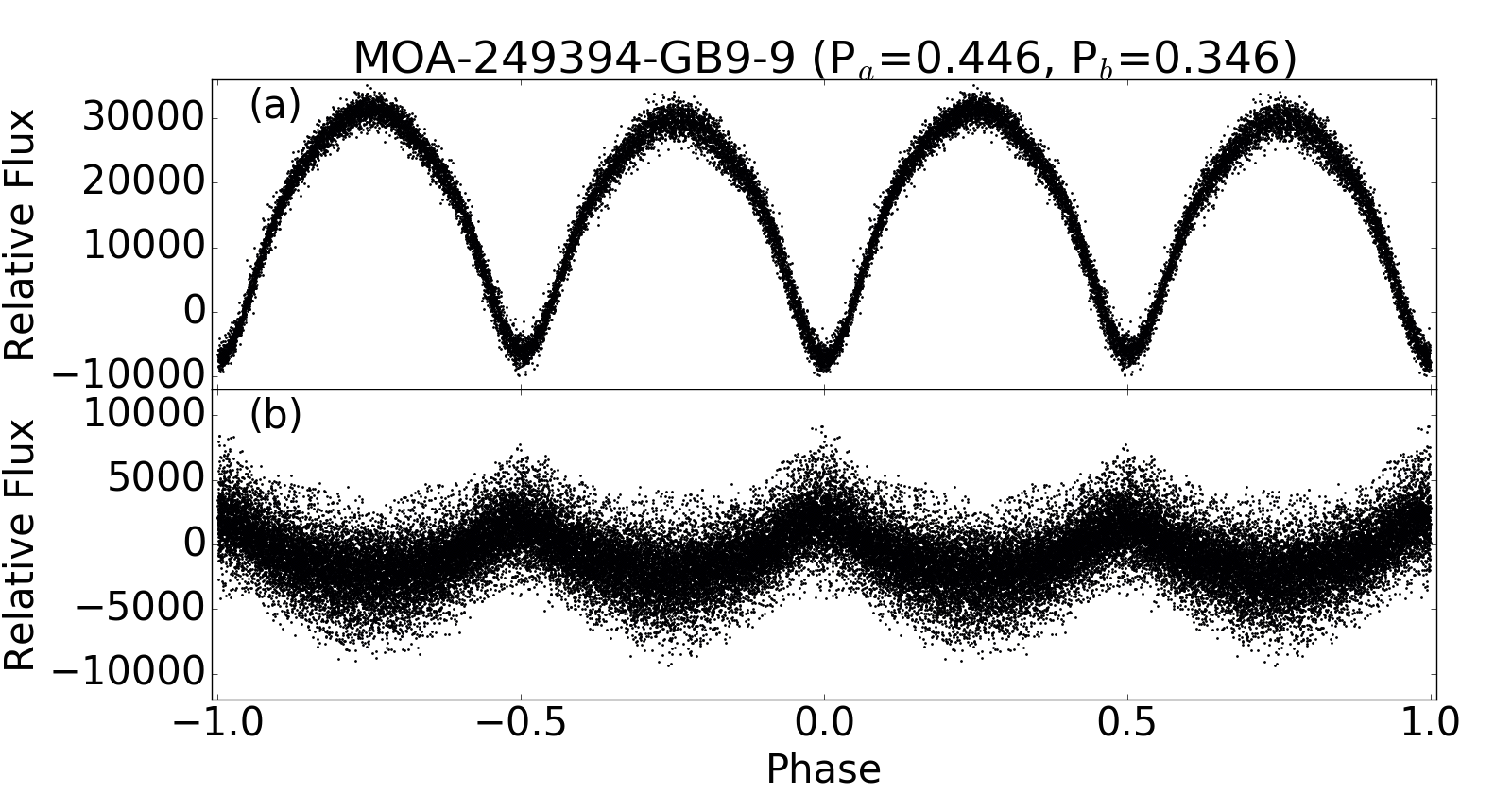}
\caption[Folded light curves of the three EBs in our samples with additional regular periodic signals.]{Folded light curves of the three EBs in our MOA samples from the GB9-9 and GB10-1 fields which were discovered to have additional regular periodic signals under their main eclipsing signals. (a) The main eclipsing signals. (b) The additional periodic signals. The secondary periodic signals in MOA-330424-GB9-9 and MOA-89673-GB9-9 are the eclipsing and ellipsoidal variation curves, respectively, likely associated with EBs near them. However, the source of the secondary periodic signal in MOA-249394-GB9-9 is uncertain. We suspect that it might be an artifact from imperfect image subtraction in difference image analysis owing to a bright variable star close to the EB in the images.}
\label{fig:double_ebs}
\end{figure*}

\section{Analysis}
\label{sec:analysis}
\subsection{Period Analysis}
\label{sec:pd_analysis}
In the beginning, all the light curves of our EB samples were cleaned following the light curve procedure same as in \citet{2017MNRAS.470..539L}. In a nutshell, we discarded outliers that are \(4.0\sigma\) above or \(9.0\sigma\) below the relative flux mean as well as detrended the light curves via linear regression. This cleaning procedure was iterated twice before going into the light curve analysis. Meanwhile, we corrected the times from Julian Day (JD) to Barycentric Julian Day (BJD). Despite the time span of over 9 years, we did not divide the light curves into segments with shorter time spans in general, except several cases for which careful treatments in eclipse time measurement were needed (see Section \ref{sec:etv_measure}).

Since the MOA fields towards which our EBs are located are densely populated, blending with nearby stars might be present. On the other hand, stellar pulsations might be present in our EBs; particularly, a component being Cepheid or RR Lyrae, which would pulsate regularly with a period comparative to the eclipse duration, will distort an eclipse shape, causing the measurement of the time of eclipse minimum to be inaccurate. Because of these problems, we attempted first to search for an additional eclipsing or hidden pulsation signal under the main eclipsing signal. To do so, we first determined the average eclipsing period with which the 9.5 year light curve could be properly folded using conditional entropy with trial periods $P_{s}^{\prime}\pm0.01$, where $P_{s}^{\prime}$ is the average eclipsing period over two MOA observational seasons provided in the MOA EB catalogue. Once the new average eclipsing period was determined, and after checking the resultant folded light curve by eye to see if it was folded properly, we binned the folded light curve in 200 bins, and created an approximate curve by calculating the mean flux value in each bin. We then produced the residual curve by subtracting the approximate curve from the folded light curve and unfolded it afterwards. The residual curve was then put through the period analysis by the condition entropy algorithm with trial periods ranging from 0.05 to 600 days. The residual curve folded with the output period was inspected by eye afterwards. 

\begin{figure*}
\centering
\includegraphics[width=0.32\textwidth]{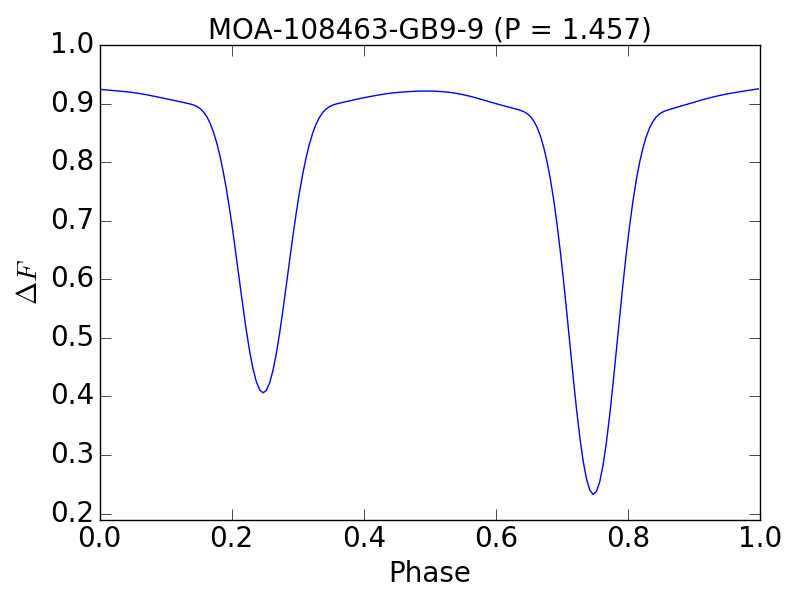}
\includegraphics[width=0.32\textwidth]{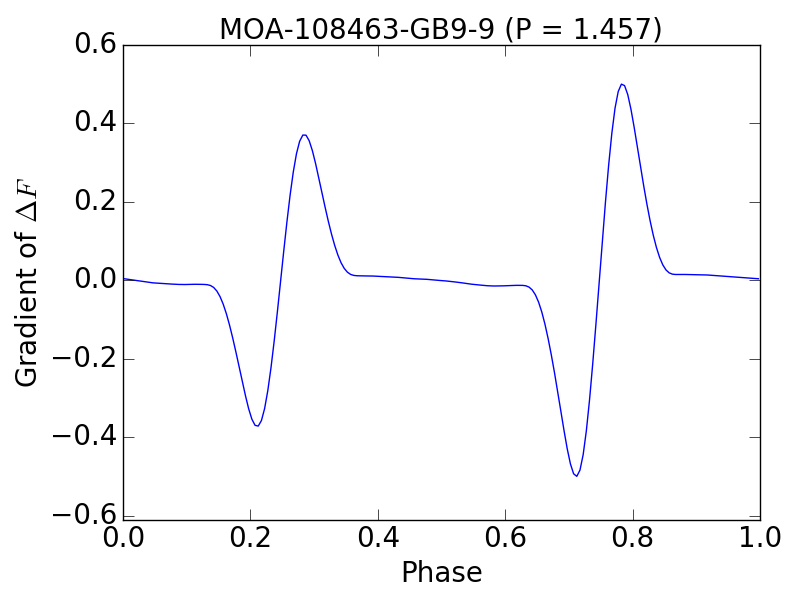}
\includegraphics[width=0.32\textwidth]{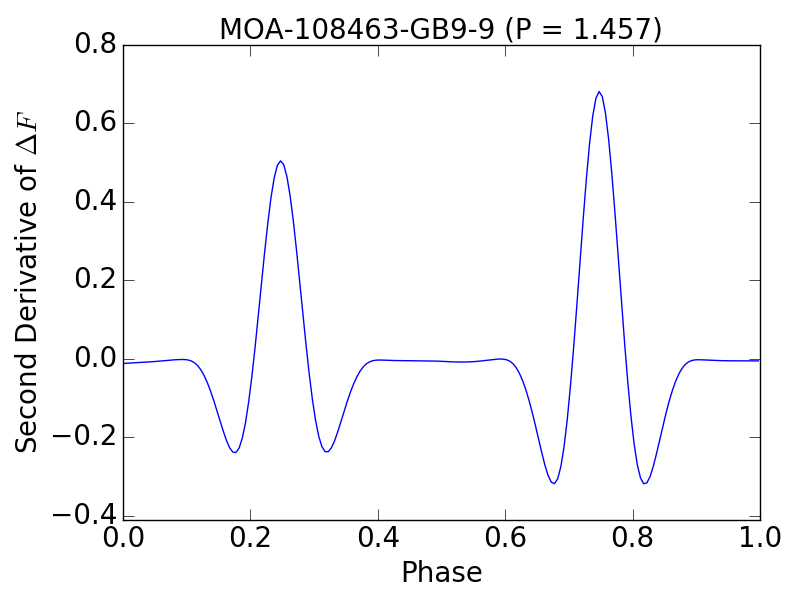}
\includegraphics[width=0.32\textwidth]{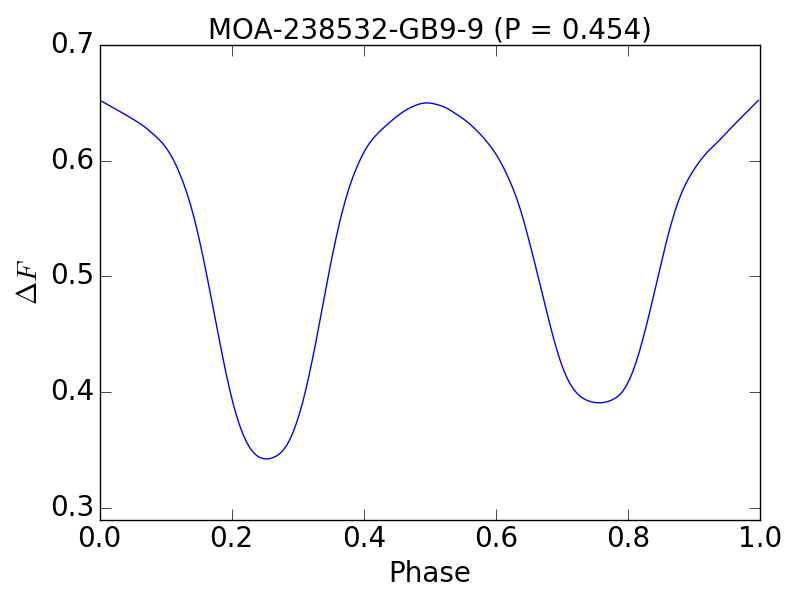}
\includegraphics[width=0.32\textwidth]{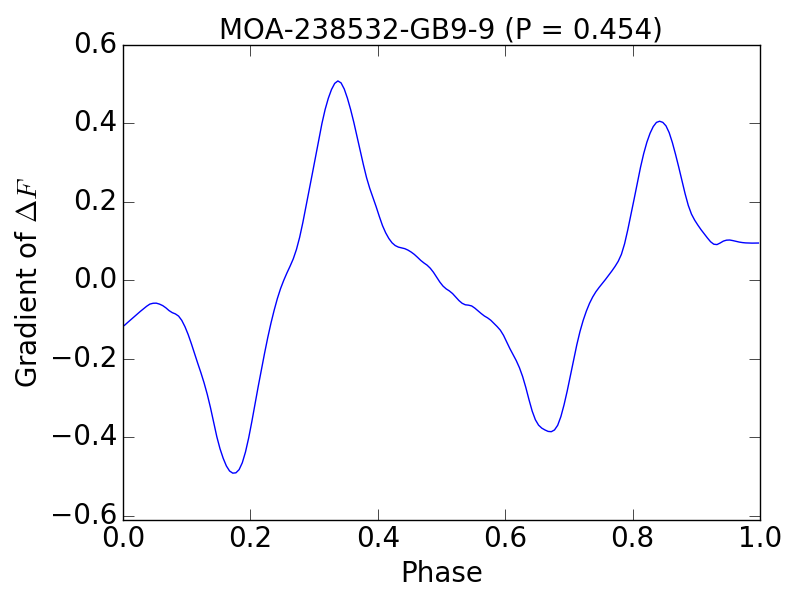}
\includegraphics[width=0.32\textwidth]{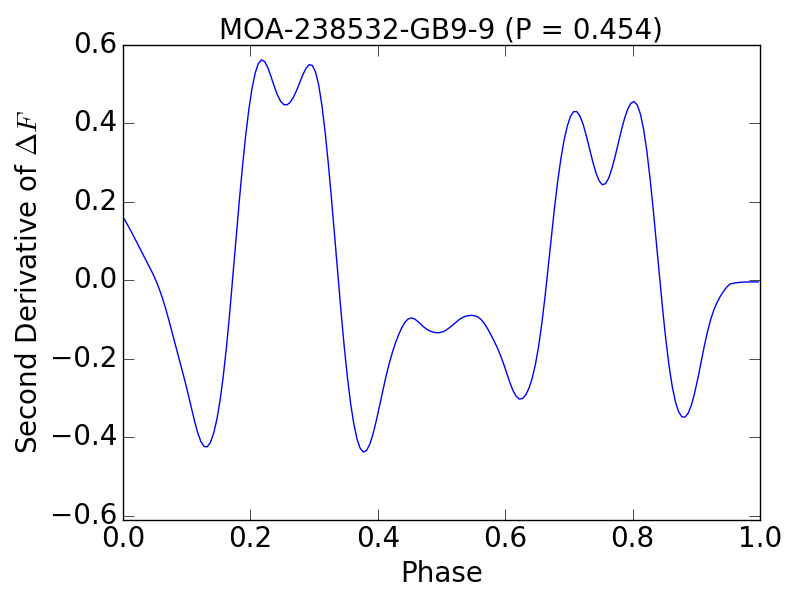}
\includegraphics[width=0.32\textwidth]{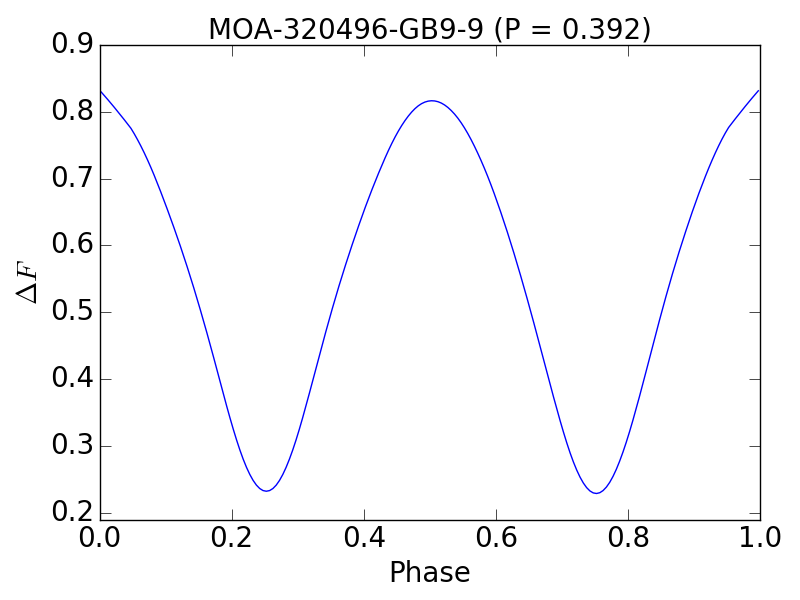}
\includegraphics[width=0.32\textwidth]{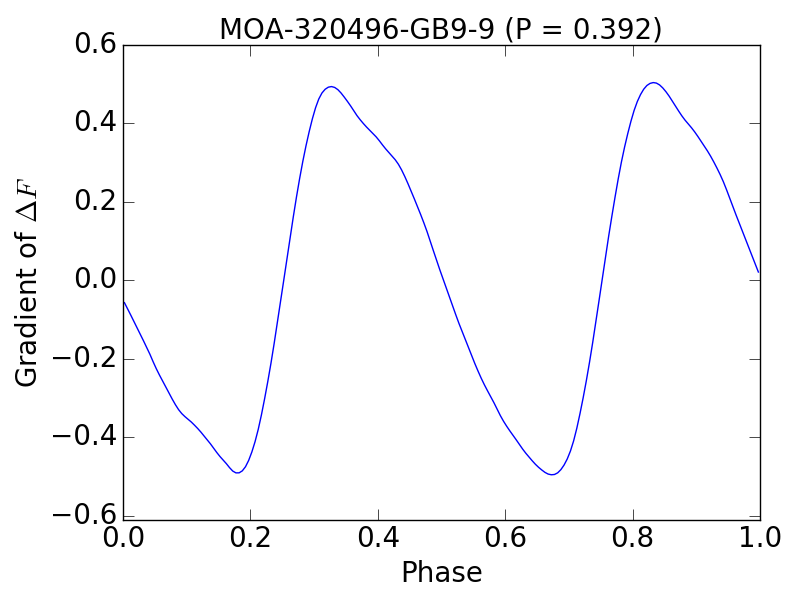}
\includegraphics[width=0.32\textwidth]{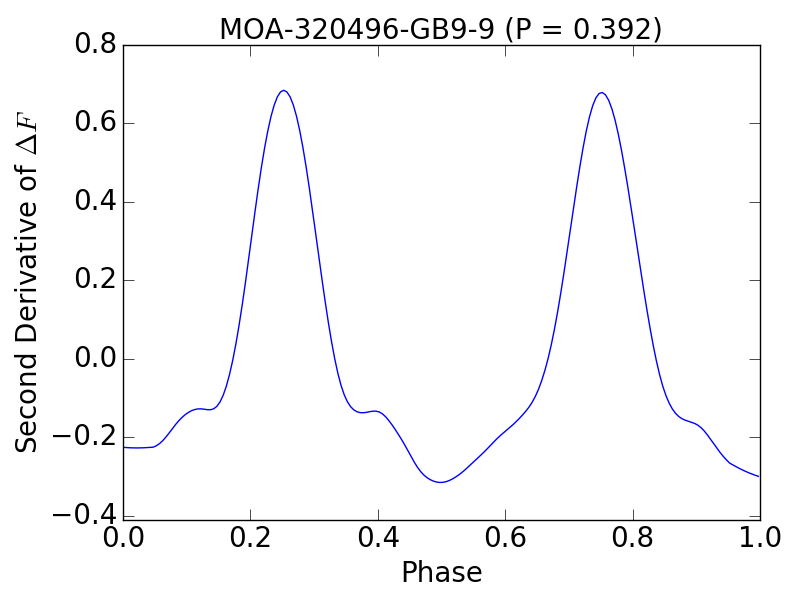}
\caption[Mean light curves, first derivative curves, second derivative curves]{Mean light curves (left), first derivative curves (middle) and second derivative curves (right) of three representative MOA EBs. The points of ingress and egress of an eclipse are determined by calculating a pair of minima in the second derivative curve that can well define the boundaries of the eclipse; i.e. the pair of minima of the Mexican hat feature in the second derivative curve. In few cases, no pair of minima in the second derivative curves corresponding to the ingress and egress of their primary or secondary eclipses could be derived and we used the maxima and minima in the first derivative curves to define the eclipse regions; for example, in the case of MOA-320496-GB9-9, we could not derive a pair of minima in the second derivative curve that could be used to define the boundaries of the primary eclipse, and thus the phases of the maximum and minimum in the first derivative curve were used to represent the primary eclipse's boundaries.}
\label{fig:deriv_curve}
\end{figure*}

In this manner, we discovered three residual light curves that exhibited periodic signals. Figure~\ref{fig:double_ebs} shows the main eclipsing and additional periodic signals of these three EBs. The additional periodic signals in MOA-330424-GB9-9 and MOA-89673-GB9-9 are obviously associated with an EB with period of 0.421 days and an ellipsoidal binary with period of 1.316 days, respectively, while we suspect the additional periodic signal in MOA-249394-GB9-9 was an artifact due to contamination by a nearby pulsating bright star or a bright EB that caused imperfect image subtraction in the DIA. The detected additional signals were subtracted from the original light curves, and the average eclipsing periods were recalculated after subtraction.

\subsection{Eclipse Time Measurement}
\label{sec:etv_measure}
To measure the times of eclipse minima, the template method using \citet{2015A&A...584A...8M}'s model to generate the eclipse templates was applied. The corresponding eclipse regions were determined by calculating the pairs of minima of the second derivative of the folded light curve which corresponds to the ingress and egress phases of the eclipses. If no valid minima were obtained from the second derivative curve, we took a pair of minima of the first derivative curve between which the eclipse minimum is located as the boundaries of the eclipse region. The procedures for identifying the ingress and egress of an eclipse are as follows:

\begin{enumerate}
  \item Derive the mean light curve by binning the folded light curve into 20 bins and calculating the mean flux in each bin.
  \item Derive the first derivative curve by calculating the gradient of the mean light curve using the function \texttt{gradient()} in \texttt{numpy}.
  \item Derive the second derivative curve by calculating the gradient of the first derivative curve using the function \texttt{gradient()} in \texttt{numpy}.
  \item Smooth the curves in each step above using the method of locally weighted scatterplot smoothing (LOWESS) provided in \texttt{statsmodels}, a Python module in statistics, and calculate the phases of maximum and minimum points using the function \texttt{argrelextrem()} in \texttt{scipy}.
  \item Take the pair of minimum points in the second derivative curve that contain the eclipse minimum as the points corresponding to ingress and egress of the eclipse.  
\end{enumerate}

To illustrate the situations for different types of EB light curves, we take MOA-108463-GB9-9, MOA-238532-GB9-9 and MOA-320496-GB9-9 as examples. Their mean light curves and first and second derivative curves are shown in Figure~\ref{fig:deriv_curve}. MOA-108463-GB9-9 is an Algol-type EB as the turning points in its mean light curve corresponding to the ingress and egress of both eclipses can be easily recognized by eye. Its first derivative curve resembles an electrocardiogram, while its second derivative curve contains two Mexican hat features associated with the primary and secondary eclipses. In the case of MOA-238532-GB9-9, instead of yielding typical Mexican hat features which have single peaks at the middles for both eclipses in the second derivative curve, double peaks were produced, indicating the presence of four contact points, which would be present for total eclipsing, in both eclipses. Unsurprisingly, the points of ingress and egress of their eclipses can be easily determined by calculating pairs of minima in their second derivative curves that the minima of their eclipses lie in between accordingly. In the case of MOA-320496-GB9-9 (which seems to be a W UMa binary), we can see that its second derivative curve fails to yield a proper Mexican hat feature for the primary eclipse having only single minimum instead of a pair of minima that would allow us to determine the ingress and egress of the eclipse. Therefore, we took the phases of the maximum and minimum in the first derivative curve instead as the boundary points of the region of its primary eclipse\footnote{In our sample of 542 MOA EBs, there were only a few cases in which we failed to find the ingress and egress of eclipses from the second derivative curves, and they all seemed to be either W UMa EBs or ellipsoidal binaries after inspecting their folded light curves by eye. For them, it might be more appropriate to use the light curves' maxima to define the boundaries of their eclipse regions. Nevertheless, we did not find it would significantly impact the accuracy of our eclipse timing.}. 

\begin{figure*}
    \centering
	\includegraphics[width=0.32\textwidth]{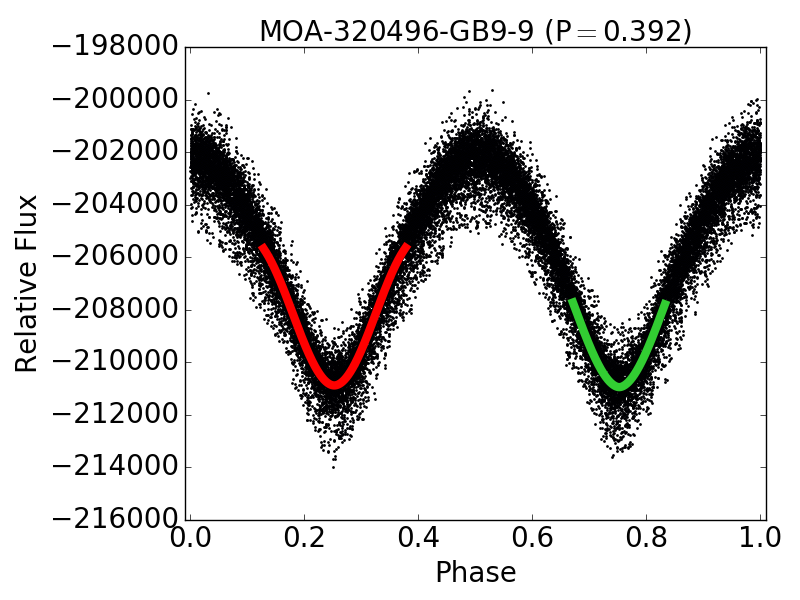}
    \includegraphics[width=0.32\textwidth]{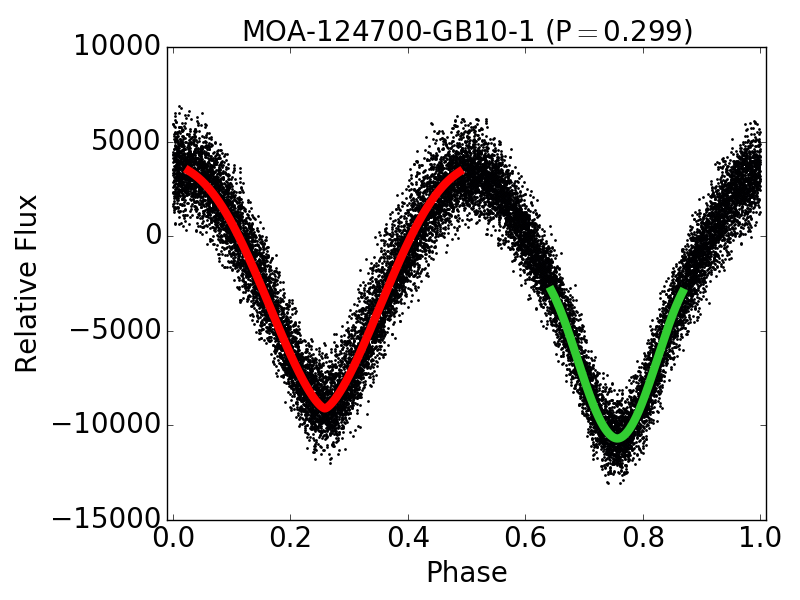}
    \includegraphics[width=0.32\textwidth]{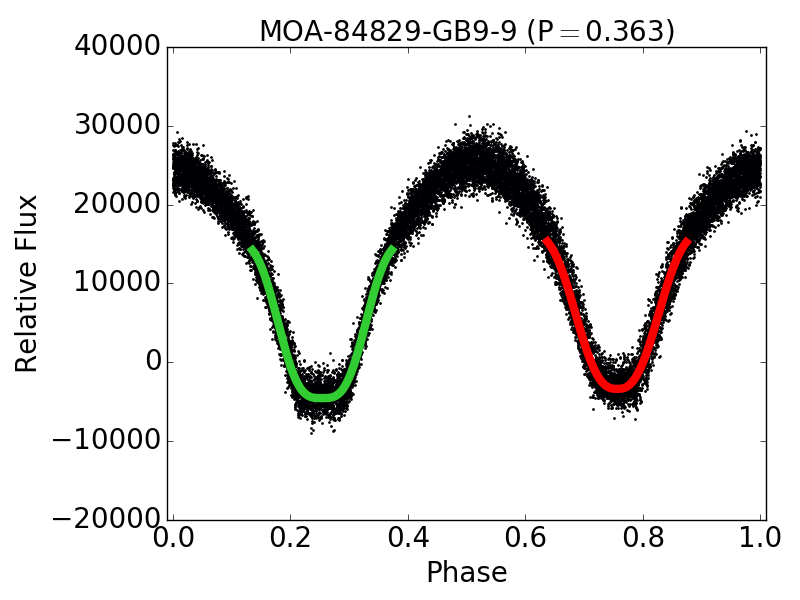}
    \includegraphics[width=0.32\textwidth]{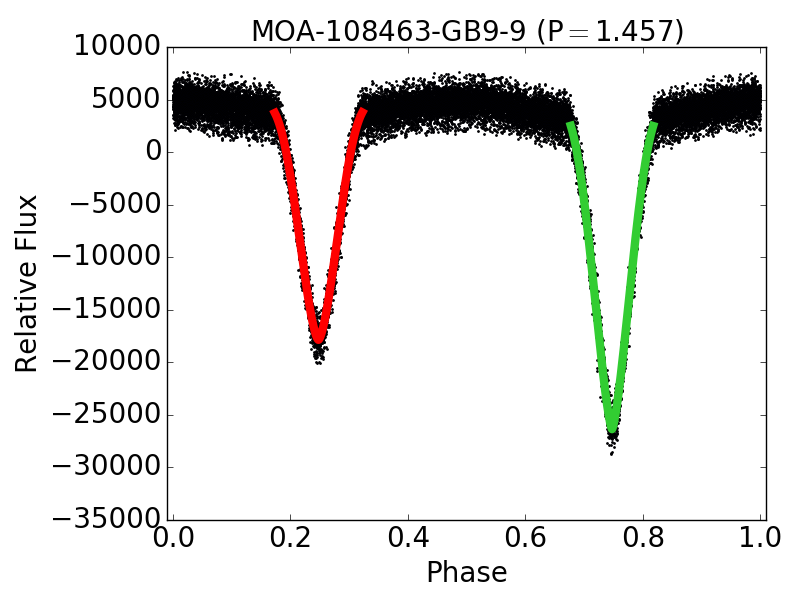}
    \includegraphics[width=0.32\textwidth]{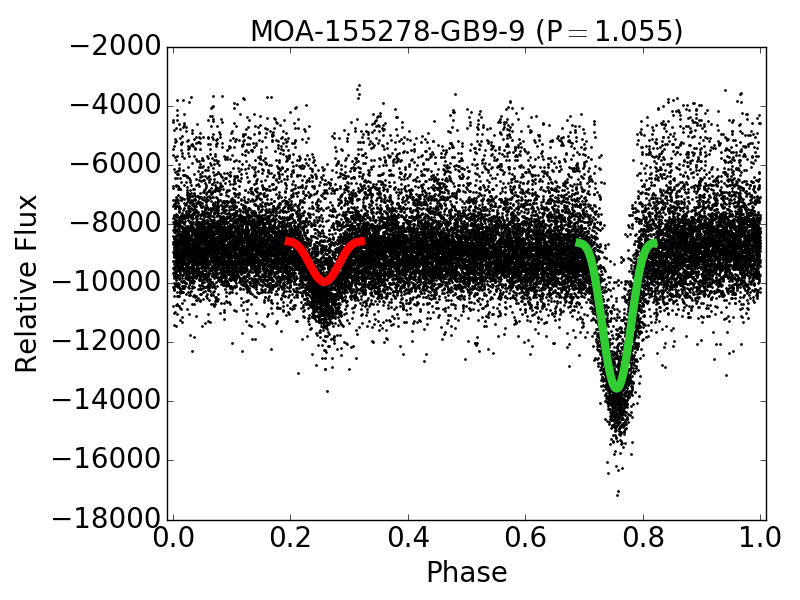}
    \includegraphics[width=0.32\textwidth]{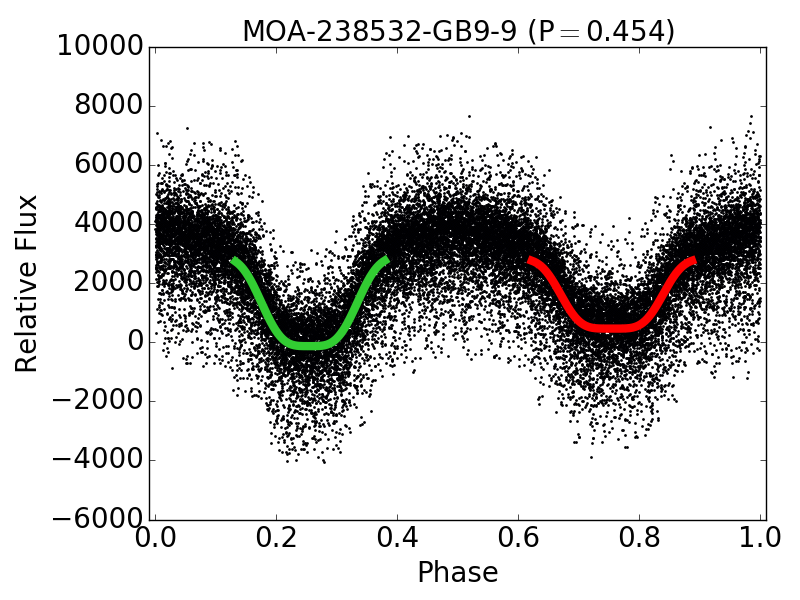}
    \caption[Templates of representative MOA EB light curves]{Templates of representative MOA EB light curves. The green and red curves represent the templates for primary and secondary eclipses, respectively. The templates were generated by fitting eq.(\ref{eq:miku}) to the eclipse regions of the folded light curves.}
    \label{fig:template_examples}
\end{figure*}

\begin{figure*}
\centering
\includegraphics[width=.45\textwidth]{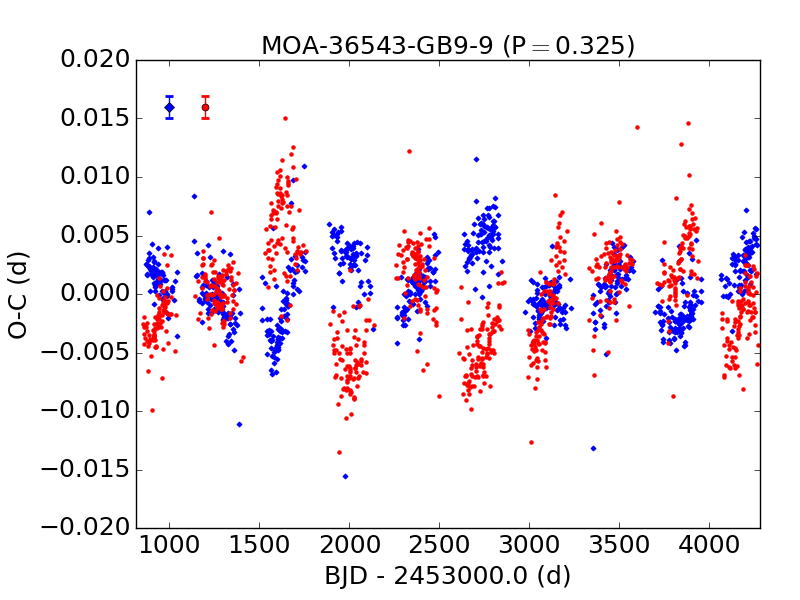}
\includegraphics[width=.45\textwidth]{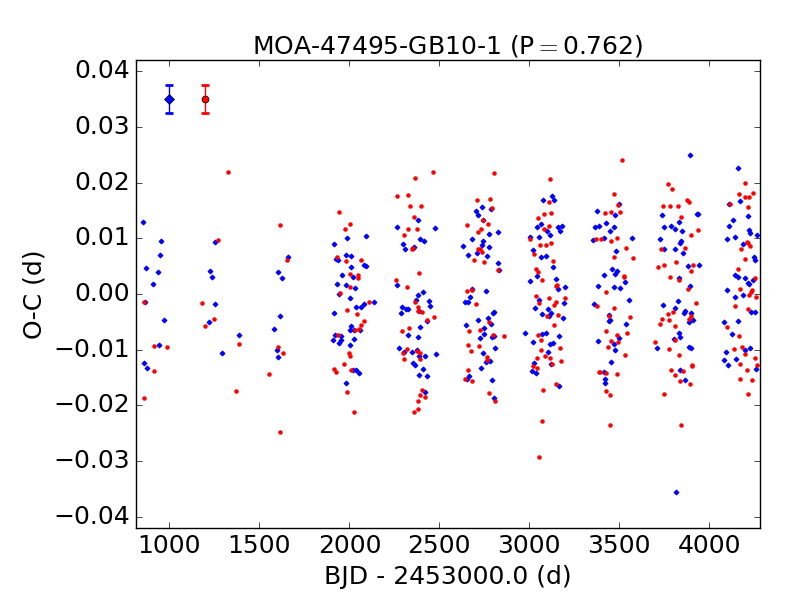}
\caption[ETV curves of MOA-289148-GB9-9 and MOA-351777-GB9-9]{Examples of the O$-$C diagrams of the MOA EBs with spurious ETVs. (left) The O$-$C diagram of MOA-36543-GB9-9, in which the ETV curves for the primary and secondary eclipses vary cyclically and behave anti-correlated to each other. (right) The O$-$C diagram of MOA-47495-GB10-1, which appears to have two separate curves or rapidly oscillating ETVs. The blue (diamond) points are the ETV measurements of the primary eclipses and the red (circle) points are those of the secondary eclipses. The average uncertainties for the primary and secondary eclipses are represented by the red and blue error bars, respectively, on the top-left corner of each figure. Note that the periods are in days.}
\label{fig:etv_antic_oscil}
\end{figure*}

Once the boundaries of the eclipse region were determined, the best-fit template was derived by fitting \citeauthor{2015A&A...584A...8M}'s model, i.e., eq.(\ref{eq:miku}) to the portion of the folded light curve between the boundaries using \texttt{emcee}, a Python implementation of the affine-invariant ensemble sampler for Markov chain Monte Carlo (MCMC) \citep{2013PASP..125..306F}. The reference epoch, $T_{0}$ in eq.(\ref{eq:etv}), in regard to the derived template, was thus defined accordingly as $T_{0}=\phi_{0}P_{s}+\tau_{0}$, where $\phi_{0}$ is the phase of the template minimum with respect to the time zero, $\tau_{0}$, which we set to be $\tau_{0}<t_{\text{obs}}$, where $t_{\text{obs}}$ is an observation time, such that both primary and secondary eclipses are not cut in phase when we folded the light curve with respect to $\tau_{0}$. The best-fit templates of six EBs in our MOA samples are shown in Figure~\ref{fig:template_examples} as examples to demonstrate the usefulness of our template generating method for different shapes of eclipses. Although we adopted \citeauthor{2015A&A...584A...8M}'s model, i.e. eq.(\ref{eq:miku}) is supposed to work for eclipse portions of detached EBs' light curves, it still generated templates which represent eclipses of contact binaries practically well even if the eclipse boundaries derived by our algorithm turned out to be located at or close to the maxima of the light curves, as verified by the case of MOA-124700-GB10-1 (see Figure~\ref{fig:template_examples}). Once the template was generated, we unfolded the light curve, and then fitted each eclipse which had at least four data points across the eclipse minimum with the template. The general idea of the template method is to obtain the time of the eclipse minimum by shifting the template horizontally until the template best fits the eclipse. In reality, however, the brightness of a star may vary over time, and hence the template parameters $t_{0}$ as well as $\alpha_{0}$ and $\alpha_{1}$ were required to vary as well to search for the best fit. Again, the best-fit parameter search was executed using \texttt{emcee}. As a result, the time of the eclipse minimum was determined by the median of the projected posterior on $t_{0}$. The uncertainty in $t_{0}$ was taken as the 1-$\sigma$ confidence interval from the median. 

The eclipse timing process described in the previous paragraph worked properly for most of the MOA EBs we studied. There are, however, six MOA EBs we could never derive periods with which they could be folded satisfactorily. Such a problem indicated that these EBs suffered significant ETVs. As improper folding could induce significant errors in the derived eclipse templates, we thus divided each of these EBs' light curves into three segments in which the first two segments evenly span the first seven years and the third segment spans the last 2.5 years. We then calculated the average eclipsing period for each segment and generated the corresponding templates, and performed the eclipse time measurement following the same process as mentioned in the previous paragraph.

We ignored the measurement points with very large uncertainties and inspected the resultant ETV curves afterwards. As expected, the ETV curves of shorter period EBs are generally denser than those of longer period EBs given that there were more eclipsing cycles for the shorter period EBs. Cyclic or quasi-cyclic variations on the time scale shorter than two MOA observation seasons ($\approx 2$yr) were noticeable in the O$-$C diagrams of several samples. A few of them exhibit quasi-periodic ETVs for their primary and secondary eclipses which are highly anti-correlated, for example, MOA-36543-GB9-9 (see Figure~\ref{fig:etv_antic_oscil}). Such anti-correlated ETVs were supposed to be due to the presence of active star spots \citep{2013ApJ...774...81T}. Scattering of the ETV points comparable to the average error bars on the time scale of one MOA observational season was common in the O$-$C diagrams of our sample. In some cases, the ETV points dispersed such that the O$-$C diagrams seemed to have two separate curves, or exhibit very rapid oscillations on very short time scales ($<100$ days), e.g., MOA-47495-GB10-1 (see Figure~\ref{fig:etv_antic_oscil}). Orbital perturbations due to very short period tertiary companions could produce oscillating ETVs, which have been observed in the \textit{Kepler} triples \citep{2016MNRAS.455.4136B}. However, we also suspect that this kind of ETVs might be spurious arising from stellar oscillations or pulsations \citep{2014MNRAS.443.3068B}. Given the frequency of and accuracy in eclipse timing from the MOA data, the proper coverage of a ETV cycle shorter than 200 days was expected to be unachievable. In order to avoid false detection of short period LTTE cycles due to over-fitting scattering of ETV points or possible spurious ETVs, we restricted the search for the LTTE cycles which are of periods longer than 200 days. We also ignored the EBs in the sample with highly anti-correlated ETV cycles without any evident long-term variation for the LTTE analysis. Further discussion about this issue is presented in Section~\ref{sec:ltte_results}.

\subsection{LTTE Analysis}
\label{sec:ltte_analy}
The O$-$C diagram of each EB was constructed according to eq.(\ref{eq:etv}) with $P_{s}$ being the average eclipsing periods over the full data time span and $T_{0}$ being the time of the eclipse template minima, except the six special EBs mentioned in Section \ref{sec:etv_measure} that the average eclipsing periods and the times of the eclipse template minima associated with the segments of the first 3.5 years were used instead. After the preliminary inspection, we decided to discard the ETV measurement points with uncertainties $>0.01$ days (except MOA-222739-GB9-9 for which we accepted the ETV measurement points with uncertainties up to $0.02$ days instead). Then we fitted the LTTE model, including or excluding the quadratic term of $E$, to the primary and secondary ETV curves simultaneously, using \texttt{pymc} \citep{2013MNRAS.429.1981B}, another Python module of MCMC fitting algorithms. For consistency between the calculations of the polynomial terms of $E$ in eq.(\ref{eq:ltte}) for primary and secondary eclipses, we added the phase difference between the minima of primary and secondary eclipse templates to the cycles, $E$, of the eclipse which is located in the second half of the folded light curve. The LTTE term in eq.(\ref{eq:ltte}) depends implicitly on $P_{2}$ and $\tau_{2}$ via the true anomaly, $\nu_{2}$, which must be calculated by solving Kepler's equation iteratively using a numerical method. We used Halley's method (see, e.g., \citealt{2009ebs..book.....K}) in our study. The calculation of $\nu_{2}$ in fact caused serious speed issues in the parameter search using \texttt{pymc}. To improve the computational speed, the calculation of $\nu_{2}$ was done using the code written in \texttt{Cython} instead of \texttt{Python}/\texttt{numpy}.

\begin{table}
\centering
	\caption[Boundaries of the parameters of the ETV model]{Boundaries of the parameters of the ETV model eq.(\ref{eq:ltte}). Note that $d$ is day, $d/c$ is day/cycle and AU is astronomical unit.}
    \bgroup
    \def\arraystretch{0.95}
    \small
	\begin{tabular}{lcc} 
		\hline
        Parameter (unit) & Lower & Upper\\
        \hline
        $c_{0}$ $(d)$ & -0.1 & 0.1 \\
        $c_{1}$ $(d)$& -0.1 & 0.1 \\
        $c_{2}$ $(d/c)$ & -0.1 & 0.1 \\
        $\log(P_{2})$ $(d)$ & $\log(P_{2}/2)$ & $\log(2P_{2})$\\
        $e_{2}$ & 0 & 0.999\\
        $\omega_{2}$ & 0 & 2$\pi$\\
        $\tau_{2}$ & 0 & 1\\
        $a_{\text{AB}}$ $(\text{AU})$ & 0 & 100\\
        \hline
		\hline
	\end{tabular}
    \egroup
    \label{tab:bd_para}
\end{table}

\begin{figure*}
\centering
\includegraphics[width=.45\textwidth]{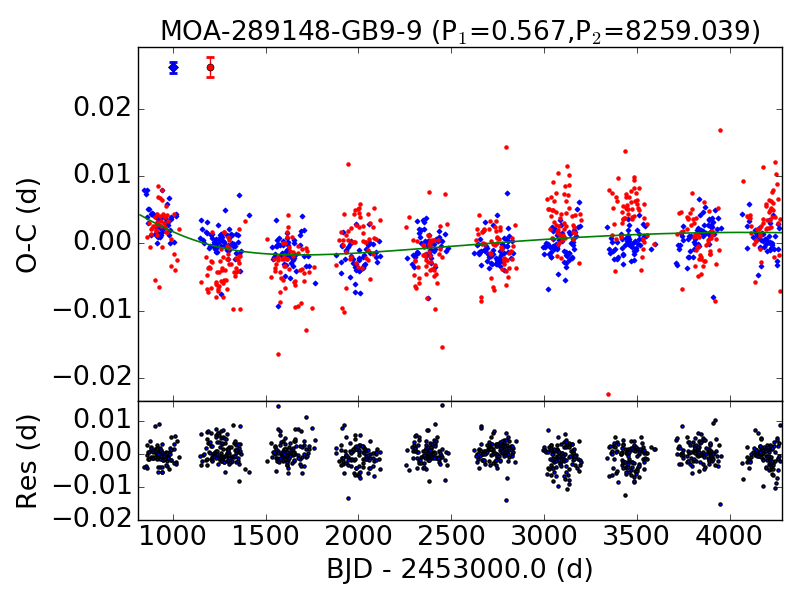}
\includegraphics[width=.45\textwidth]{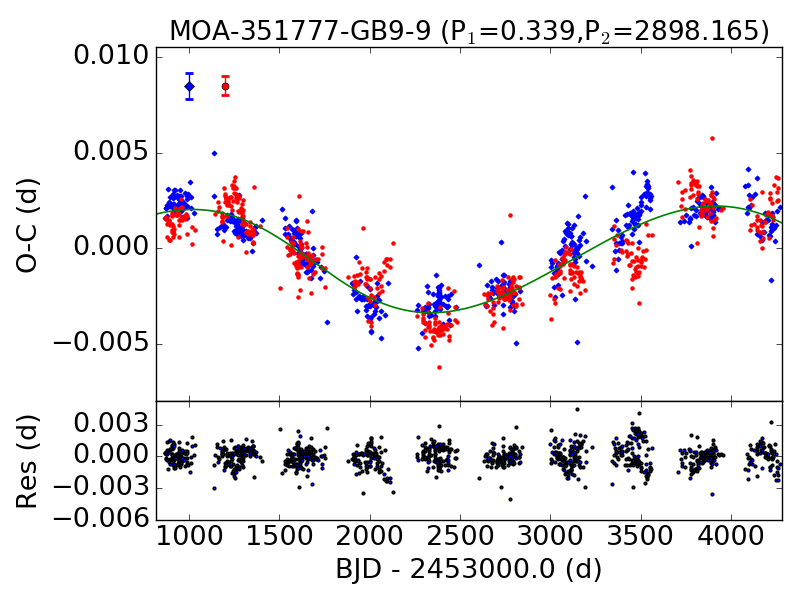}
\caption[ETV curves of MOA-289148-GB9-9 and MOA-351777-GB9-9]{ETV curves of MOA-289148-GB9-9 and MOA-351777-GB9-9. $P_{1}$ is the period of the inner binary determined by the conditional entropy method, while $P_{2}$ is the period of the tertiary companion given by the LTTE solution. The blue (diamond) points are the ETV measurements of the primary eclipses and the red (circle) points are those of the secondary eclipses, while the green lines represent the best fits of the ETV model defined by eq.(\ref{eq:ltte}). The bottom panels show the residual curves. Anti-correlated behaviours between the primary and secondary ETV curves are seen on the time scale of a year, while the long-term trends of both curves are consistent. As can be seen in the residual curve of MOA-351777-GB9-9, for example, there is significant deviation between the trends of the primary and secondary ETV curves during the eighth MOA observational season; however, the ETV curves also exhibit consistent long-term variations. The uncertainties for the primary and secondary eclipses are represented by the red and blue error bars, respectively, on the top-left corner of each figure. Note that the periods are in days.}
\label{fig:etv_spot_eg}
\end{figure*}

The Metropolis-Hastings algorithm is used in \texttt{pymc} for distribution sampling. We adopted the built-in normal distribution function in \texttt{pymc} as the likelihood function and assumed a uniform prior for each parameter over the boundaries that we assumed to be appropriate (see Table~\ref{tab:bd_para}). After testing the model fitting algorithm, we realized that the likelihood function might not be able to converge, or it might converge incorrectly to a local minimum, if the initial guess of the outer period value was not close to the true value. The difficulty in having a good guess of $P_{2}$ happened particularly when only the partial LTTE cycle was observed. Concerning these problems, and with the usage of the New Zealand eScience Infrastructure (NeSI) high performance computing facilities, the parameter search was carried out over a set of initial values of $P_{2}$ as long as we had no confident estimation of the value of $P_{2}$ by eye. For a ETV curve with a potential LTTE signal of period longer than 3000 days, for example, we ran the model fitting with initial values of $P_{2}$ from 2000, 3000, 4000, 5000, 6000, 7000, 8000 and 10000 days, respectively. For convenience, the search was over the $\log(P_{2})$ space instead, bounded between $\log(P_{2}/2)$ and $\log(2P_{2})$. The initial guess values of other orbital parameters including $e_{2}$, $\omega_{2}$ and $\tau_{2}$ were taken to be the middles of their boundaries in principle, while the projected semi major axis of the absolute orbit of a tertiary companion, i.e., $a_{\text{AB}}\sin i_{2}$, in AU was set to be 0.5 as the initial guess value based on the properties of the \textit{Kepler} triple candidates discovered by \citeauthor{2016MNRAS.455.4136B} (\citeyear{2016MNRAS.455.4136B}) which are typically of $P_{2}<4$ years and $a_{\text{AB}}\sin i_{2}$ < 1$\,$AU.

\begin{figure}
\centering
\includegraphics[width=0.45\textwidth]{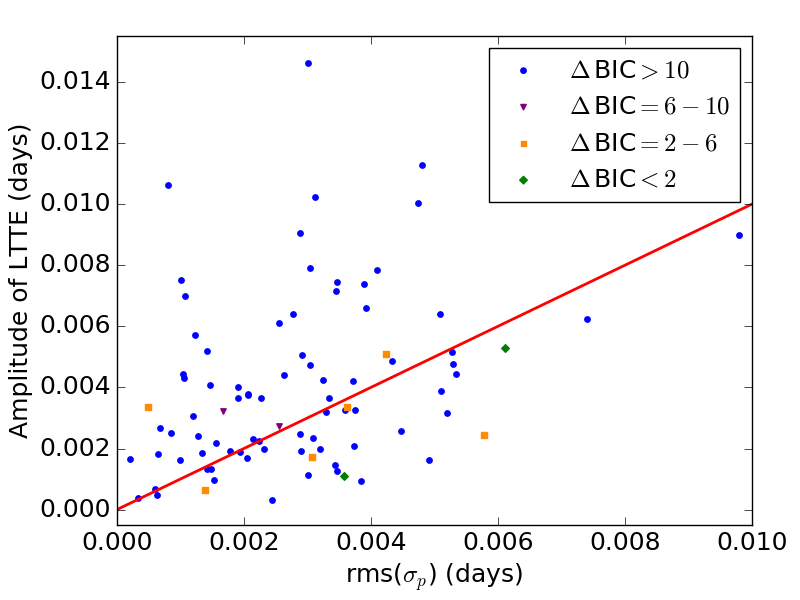}
\caption[Amplitude of LTTE vs root mean square of uncertainty in eclipse timing for primary eclipses]{Amplitude of LTTE, $A_{\text{LTTE}}$, vs root mean square of uncertainty in eclipse timing for primary eclipses, rms($\sigma_{p}$), for 91 triple candidates identified in the MOA EB sample of periods $<2$ days from the GB9-9 and GB10-1 fields. The red line represents $A_{\text{LTTE}}$ equal to rms($\sigma_{p}$). About half of these 91 triple candidates have $A_{\text{LTTE}}$ larger than rms($\sigma_{p}$). The BIC was used to decide whether the ETV model with the LTTE was accepted or not. Note that $\Delta\,BIC$ = $BIC(\textbf{P}(E))-BIC(\textbf{P}(E)+LTTE)$, where $\textbf{P}(E)$ represents the polynomial of $E$ in eq.(\ref{eq:ltte}) and $E$ is cycle. 81 of these triple candidates have $\Delta\,BIC$ > 10, indicating the best fits of the ETV model with the LTTE were strongly preferable. These 81 triple candidates include all those with inner periods $P_{1}<0.26$ days and all those with outer eccentricities $e_{2}>0.9$.}
\label{fig:amp_rms}
\end{figure}

Although the mass transfer would happen in contact and semi-detached binaries, the reliability of the best-fit solution from the LTTE model incorporating the quadratic term of $E$ in eq.(\ref{eq:etv}) might be questionable because such a combination can easily produce a satisfactory fit to a long-term ETV curve that leads to false positive LTTE detection. Therefore, we always preferred the best-fit solution of the LTTE model without the quadratic term unless the BIC value of the best fit with the quadratic term was lower than that without the quadratic term by at least 10, indicating the best fit with the quadratic term is highly favourable. In addition, the detection of the LTTE was accepted to be genuine only if the BIC value of the best-fit LTTE solution was lower than that of the best-fit solution of the quadratic equation of $E$, which was also derived using \texttt{pymc}. 

\begin{figure}
\centering
\includegraphics[width=0.45\textwidth]{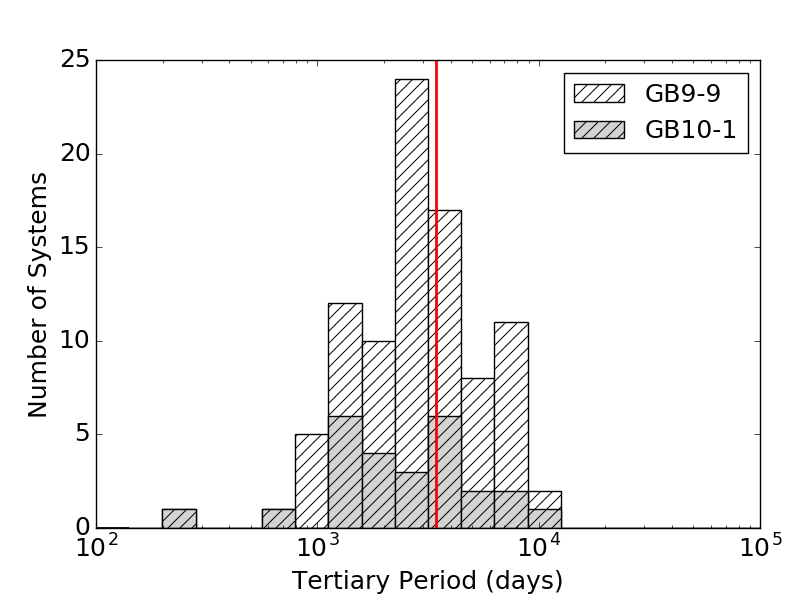}
\caption[Distribution of tertiary periods of all triple candidates in the whole MOA EB sample]{Distribution of the tertiary periods ($P_{2}$) of 91 triple candidates identified in our MOA EB sample of periods $<2$ days from the GB9-9 (light) and GB10-1 (dark) fields. The distribution for the whole sample peaks at 2660 days. However, the separate distributions are not consistent with each other. The distribution for the sample from the GB9-9 field has the sharp peak at 2660 days, while the distribution for the sample from the GB10-1 field appears to be bimodal with one peak at $P_{2}\approx 3700$ days and the other at $P_{2}\approx 1300$ days. Note that the tertiary periods are binned in logarithmic bins of 20 from $10^{2}$ days to $10^{5}$ days. The red lines represent the time span of the MOA data which is about 3420 days.}
\label{fig:outer_p_all}
\end{figure}

\begin{table*}
	\centering
	\caption{Orbital elements from the LTTE solutions for 65 EBs in the GB9-9 field. Note that $P_{1}$ is the period of the inner binary determined by the conditional entropy method plus the correction, $c_{1}$, given by the best fit of eq.(\ref{eq:ltte}) to the ETV curve, and $\Delta P_{1} = 2c_{2}$, where $c_{2}$ is the second order coefficient in eq.(\ref{eq:ltte}), is the change in inner binary orbital period per orbital cycle in units of [day/cycle], and $m_\text{AB}$ was taken as $2\,M_{\odot}$ when calculating ($m_{\text{C}}$)$_{\text{min}}$.}
    \bgroup
    \def\arraystretch{1.}
    \scriptsize
	\begin{tabular}{ccccccccccc} 
		\hline
        No. & $P_{1}$ & $\Delta P_{1}$ & $P_{2}$ & $e_{2}$ & $\omega_{2}$ & $\tau_{2}$ & $a_{\text{AB}}\sin i_{2}$ & $f(\,m_{\text{C}})$ & $(m_{\text{C}})_{\text{min}}$ & $\Delta\,BIC$\\
& (d) & $10^{-10}\times$(d/c) & (d) & & (deg) & (MBJD) & AU & & $M_{\odot}$ &\\
        \hline
		349130&0.2528512(2)&--&7926(80)&0.29(1)&1(1)&53979.7(2)&7.5(1)&0.89(4)&2.7&178.07\\
284305&0.5086635(1)&--&3428(96)&0.96(4)&88(5)&54621.65(4)&1.61(6)&0.048(6)&0.7&912.89\\
155278&1.0545500(2)&--&1888(29)&0.7(1)&35(10)&54789.43(6)&1.1(1)&0.05(2)&0.7&3.35\\
145571&1.2480074(6)&--&3166(84)&0.51(2)&83(5)&56119.25(4)&2.60(7)&0.23(2)&1.39&54.25\\
108463&1.45689719(6)&--&2374(21)&0.99(1)&200(8)&55636.74(1)&2.1(8)&0.2(3)&1.38&200.73\\
19030&1.1797806(3)&--&2590(115)&0.3(1)&309(28)&56176.1(1)&0.61(6)&0.004(1)&0.29&4.43\\
84829&0.3634425(5)&--&6136(1911)&0.41(9)&312(11)&55152.6(4)&1.1(4)&0.004(5)&0.28&900.73\\
351777&0.339333891(4)&--&2898(11)&0.130(9)&4(4)&54712.16(3)&0.478(2)&0.00174(3)&0.2&92.92\\
182318&0.39390818(1)&--&3004(25)&0.47(3)&4(3)&54285.03(4)&0.327(6)&0.00052(3)&0.13&387.54\\
227115&0.86548591(6)&--&2374(20)&0.98(2)&144(14)&54869.39(3)&0.8(3)&0.01(1)&0.41&9.28\\
217605&0.7081659(3)&-15(1)&1617(10)&0.30(4)&142(7)&54907.54(3)&1.36(3)&0.130(9)&1.07&461.82\\
360672&0.35884830(2)&-2.97(3)&948(1)&0.60(2)&105(2)&53915.18(4)&0.327(3)&0.0052(2)&0.3&984.98\\
34057&1.1061859(2)&--&2752(21)&0.99(1)&320(14)&54349.35(1)&2.1(7)&0.2(2)&1.21&13.51\\
72704&0.97866744(4)&--&1396(5)&0.51(3)&115(5)&55016.74(2)&0.68(2)&0.022(1)&0.52&36.79\\
17921&0.44477494(3)&--&2720(169)&0.5(1)&320(12)&55506.97(10)&0.19(2)&0.00011(4)&0.08&87.01\\
356144&0.7865492(2)&--&4424(84)&0.70(4)&6(3)&54208.43(5)&1.59(5)&0.027(3)&0.56&209.79\\
250567&0.34603681(2)&-2.01(4)&2498(8)&0.89(2)&173(2)&54258.05(2)&0.26(2)&0.00039(10)&0.12&380.84\\
157806&0.5361466(6)&--&3481(736)&0.21(7)&282(30)&56603.5(3)&0.4(2)&0.0008(10)&0.15&105.52\\
67484&0.24226037(8)&-2.0(1)&3021(760)&0.74(8)&193(5)&56378.7(3)&0.34(9)&0.0006(6)&0.14&110.26\\
182430&0.35978161(1)&--&2714(26)&0.58(5)&172(2)&53826(1)&0.37(2)&0.0009(2)&0.16&105.52\\
187318&0.43288552(6)&5.1(1)&844(2)&0.37(3)&236(10)&54146.46(7)&0.335(7)&0.0071(5)&0.34&759.87\\
220479&0.4423801(1)&--&6326(587)&0.70(4)&3(3)&54182.0(1)&1.09(8)&0.004(1)&0.28&143.13\\
249030&0.69573952(10)&--&2793(62)&0.77(4)&288(3)&55382.11(3)&1.87(5)&0.112(9)&1.0&240.62\\
155668&0.2972139(1)&-1.7(2)&1687(43)&0.8(1)&177(8)&54362.5(1)&0.4(2)&0.004(5)&0.26&51.5\\
109391&0.30558171(3)&--&3074(141)&0.30(6)&277(16)&55169.33(8)&0.40(2)&0.0009(1)&0.16&250.23\\
380523&0.34710273(8)&-2.9(2)&1325(9)&0.96(5)&100(17)&54328.05(3)&0.43(5)&0.006(2)&0.32&88.28\\
65718&0.3320172(1)&--&8436(562)&0.64(4)&36(4)&55843.9(1)&1.27(9)&0.0038(9)&0.27&138.62\\
22226&0.3603168(1)&-4.4(3)&2460(29)&0.95(5)&354(4)&54028.1(1)&0.8(4)&0.01(2)&0.39&154.63\\
361861&0.2794994(1)&--&7617(1234)&0.78(3)&148(4)&54606.1(2)&0.9(1)&0.0017(9)&0.2&72.07\\
256806&0.2665336(3)&--&10918(4822)&0.69(9)&179(8)&54488.1(6)&1.6(6)&0.005(7)&0.29&10.47\\
303209&0.28954488(7)&--&3687(127)&0.37(4)&163(5)&57488.72(4)&0.69(5)&0.0033(7)&0.26&57.64\\
159607&0.31115756(7)&5.7(1)&1247(4)&0.78(2)&329(2)&54001.20(4)&1.13(5)&0.12(2)&1.05&654.83\\
322149&0.4825322(2)&--&4788(200)&0.57(3)&83(4)&56622.07(6)&1.41(8)&0.016(3)&0.46&278.28\\
238532&0.45366092(7)&--&1093(5)&0.81(9)&137(7)&54569.77(2)&1.4(2)&0.3(2)&1.6&83.34\\
67250&0.4252117(1)&--&8609(1761)&0.65(7)&291(9)&55993.2(3)&0.9(1)&0.0014(8)&0.19&45.69\\
135452&0.5006658(9)&--&7711(2664)&0.69(9)&335(5)&55539.7(5)&2.6(8)&0.04(4)&0.65&238.14\\
101545&0.26379918(4)&--&1859(39)&0.6(1)&28(14)&54933.39(9)&0.69(9)&0.013(5)&0.42&209.86\\
83874&0.3484025(3)&--&7974(3098)&0.7(1)&146(11)&55161.9(5)&0.9(3)&0.002(2)&0.2&42.38\\
7772&0.31983968(5)&--&1967(252)&0.8(2)&110(45)&53972.2(7)&0.21(5)&0.0003(2)&0.11&0.05\\
315321&0.4189762(1)&--&3324(465)&0.8(1)&134(17)&55793.6(2)&1.0(3)&0.011(10)&0.4&32.31\\
367657&0.4167100(1)&2.4(4)&1373(12)&0.76(8)&152(6)&55080.16(2)&0.8(1)&0.03(1)&0.61&225.18\\
306577&0.448403(1)&--&4623(1106)&0.75(9)&121(11)&53919.0(7)&1.9(6)&0.05(5)&0.69&18.2\\
180438&0.23953468(5)&--&3569(297)&0.6(1)&170(11)&56479.3(1)&0.69(8)&0.004(1)&0.26&33.66\\
43392&0.7782322(2)&--&2350(39)&0.25(9)&8(9)&54525.98(9)&1.14(4)&0.036(4)&0.63&167.38\\
357126&0.4173683(1)&--&2329(68)&0.8(2)&276(16)&56126.51(3)&0.9(1)&0.019(9)&0.49&26.68\\
157098&0.4248364(1)&-7.6(3)&1355(12)&0.8(2)&55(12)&53863.2(5)&0.38(8)&0.004(3)&0.27&17.21\\
372358&0.43437314(9)&--&2049(54)&0.7(1)&344(9)&55682.72(5)&1.1(2)&0.04(2)&0.68&228.98\\
146280&0.4860822(4)&--&2773(159)&0.4(2)&155(29)&54747.1(2)&1.7(3)&0.09(4)&0.92&18.23\\
333535&0.7721721(1)&--&2940(242)&0.64(8)&205(8)&55231.6(1)&0.42(4)&0.0011(4)&0.17&12.98\\
325649&0.4836958(3)&11.0(9)&924(10)&0.3(1)&11(11)&54545.35(6)&0.73(6)&0.06(2)&0.78&130.24\\
289148&0.5670889(6)&--&8259(350)&0.50(3)&56(3)&53945.4(9)&2.0(3)&0.015(8)&0.45&91.59\\
238768&0.30033613(3)&--&2711(27)&0.99(1)&172(4)&54001.81(6)&1.7(9)&0.08(14)&0.89&116.48\\
265355&0.3747563(8)&10(1)&2637(700)&0.6(2)&252(19)&55911.1(4)&0.36(9)&0.0009(8)&0.16&47.26\\
367659&0.3096533(2)&--&8107(1308)&0.67(5)&329(10)&56234.8(2)&1.6(2)&0.008(4)&0.35&127.32\\
117331&0.4156716(4)&--&5959(672)&0.2(1)&169(47)&54583.2(8)&0.9(2)&0.003(2)&0.24&920.65\\
222500&0.3150402(2)&--&3296(589)&0.2(2)&301(62)&56634.9(3)&0.3(1)&0.0004(4)&0.12&4.36\\
137966&0.4018154(3)&--&3461(425)&0.2(2)&207(141)&55968.0(6)&0.4(1)&0.0010(8)&0.16&4.33\\
59639&1.1599840(3)&--&3842(151)&0.06(6)&233(85)&54913.9(8)&0.65(3)&0.0025(4)&0.23&19.08\\
40690&0.4916289(5)&--&4568(616)&0.2(1)&247(57)&54707.4(7)&0.9(2)&0.005(3)&0.31&0.3\\
129541&0.5992140(4)&--&5593(1810)&0.44(10)&216(24)&56753.8(5)&0.5(2)&0.0004(6)&0.12&273.27\\
120170&0.42183406(7)&--&2968(462)&0.5(4)&280(55)&55105.8(3)&0.17(4)&0.00007(6)&0.068&28.44\\
296972&0.34078361(3)&-4.38(5)&1312(1)&0.038(7)&0.3(4)&54144.096(6)&1.247(4)&0.150(2)&1.14&3631.27\\
146665&0.31797683(10)&--&4157(67)&0.007(10)&136(45)&57806.7(1)&2.95(6)&0.20(1)&1.29&576.0\\
218937&0.4387632(2)&-5.1(4)&1026(5)&0.03(3)&355(4)&53835.0(8)&0.76(2)&0.055(5)&0.75&368.45\\
$249394^{*}$&0.44615900(2)&--&2310(4)&0.98(1)&151(9)&53825(1)&1.8(6)&0.2(2)&1.14&1720.68\\
		\hline
	\end{tabular}
    \egroup
    \label{tab:ltte_solutions_gb9}
    \begin{tablenotes}
         \footnotesize
        \item * - with additional periodic signal
        \end{tablenotes}
\end{table*}
\begin{table*}
	\centering
	\caption{Orbital elements from the LTTE solutions for 26 EBs in the GB10-1 field. Note that $P_{1}$ is the period of the inner binary determined by the conditional entropy method plus the correction, $c_{1}$, given by the best fit of eq.(\ref{eq:ltte}) to the ETV curve, and $\Delta P_{1} = 2c_{2}$, where $c_{2}$ is the second order coefficient in eq.(\ref{eq:ltte}), is the change in inner binary orbital period per orbital cycle in units of [day/cycle], and $m_\text{AB}$ was taken as $2\,M_{\odot}$ when calculating ($m_{\text{C}}$)$_{\text{min}}$.}
    \bgroup
    \def\arraystretch{1.}
    \scriptsize
	\begin{tabular}{ccccccccccc} 
		\hline
        No. & $P_{1}$ & $\Delta P_{1}$ & $P_{2}$ & $e_{2}$ & $\omega_{2}$ & $\tau_{2}$ & $a_{\text{AB}}\sin i_{2}$ & $f(\,m_{\text{C}})$ & $(m_{\text{C}})_{\text{min}}$ & $\Delta\,BIC$\\
& (d) & $10^{-10}\times$(d/c) & (d) & & (deg) & (MBJD) & AU & & $M_{\odot}$ &\\
        \hline
		136114&0.43213930(7)&--&10211(767)&0.73(3)&255(5)&55357.7(1)&0.80(4)&0.0007(1)&0.14&105.99\\
64799&0.29455691(1)&--&2272(9)&0.35(2)&266(4)&54645.16(3)&0.724(8)&0.0098(3)&0.38&720.01\\
33369&0.234584737(3)&--&1565(12)&0.55(7)&159(5)&54014.0(2)&0.077(5)&0.000025(5)&0.047&28.18\\
73826&0.365257839(5)&--&1982(17)&0.39(5)&108(10)&53915.6(5)&0.088(3)&0.000023(2)&0.046&75.2\\
124700&0.29947914(3)&-0.72(5)&1319(4)&0.96(4)&29(10)&54272.92(2)&0.4(2)&0.006(7)&0.32&105.24\\
94453&0.3549125(4)&3.8(8)&5528(674)&0.41(7)&335(18)&56547.4(2)&0.6(1)&0.0010(7)&0.17&7.44\\
129173&0.560314938(4)&--&247.10(6)&0.04(1)&192(1)&54070.025(3)&0.296(1)&0.0566(8)&0.75&569.8\\
106715&0.4316864(1)&--&4375(415)&0.24(4)&249(18)&54865.5(1)&0.54(5)&0.0011(4)&0.17&162.17\\
15762&0.5102986(4)&-6.0(4)&3421(298)&0.61(3)&349(1)&55356.9(1)&1.7(1)&0.05(2)&0.74&1422.64\\
101793&0.38895230(4)&2.29(6)&745(1)&0.13(2)&241(23)&54471.29(7)&0.446(5)&0.0213(7)&0.51&533.97\\
89558&0.40350925(9)&--&1963(67)&0.8(2)&288(25)&54798.1(1)&0.39(9)&0.002(1)&0.21&17.79\\
58083&0.2510848(3)&-5.0(3)&2802(111)&0.48(6)&301(8)&54569.03(9)&0.67(6)&0.005(1)&0.3&70.96\\
195663&0.3003945(2)&--&6545(680)&0.90(7)&204(12)&54643.1(2)&2.3(8)&0.04(4)&0.64&14.54\\
181626&0.37224225(6)&20.4(1)&1221(3)&0.57(3)&77(3)&54491.62(1)&0.72(1)&0.034(2)&0.61&496.58\\
77420&0.66270095(8)&--&1484(356)&0.6(4)&156(117)&54350(1)&0.07(4)&0.00002(4)&0.042&61.24\\
181398&0.4134596(2)&--&1708(130)&0.4(3)&278(58)&54639.6(4)&0.5(1)&0.004(3)&0.28&151.49\\
93468&0.49273427(6)&--&3202(312)&0.6(2)&285(28)&53983.6(8)&0.20(3)&0.00011(6)&0.08&38.19\\
41908&0.42916153(3)&--&3881(336)&0.96(4)&131(14)&55977.8(1)&0.5(1)&0.0011(9)&0.17&15.49\\
124463&0.8823691(2)&--&1565(40)&0.8(2)&3(4)&55293.87(5)&0.9(3)&0.03(4)&0.61&23.15\\
89172&0.2596522(1)&--&6644(2795)&0.7(1)&338(13)&56438.1(6)&0.6(2)&0.0006(9)&0.14&34.29\\
95682&0.38272308(9)&--&2242(220)&0.2(2)&276(91)&54879.1(6)&0.29(4)&0.0006(3)&0.14&27.47\\
63896&0.32696380(7)&11.5(1)&1370(8)&0.73(5)&170(2)&55179.94(1)&0.48(4)&0.008(2)&0.35&124.83\\
174776&0.4829741(8)&--&4804(489)&0.22(5)&151(11)&58491.4(1)&3.4(5)&0.2(1)&1.36&261.24\\
102925&0.3926504(1)&--&3480(32)&0.383(2)&242.1(7)&56860.06(1)&3.19(4)&0.36(1)&1.7&3119.59\\
$63946^{*}$&0.41931334(2)&--&1827(51)&0.9(1)&1(2)&54515.31(10)&0.23(10)&0.0005(6)&0.13&5.76\\
$69632^{*}$&0.4569038(2)&--&3971(241)&0.78(2)&65(4)&57200.09(8)&0.63(6)&0.0021(6)&0.22&4.19\\
		\hline
	\end{tabular}
    \egroup
    \label{tab:ltte_solutions_gb10}
\begin{tablenotes}
         \footnotesize
        \item * - with additional periodic signal
\end{tablenotes}
\end{table*}

\section{Results}
\label{sec:ltte_results}
\subsection{The reliability of the results}
We attempted to search for LTTE in all MOA EBs of periods $<$ 2 days in the GB9-9 and GB10-1 fields. In these two fields, there are 542 EBs within the period range we were interested in. 436 and 106 of them come from the GB9-9 and GB10-1 fields, respectively. 

Following the procedures of the ETV analysis in the precious section, there are 91 EBs for which we could derive LTTE solutions that fit their ETV curves and, thus, we catalogized them as triple candidates. 65 of these triple candidates were identified in GB9-9, while 26 were in the GB10-1 field. The derived orbital parameters of the triple candidates are shown in Table~\ref{tab:ltte_solutions_gb9} and Table~\ref{tab:ltte_solutions_gb10}, respectively. Whether the LTTE solutions are reliable is always questionable as several mechanisms can produce ETV curves that mimic the LTTE. In particular, we noticed that there are a certain number of cases in which the ETVs for their primary and secondary eclipses vary cyclically and behave anti-correlated to each other on the time scale of a year, while the long-term trends were consistent, e.g., MOA-289148-GB9-9 and MOA-351777-GB9-9 (see Figure~\ref{fig:etv_spot_eg}). The anti-correlated behaviours in the ETV curves were likely attributed to star spots present on the surfaces of the EBs' active components \citep{2013ApJ...774...81T}. Generally, averaging the ETVs of the primary and secondary eclipses might reduce the contribution of such spurious ETV. But either the primary or secondary eclipse would usually be missing in a cycle, thus averaging was not applicable for the majority of the MOA samples. Nonetheless, we recognized that the best fit obtained by \texttt{pymc} would roughly represent the solution to the mean ETV curve if we fit the ETV curves of primary and secondary eclipses simultaneously, provided that the uncertainties in ETVs for primary and secondary eclipses are comparable. 

In addition, the LTTE solution might represent the over-fitting to the ETV curve when the uncertainties in the times of eclipse minima were overall larger than the LTTE amplitude. Particularly, the model of quadratic ETV plus LTTE could easily provide a good fit to a ETV curve, leading to false positive detection of LTTE. To avoid over-fitting, we used the BIC to decide whether to accept or reject the solution from the model with more free parameters. In our ETV analysis, we accepted the solution of the LTTE model plus the quadratic term of $E$ as the best-fit only if its BIC value was lower than that excluding the quadratic term of $E$ by 10. Besides, the detection of LTTE was accepted eventually only if the BIC value of the LTTE solution was lower than that of the parabolic solution. In this way, we accepted the ETV curves of 22 samples to be best fitted by the LTTE model plus the quadratic terms, while the fits by the LTTE model without the quadratic terms were preferred for 69 samples. Figure~\ref{fig:amp_rms} shows the plot of LTTE amplitudes, $A_{\text{LTTE}}$, against root-mean-square errors in eclipse timing for primary eclipses, rms$(\sigma_{p})$. There were only about half of the detected LTTE signals with amplitudes greater than the values of rms$(\sigma_{p})$. Nonetheless, among these 91 triple candidates, 88 of them have differences between the BIC values of the LTTE and parabolic solutions larger than 10, indicating the LTTE solutions are very strongly preferable. On the other hand, there are two of them, i.e. MOA-40690-GB9-9 and MOA-7772-GB9-9, which have the BIC differences barely above 0, indicating the statistical evidence for detection of the LTTE in them is weak, although they are still included in the list of EBs with detected LTTE signals.

\begin{figure}
\centering
\includegraphics[width=0.45\textwidth]{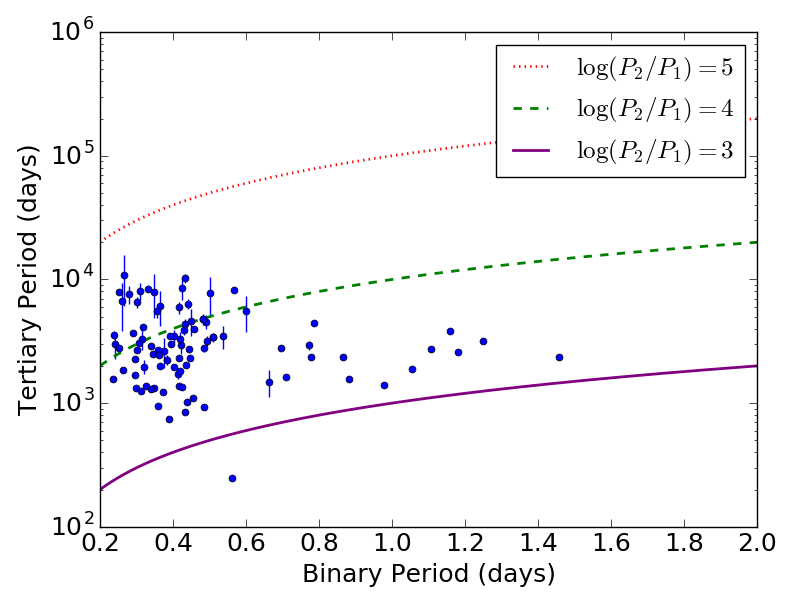}
\caption[Binary period vs tertiary period for 91 triple candidates in GB9-9 and GB10-1]{Binary period ($P_{1}$) vs tertiary period ($P_{2}$) for 91 triple candidates identified in the GB9-9 and GB10-1 fields. All the triple candidates have $\log(P_{2}/P_{1})$ between 3 and 5, except MOA-129173-GB10-1 which has a close tertiary companion of $P_{2}$ about 247 days according to its best-fit LTTE solution.}
\label{fig:P1_vs_P2}
\end{figure}

\begin{figure*}
\centering
\includegraphics[width=0.45\textwidth]{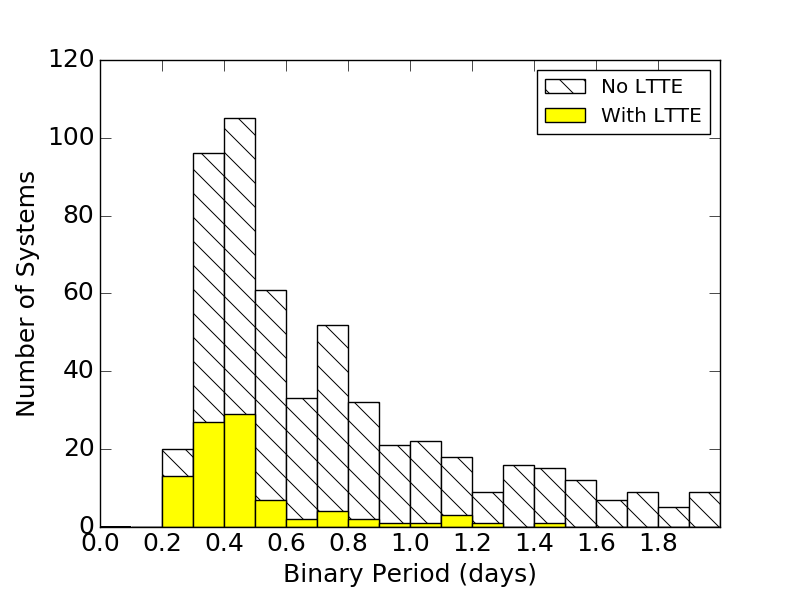}
\includegraphics[width=0.45\textwidth]{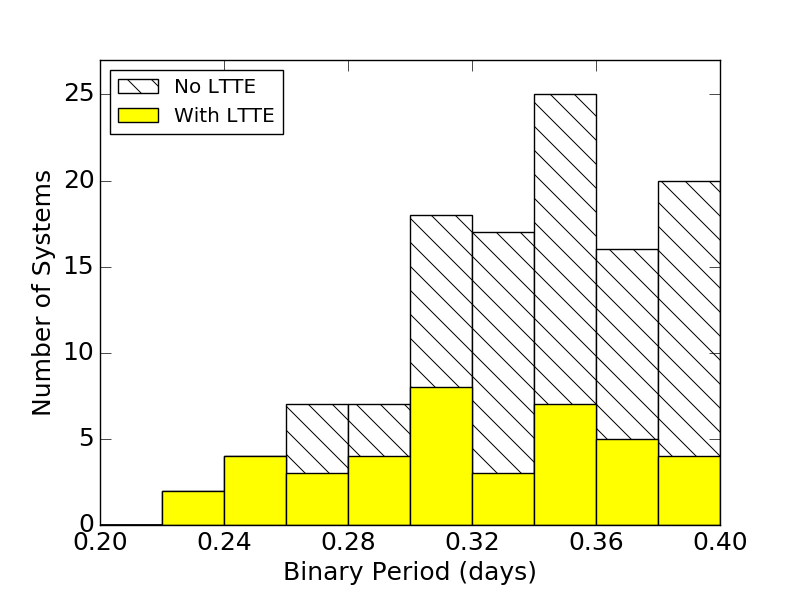}
\caption{The period histogram of the 542 MOA EB sample of periods $<2$ days (left) and the period histogram of the subsample containing all MOA EBs of periods $<0.4$ days (right) from the GB9-9 and GB10-1 fields. The portion of the whole EB sample with detected LTTE signals was filled with yellow, while the rest was hatched with diagonal lines.}
\label{fig:eb_p1_dis}
\includegraphics[width=0.45\textwidth]{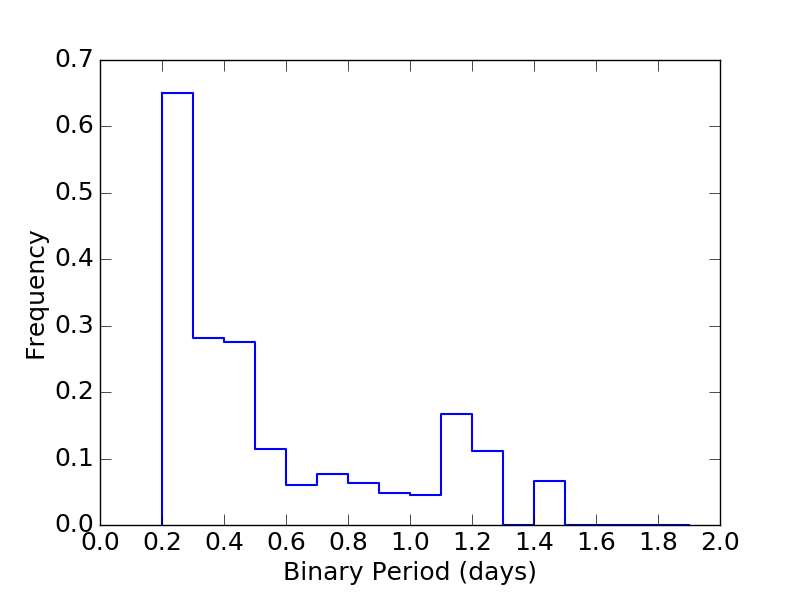}
\includegraphics[width=0.45\textwidth]{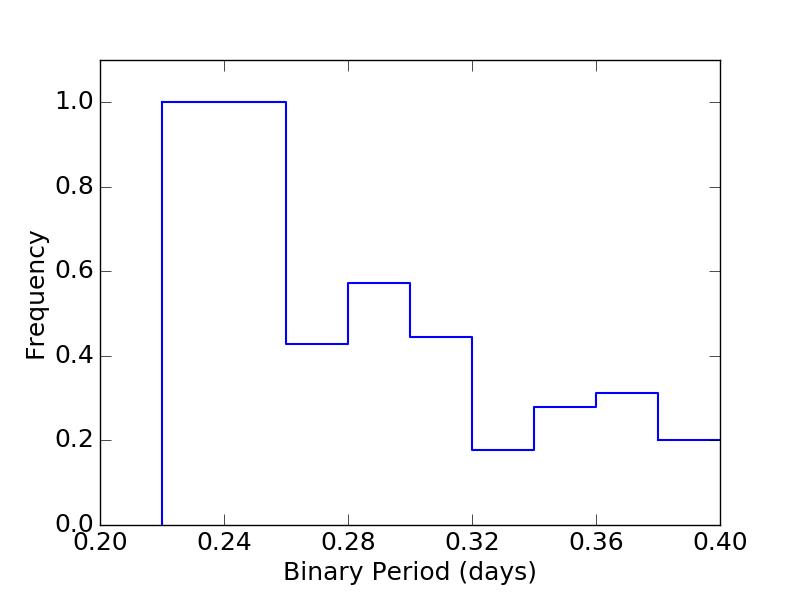}
\caption{Frequency of the MOA EBs with detected LTTE signals. Frequency is defined as the number of EBs with detected LTTE signals over the total number of EBs in each bin.}
\label{fig:outer_freq}
\end{figure*} 

\begin{figure*}
\centering
\includegraphics[width=.32\textwidth]{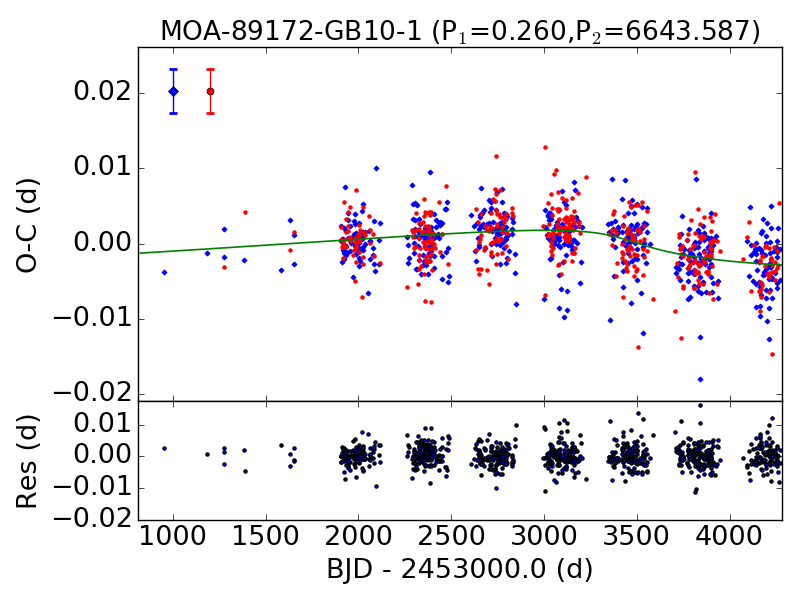}
\includegraphics[width=.32\textwidth]{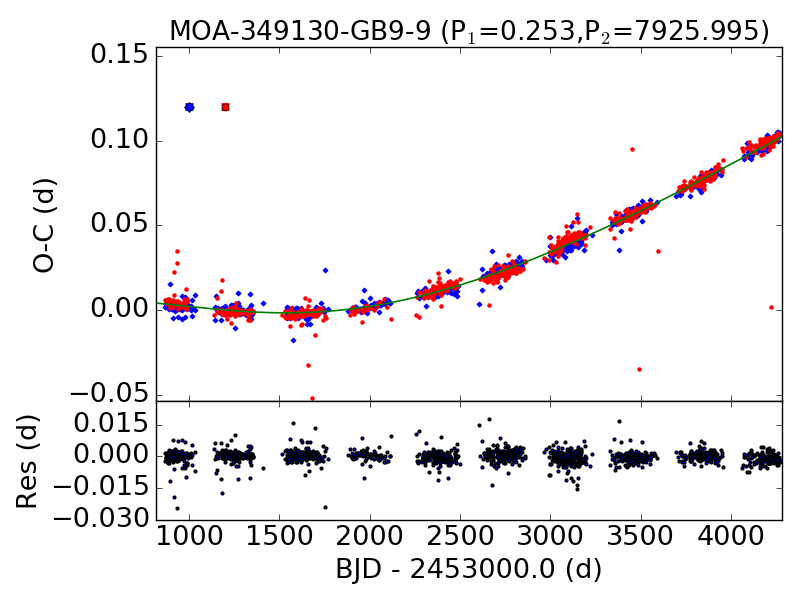}
\includegraphics[width=.32\textwidth]{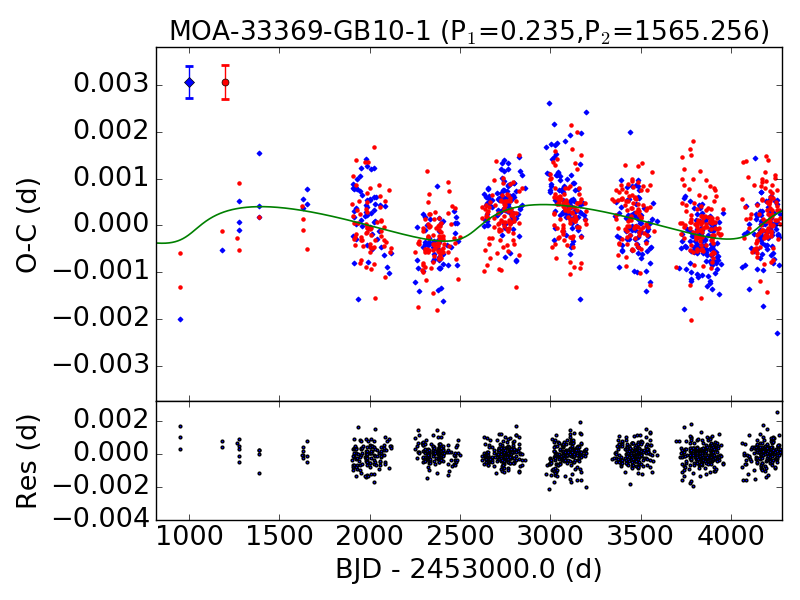}
\includegraphics[width=.32\textwidth]{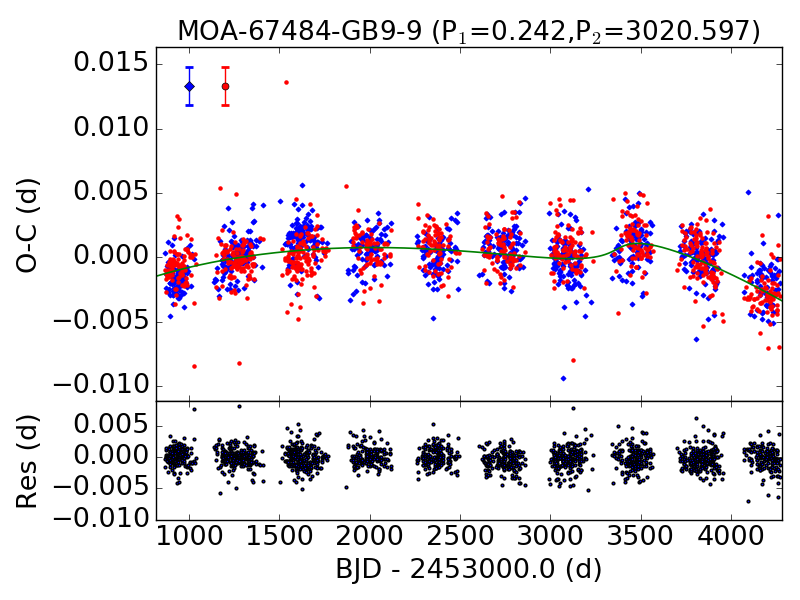}
\includegraphics[width=.32\textwidth]{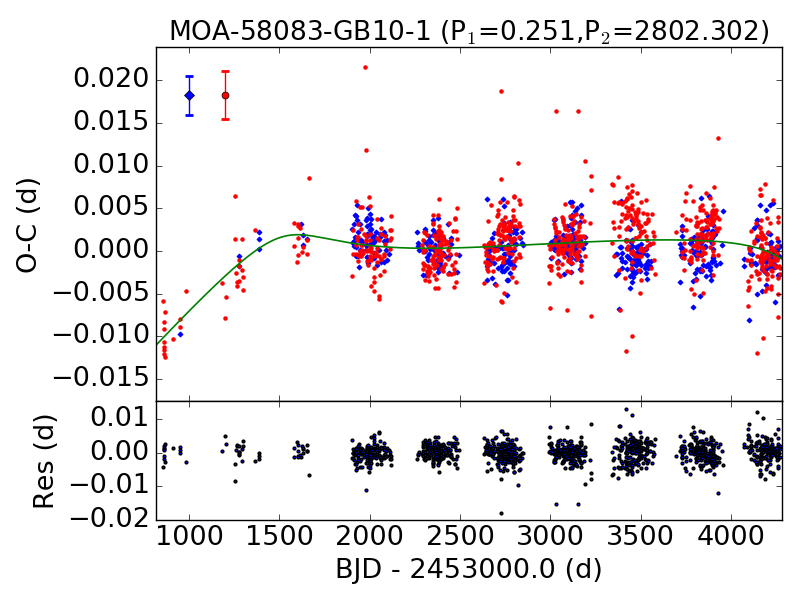}
\includegraphics[width=.32\textwidth]{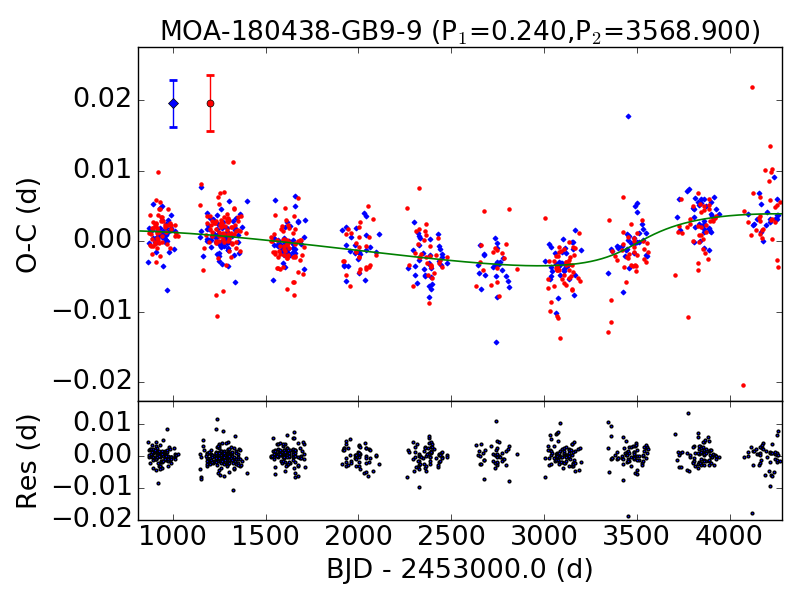}
\caption[ETV curves of the six EB candidates of periods $<0.26$ days]{ETV curves of six EBs of periods $<0.26$ days. $P_{1}$ is the period of the inner binary determined by the conditional entropy method, while $P_{2}$ is the period of the tertiary companion given by the LTTE solution. The blue (diamond) points are the ETV measurements of the primary eclipses and the red (circle) points are those of the secondary eclipses, while the green lines represent the best fits of the ETV model defined by eq.(\ref{eq:ltte}). Also, the average uncertainties for the primary and secondary eclipses are represented by the red and blue error bars, respectively, on the top-left corner of each figure. The bottom panels show the residual curves. Note that the periods are in days.}
\label{fig:etv_low_cb}
\end{figure*}

\subsection{Statistics and Distribution}
Since we selected the EBs from the subfields GB9-9 and GB10-1 only in terms of period alone, it represents a homogeneous sample of EBs of periods $<2$ days. Therefore, it is worth examining distributions and statistics of several interesting orbital parameters. 

\subsubsection{Tertiary period}
The advantage of the ETV method is that we can derive the orbital periods and eccentricities of tertiary companions from the LTTE solutions. Figure~\ref{fig:outer_p_all} shows the distribution of the tertiary period of all triple candidates in our sample as well as the distributions of the triple candidates in the GB9-9 and GB10-1 fields, respectively, for comparison. We used 20 bins to bin the tertiary period from $\log(P_{2})=0$ to $\log(P_{2})=5$. The tertiary period distribution peaked at $\log(P_{2} \approx 3.4)$, which is close to the time span of the MOA data, i.e., 3420 days. Since the LTTE signal of period longer than 3420 days would only have a portion of its cycle seen in the O-C diagram, it would usually be indistinguishable from the parabolic ETV unless the portion of the LTTE curve seen in the O-C diagram has a curvature significantly different from that of a parabolic curve. Therefore, we suspected the lack of triple candidates of longer outer periods is due to the limited time span of the data. On the other side, there is almost no detection of tertiary companions of periods $<$ 600 days. MOA-129173-GB10-1 is the only one having a tertiary companion of period shorter than 600 days\footnote{In fact, MOA-129173-GB10-1 is one of the three triple candidates discovered in the preliminary ETV investigation of \citet{2017MNRAS.470..539L}. The other two are MOA-115233-GB10-9 and MOA-360325-GB10-7 which have tertiary companions of periods 427 days and 482 days, respectively. However, concerned with the homogeneity, we did not include these two in our sample of MOA triple candidates for statistical analysis.}. 

The lack of tertiary companions of periods $<$ 600 days might be related to the general formation process of contact binaries. However, we have to also point out that the LTTE amplitude increases as the outer period increases or the mass of the tertiary companion increases, so short period and low mass tertiary companions might be simply undetectable given the uncertainties in ETV measurements from the MOA data. Also, the existence of regular gaps between two MOA observational seasons in the data always results in regular gaps in the ETV curves which in turn prevent the proper coverage of short period LTTE signals and might make the short period LTTE signals difficult to be detected. In addition, the triple candidates in the GB9-9 and GB9 follow distinctive outer period distributions. Particularly, the tertiary period distribution of the GB10-1 sample seemed to be bimodal with a peak at $P_{2} \approx 3700$ days and the other at $P_{2} \approx 1300$ days. It could be just the effect of the small sample size, but we suspected the peak at $P_{2} \approx 1300$ days resulted from the non-uniform density of the light curves of the GB10-1 sample, in which there are fewer data points in the period of the first two observational seasons because lower cadences for imaging were taken towards the GB10 field during that period. 

We also plotted the tertiary period ($P_{2}$) against the inner binary period ($P_{1}$) for the 91 triple candidates as shown in Figure~\ref{fig:P1_vs_P2}. All of the triple candidates have the period ratios $P_{2}/P_{1}$ between $10^{3}$ and $10^{5}$, except MOA-129173-GB10-1 of which the period ratio is below $10^{3}$.

\begin{figure}
\centering
\includegraphics[width=0.45\textwidth]{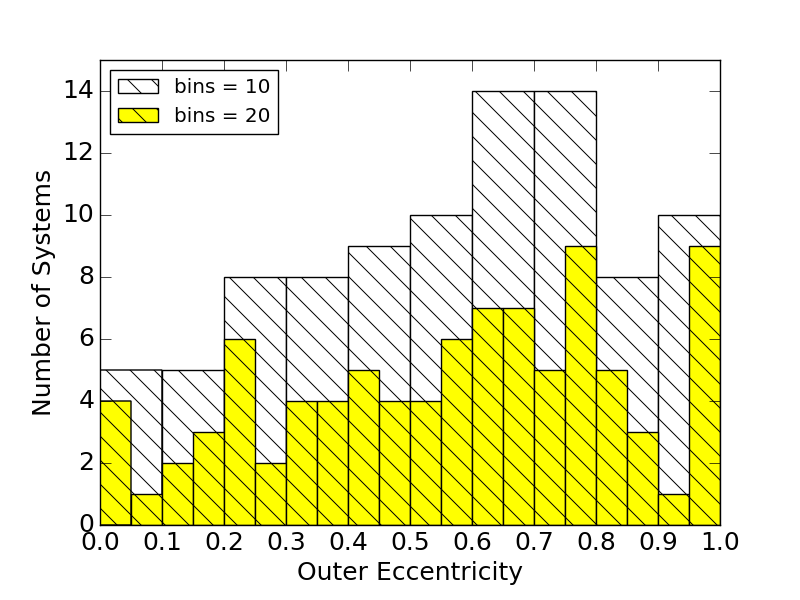}
\caption[Distribution of outer eccentricity for 91 triple candidates in GB9-9 and G10-1]{Distribution of outer eccentricity ($e_{2}$) for 91 triple candidates in the GB9-9 and G10-1 fields. The distribution is binned into 10 (white) and 20 (yellow) bins, respectively, and they are plotted on top of each other in the same graph. The distribution increases as eccentricity increases and peaks at about $e_{2} = 0.7-0.8$. The excess of outer eccentricity is observed at $e_{2}>0.9$ in the triple candidates we identified.}
\label{fig:outer_ecc_dis}
\end{figure}

\begin{figure*}
\centering
\includegraphics[width=.32\textwidth]{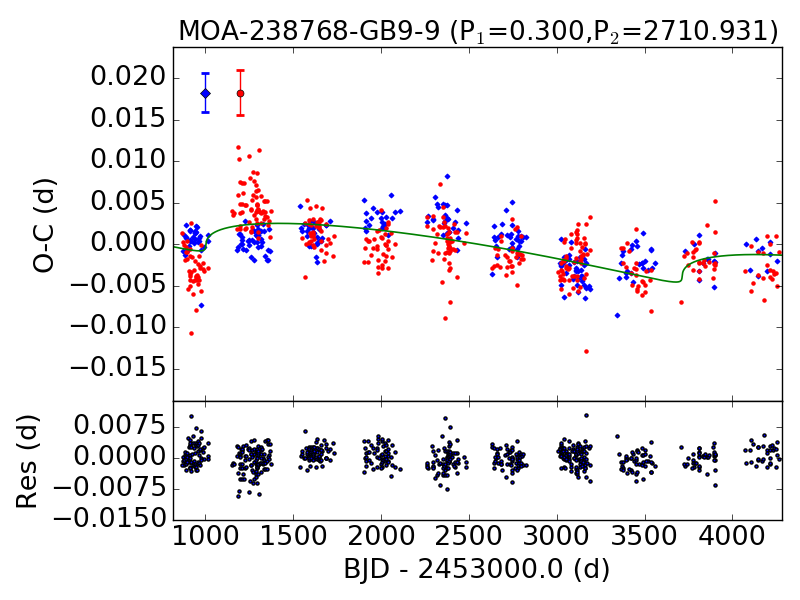}
\includegraphics[width=.32\textwidth]{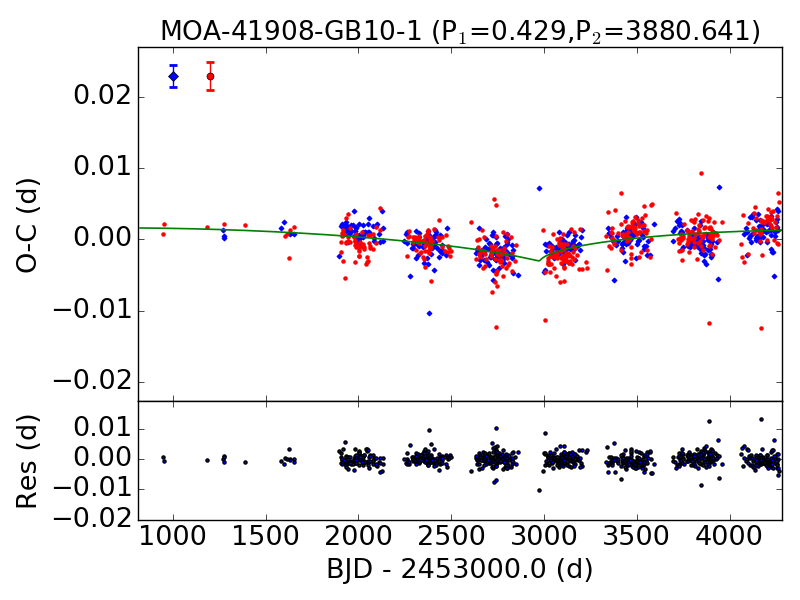}
\includegraphics[width=.32\textwidth]{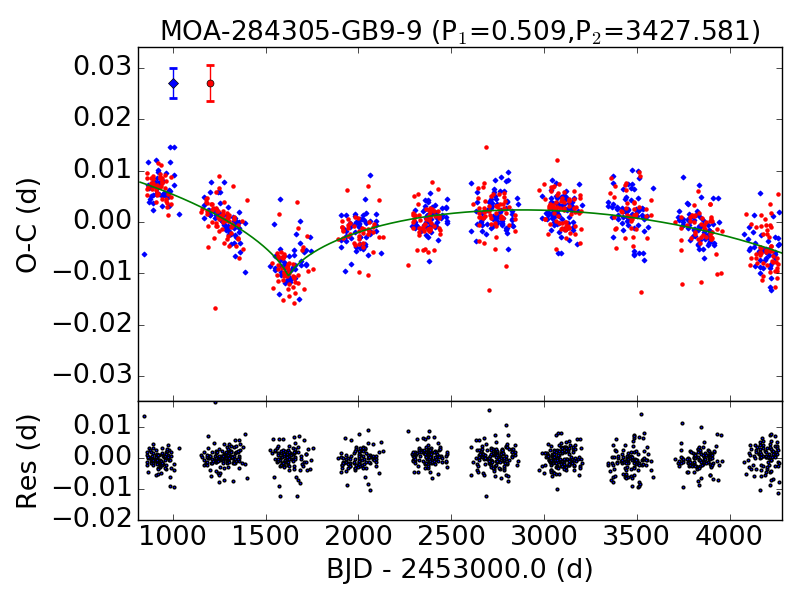}
\includegraphics[width=.32\textwidth]{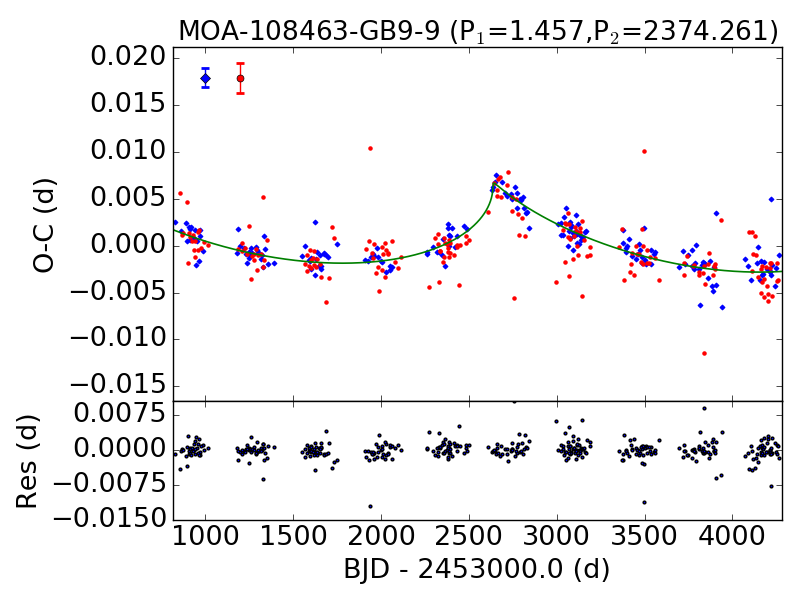}
\includegraphics[width=.32\textwidth]{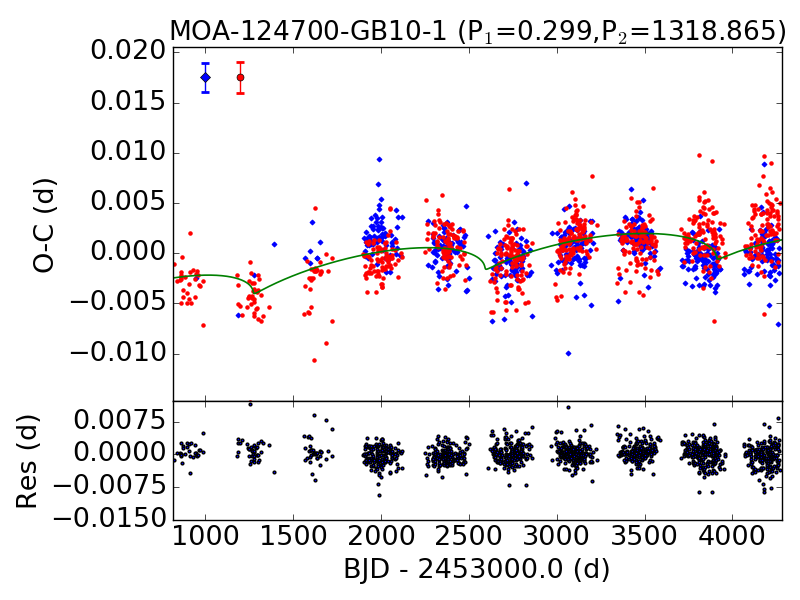}
\includegraphics[width=.32\textwidth]{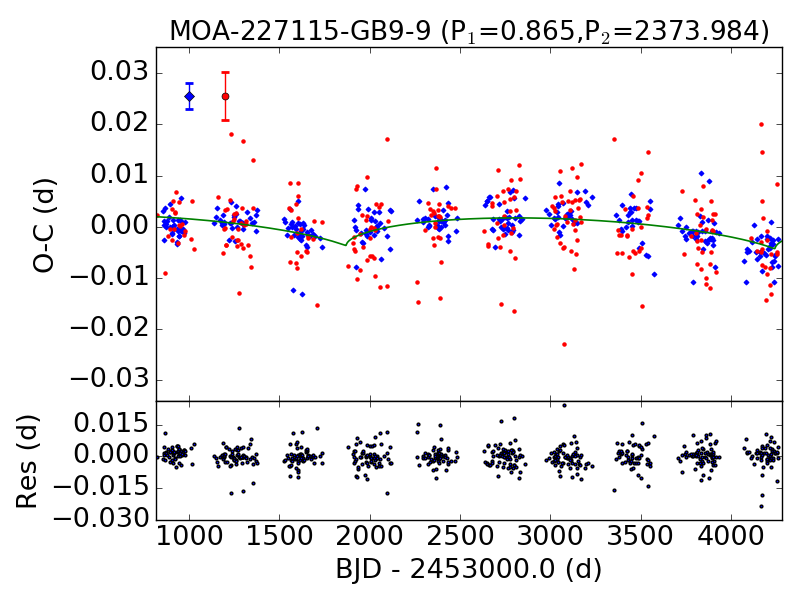}
\includegraphics[width=.32\textwidth]{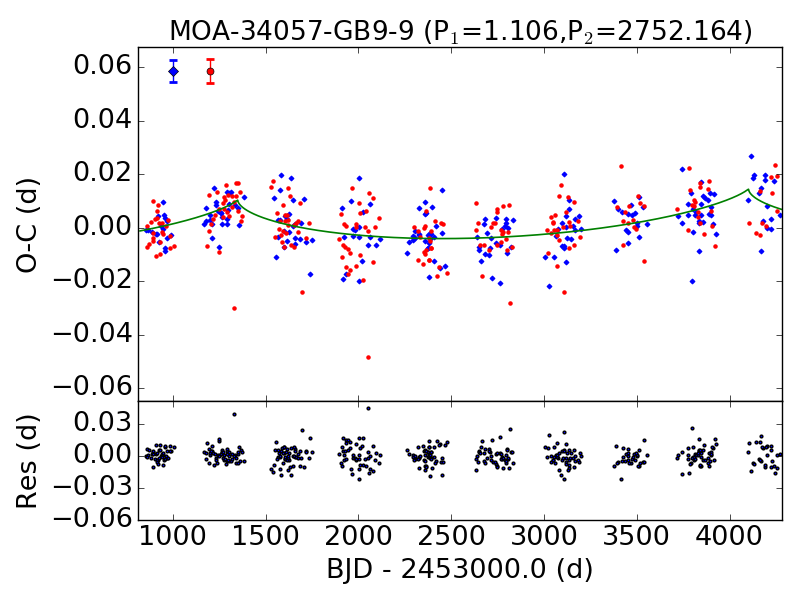}
\includegraphics[width=.32\textwidth]{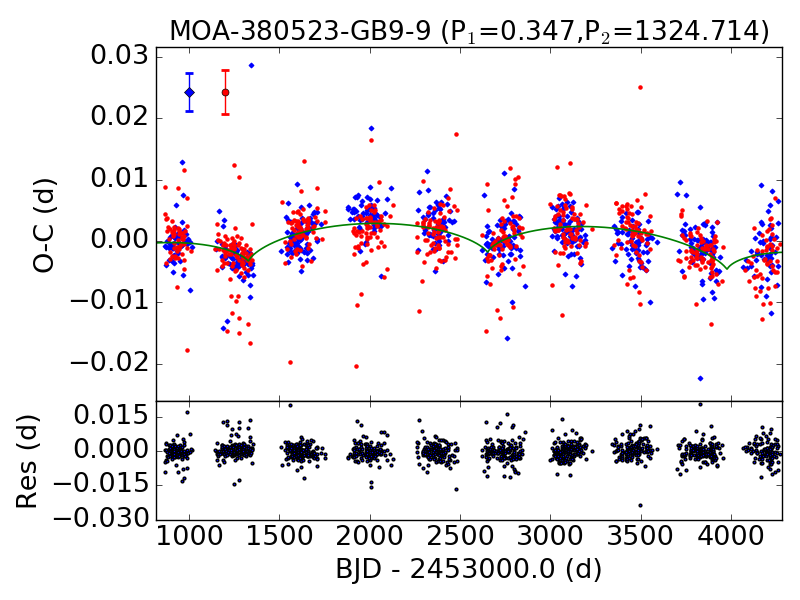}
\includegraphics[width=.32\textwidth]{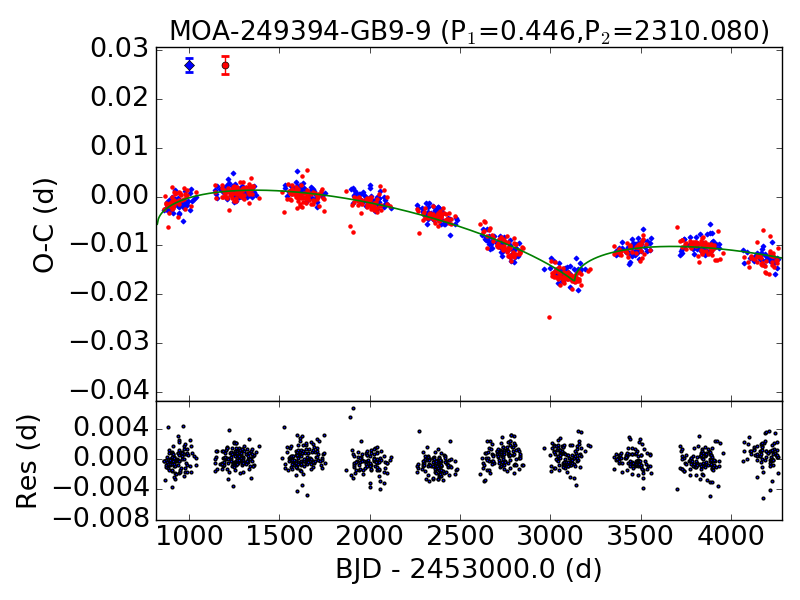}
\includegraphics[width=.32\textwidth]{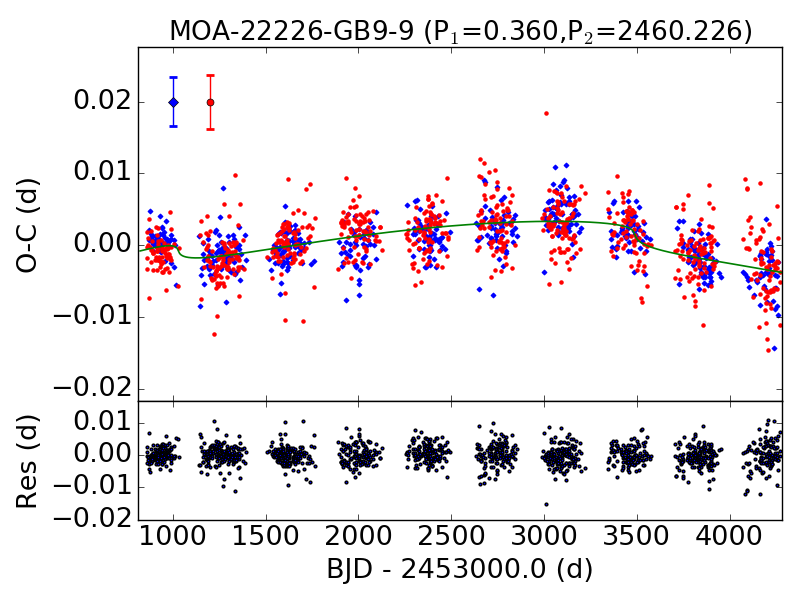}
\caption[ETV curves of ten triple candidates with outer eccentricities $>0.9$]{ETV curves of ten triple candidates with outer eccentricities $e_{2}>0.9$. $P_{1}$ is the period of the inner binary determined by the conditional entropy method, while $P_{2}$ is the period of the tertiary companion given by the LTTE solution. The blue (diamond) points are the ETV measurements of the primary eclipses and the red (circle) points are those of the secondary eclipses, while the green lines represent the best fits of the ETV model defined by eq.(\ref{eq:ltte}). Also, the average uncertainties for the primary and secondary eclipses are represented by the red and blue error bars, respectively, on the top-left corner of each figure. The bottom panels show the residual curves. Note that the periods are in days.}
\label{fig:etv_ecc_0.9}
\end{figure*}

\begin{figure}
\centering
\includegraphics[width=0.45\textwidth]{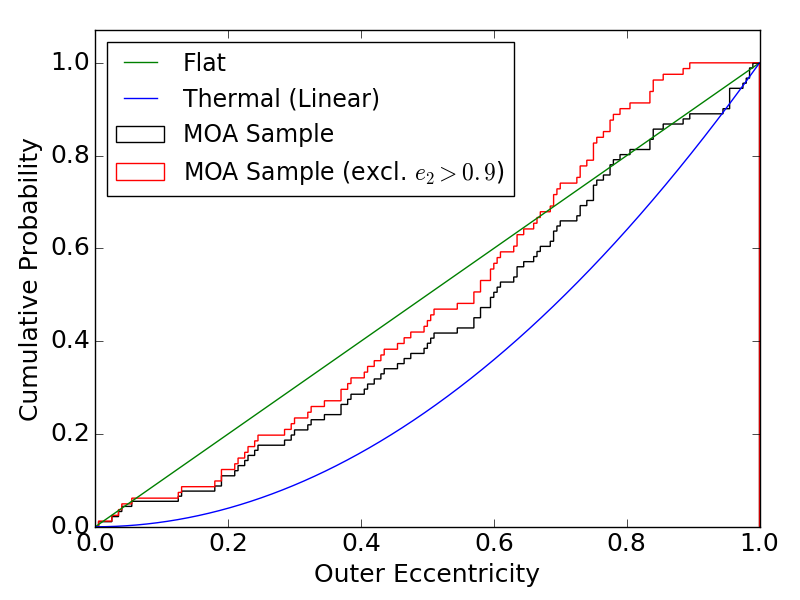}
\caption[Cumulative distribution of the outer eccentricity for all 91 triple candidates in the sample from GB9-9 and GB10-1]{Cumulative distribution of outer eccentricity for the triple candidates identified in the GB9-9 and GB10-1 fields. The green curve represents the cumulative distribution for uniform distribution of eccentricity from 0 to 1. The blue curve represents the cumulative distribution for thermal (or linear) eccentricity distribution derived by \citeauthor{1919MNRAS..79..408J} (\citeyear{1919MNRAS..79..408J}). The cumulative distributions of the outer eccentricity of the triple candidates, excluding and including the triple candidates with $e_{2}>0.9$, in our sample are represented by the red and black curves, respectively, and their underlaying distributions are distinct from each other.}
\label{fig:ecc_cpf}
\centering
\includegraphics[width=0.45\textwidth]{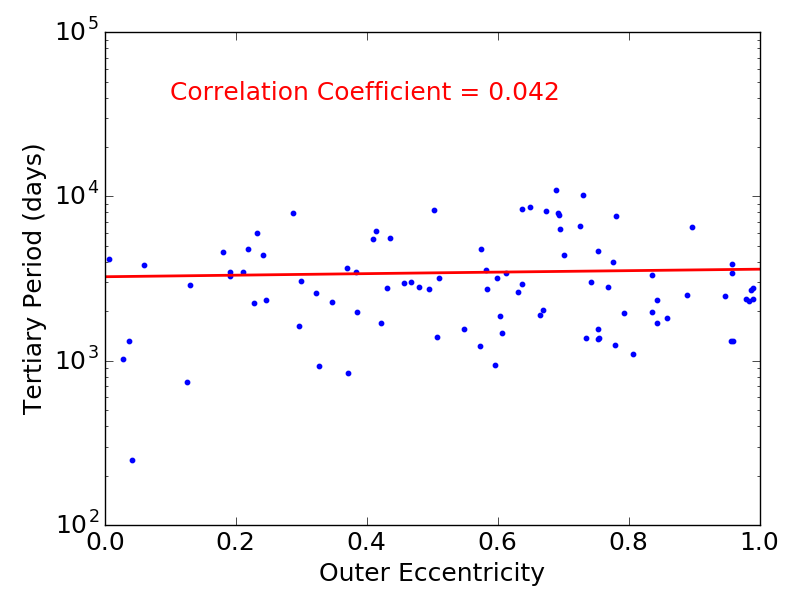}
\caption[Tertiary period vs outer eccentricity for 91 triple candidates in GB9-9 and GB10-1.]{Tertiary period ($P_{2}$) vs outer eccentricity ($e_{2}$) for 91 triple candidates in the GB9-9 and GB10-1 fields. The red curve is the best linear fit which has a correlation coefficient of 0.042, indicating there is no significant correlation between $P_{2}$ and $e_{2}$ for the group of these 91 triple candidates.}
\label{fig:e2_P2_plot}
\end{figure}

\begin{figure*}
\includegraphics[width=.32\textwidth]{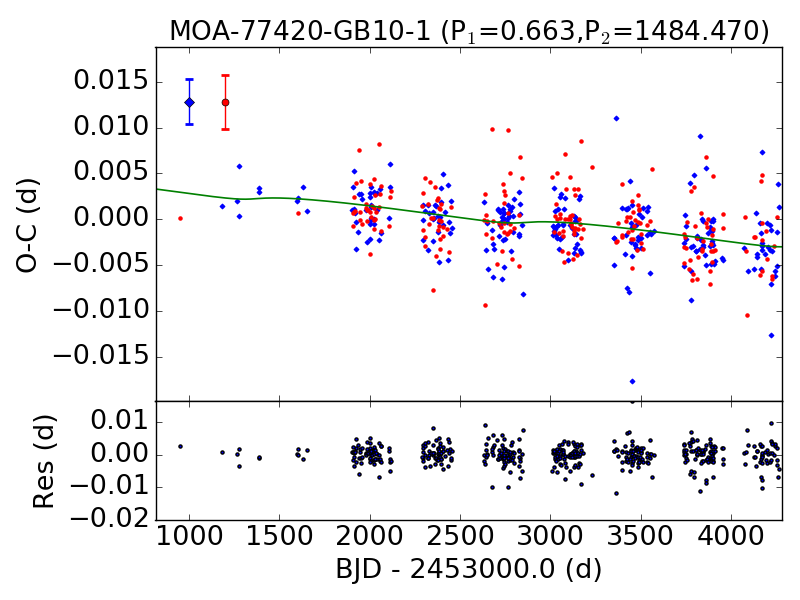}
\includegraphics[width=.32\textwidth]{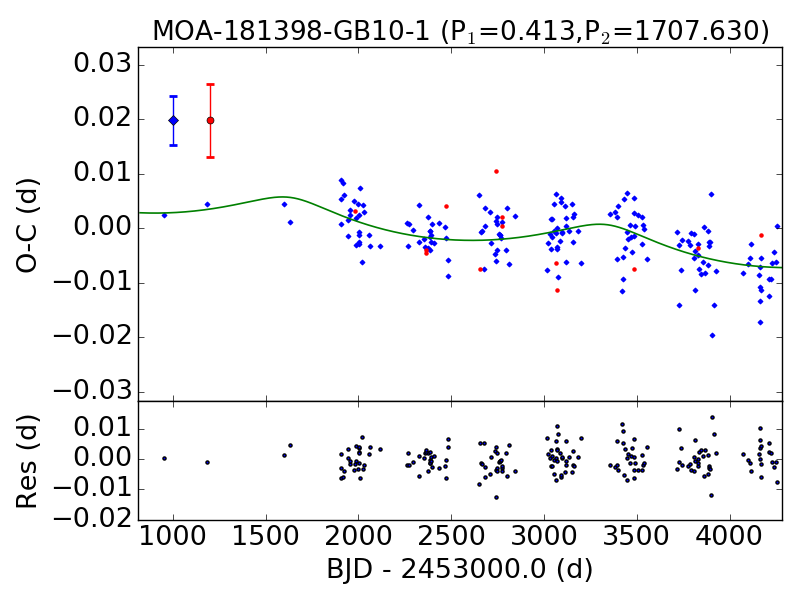}
\includegraphics[width=.32\textwidth]{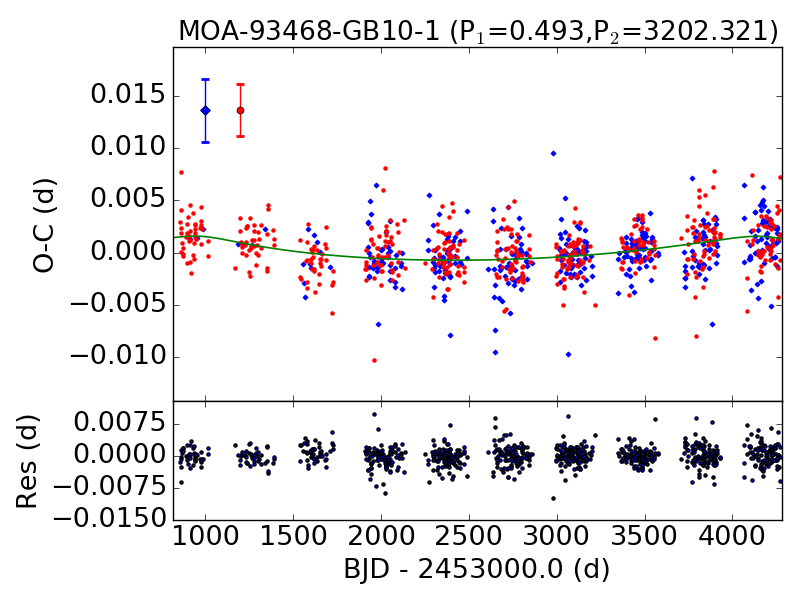}
\includegraphics[width=.32\textwidth]{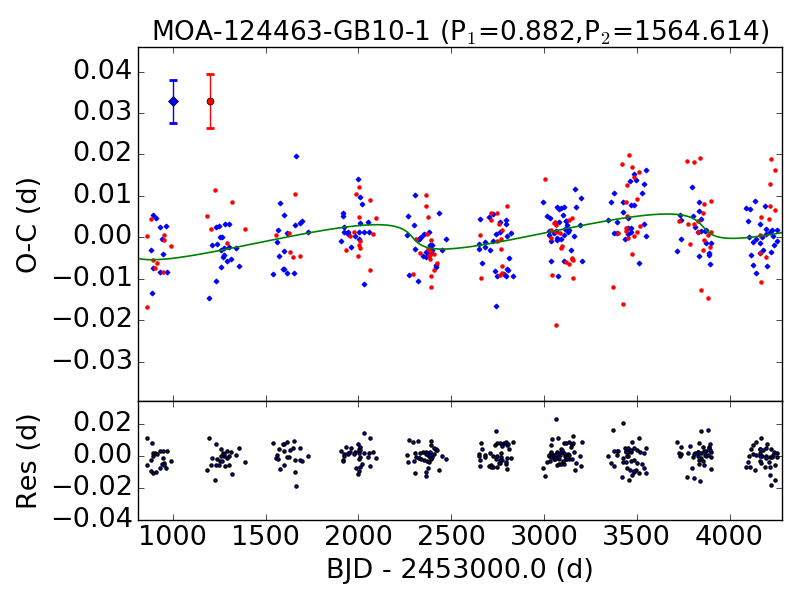}
\includegraphics[width=.32\textwidth]{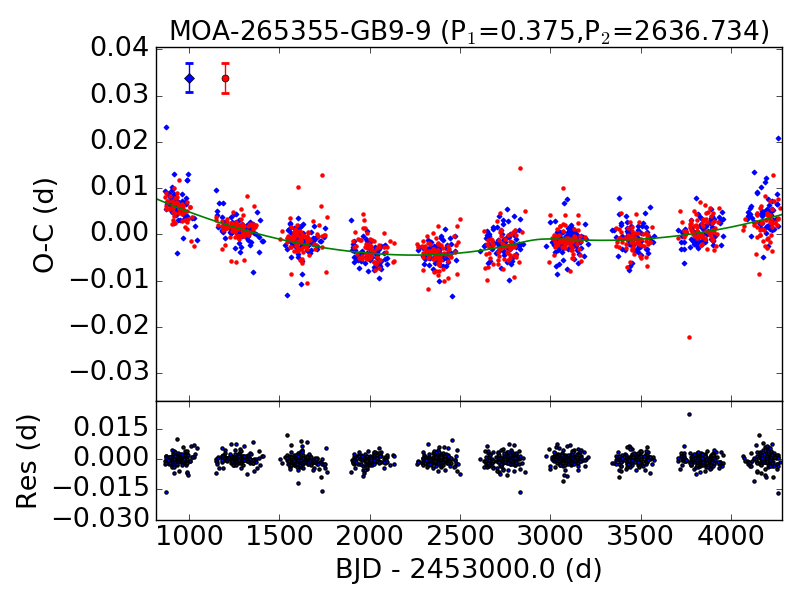}
\includegraphics[width=.32\textwidth]{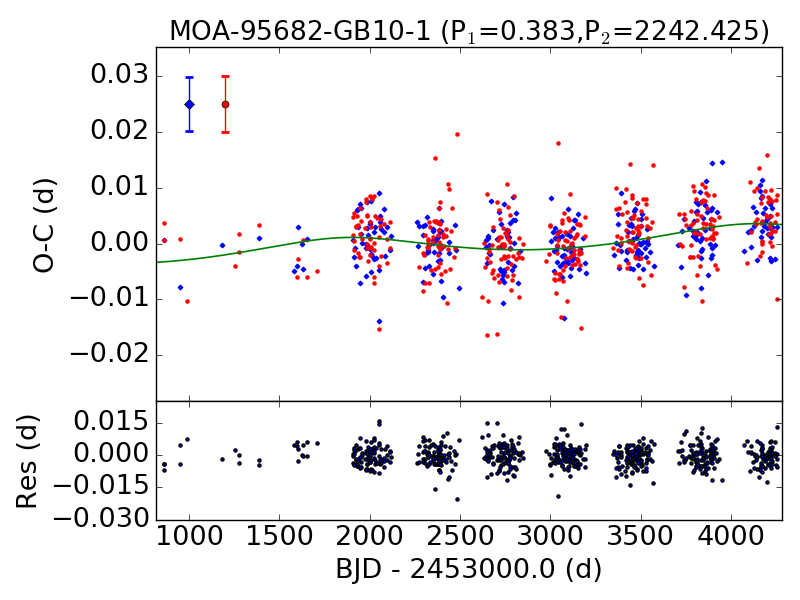}
\includegraphics[width=.32\textwidth]{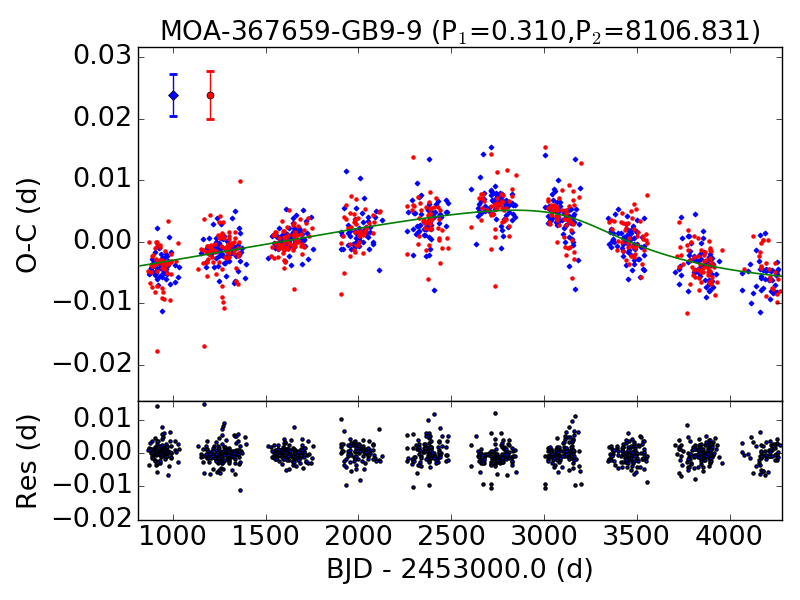}
\includegraphics[width=.32\textwidth]{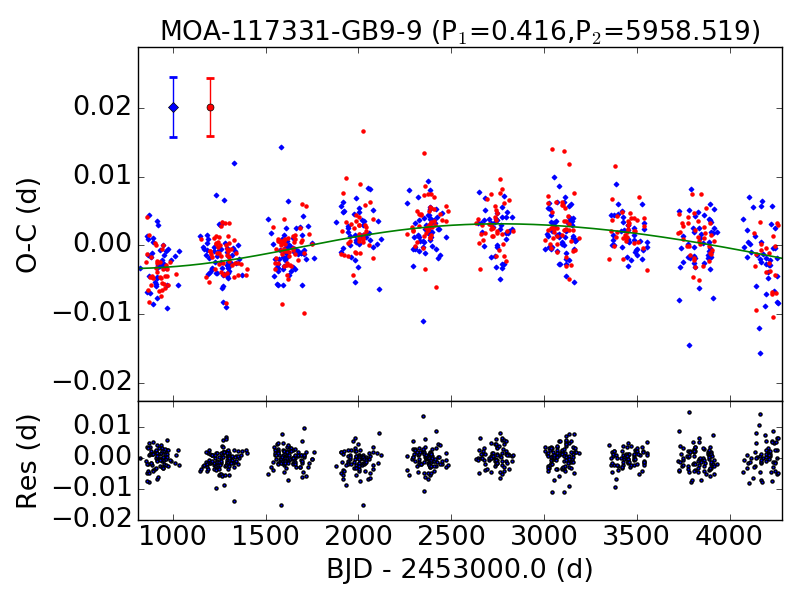}
\includegraphics[width=.32\textwidth]{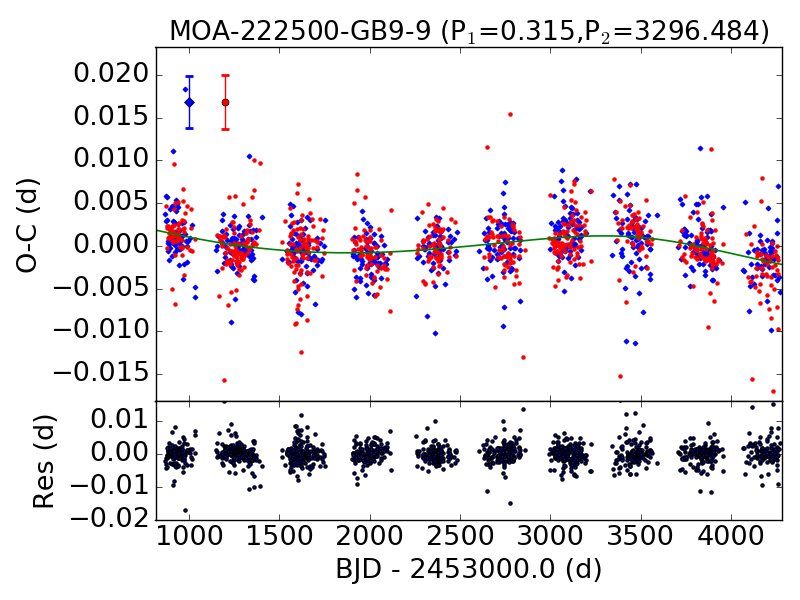}
\includegraphics[width=.32\textwidth]{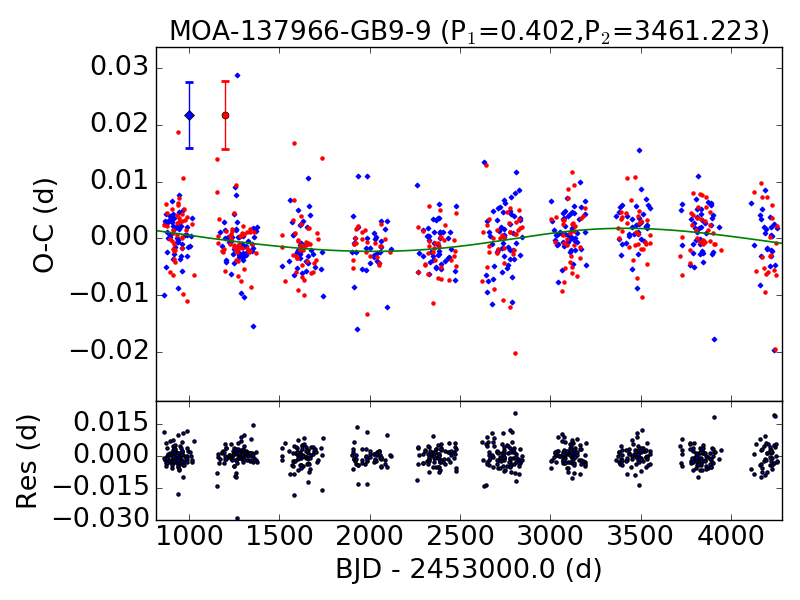}
\includegraphics[width=.32\textwidth]{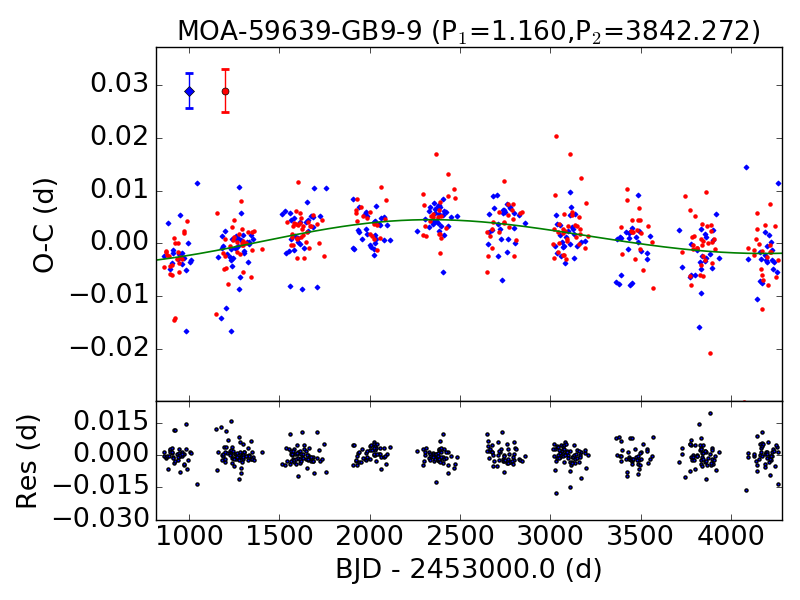}
\includegraphics[width=.32\textwidth]{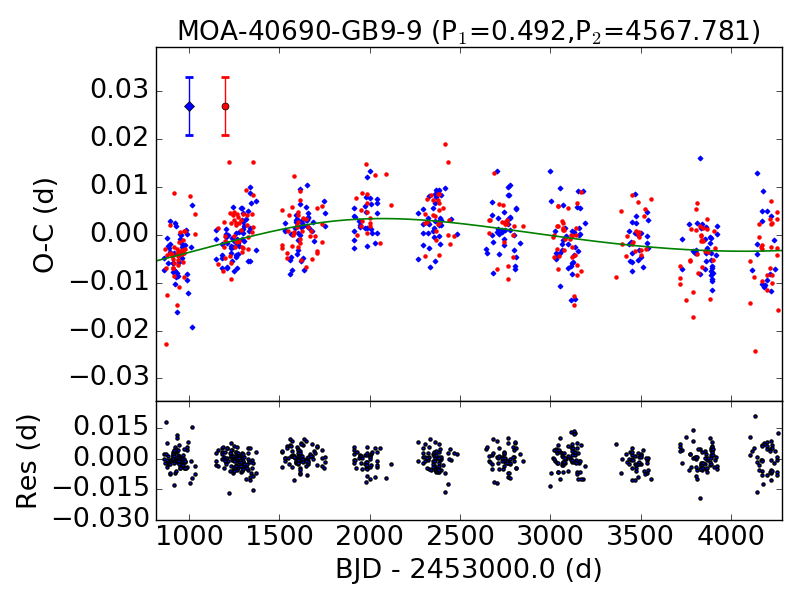}
\includegraphics[width=.32\textwidth]{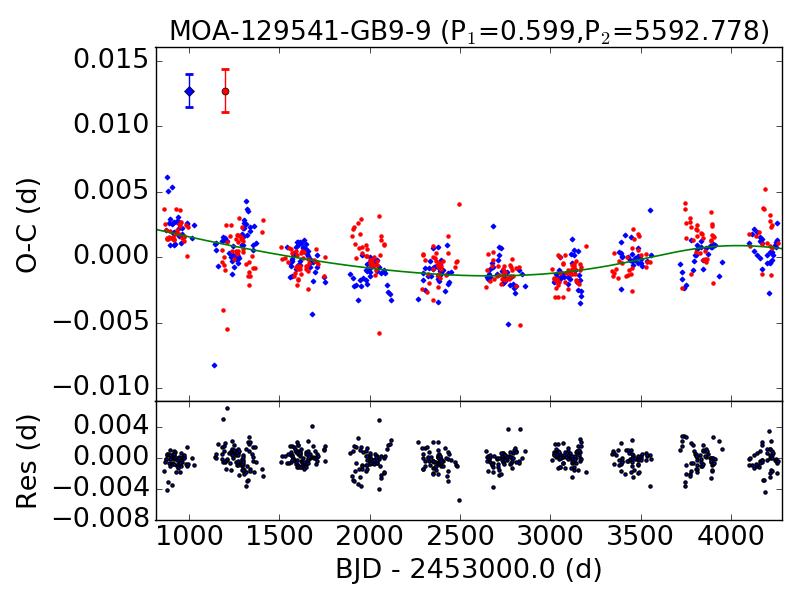}
\includegraphics[width=.32\textwidth]{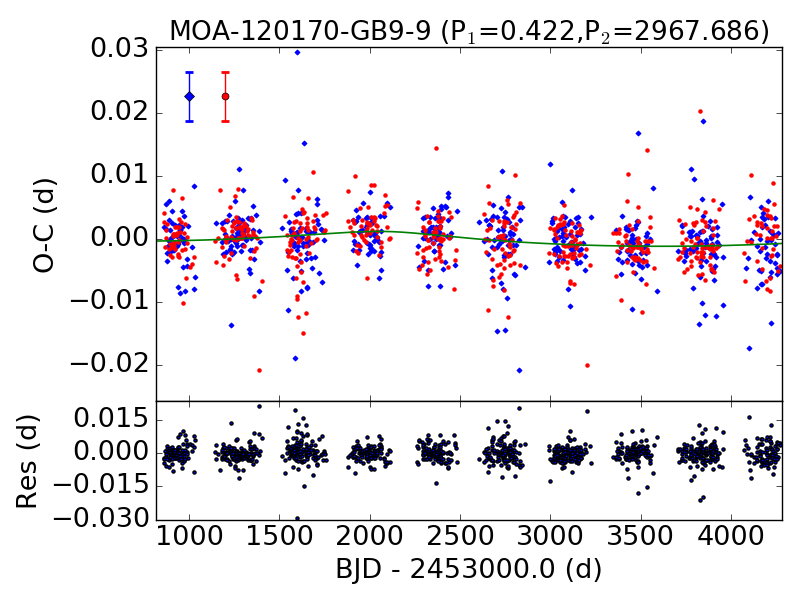}
\includegraphics[width=.32\textwidth]{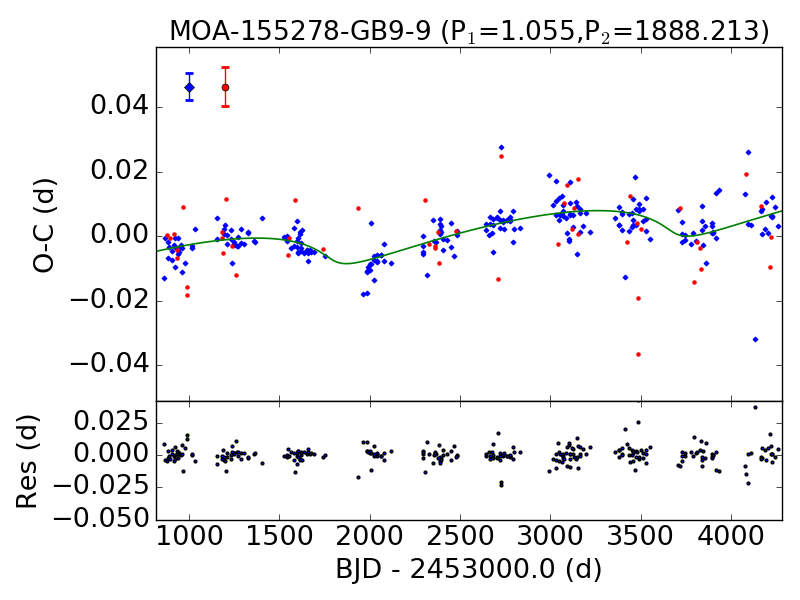}
\caption[ETV curves of the rest of MOA Triple Candidates]{ETV curves of all the other 73 MOA triple candidates.}
\label{fig:etv_17}
\end{figure*}               

\begin{figure*}
\ContinuedFloat
\includegraphics[width=.32\textwidth]{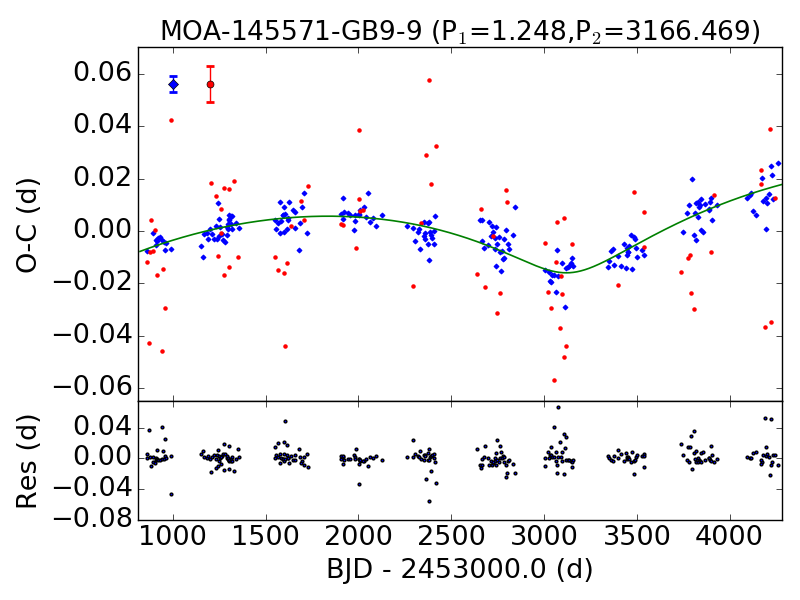}
\includegraphics[width=.32\textwidth]{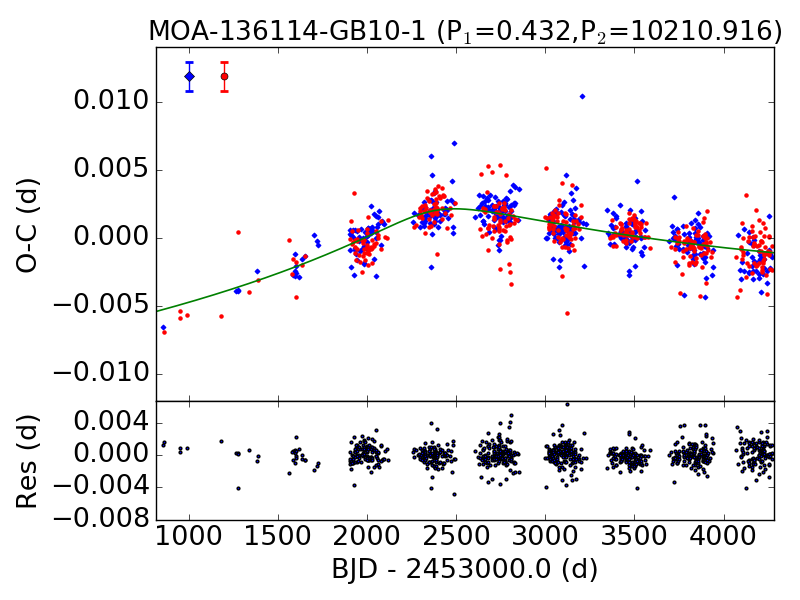}
\includegraphics[width=.32\textwidth]{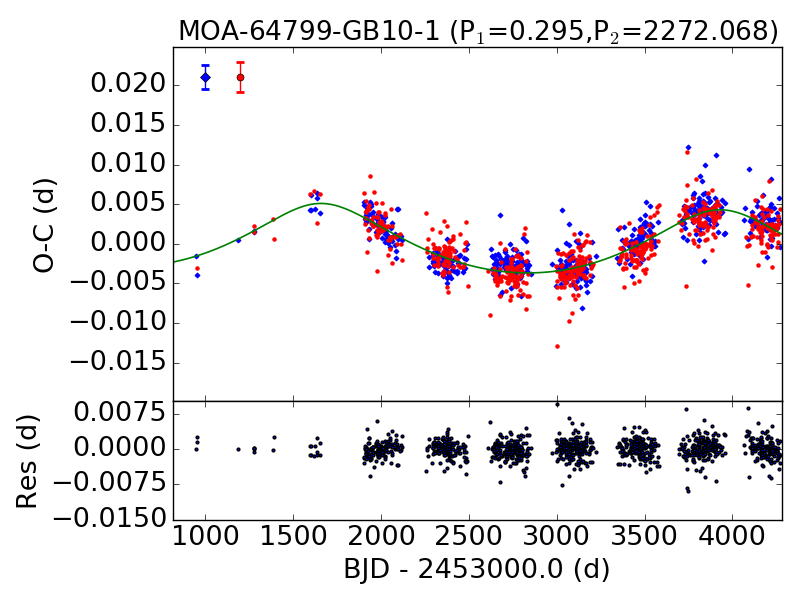}
\includegraphics[width=.32\textwidth]{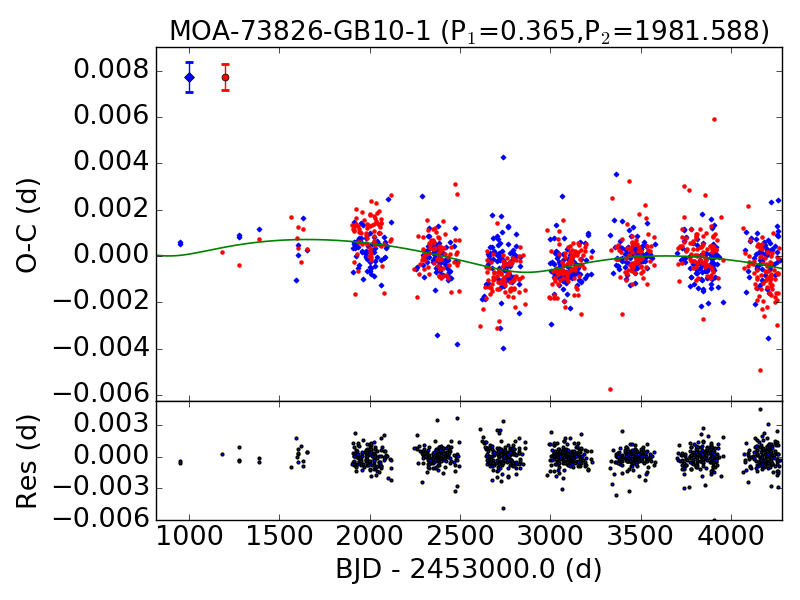}
\includegraphics[width=.32\textwidth]{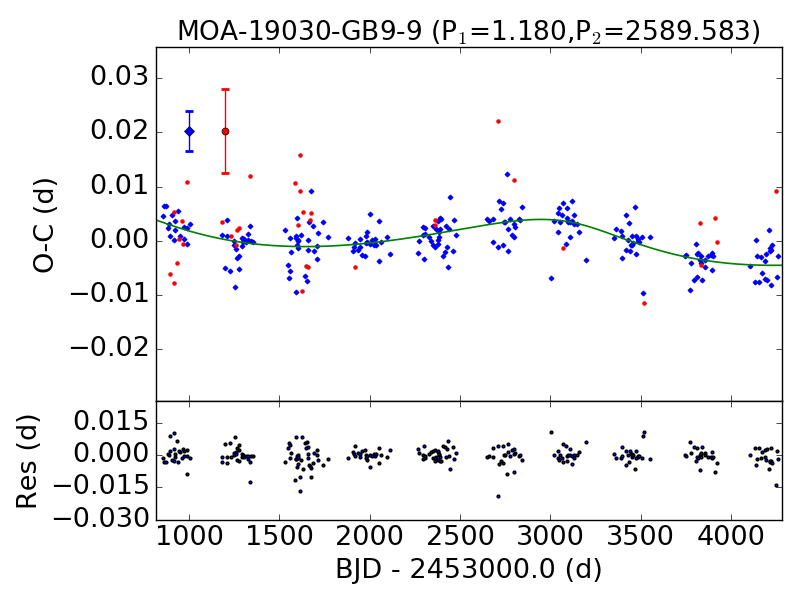}
\includegraphics[width=.32\textwidth]{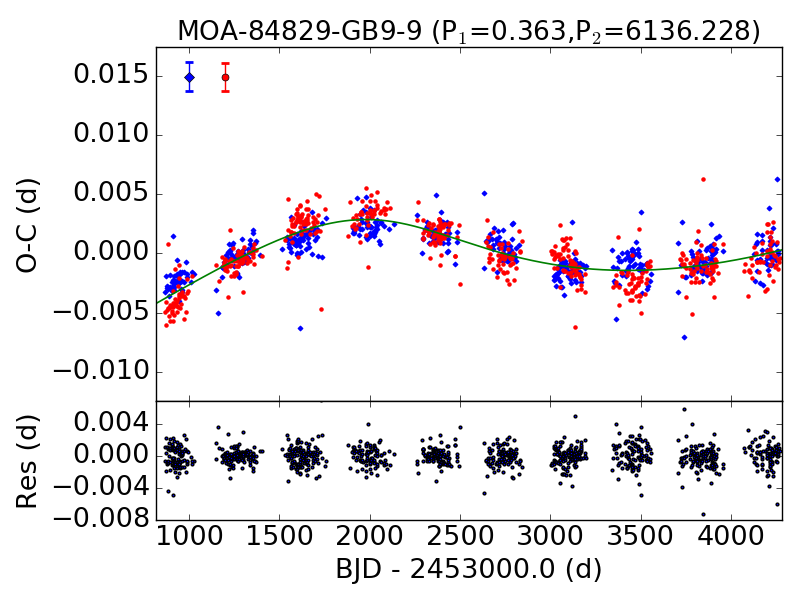}
\includegraphics[width=.32\textwidth]{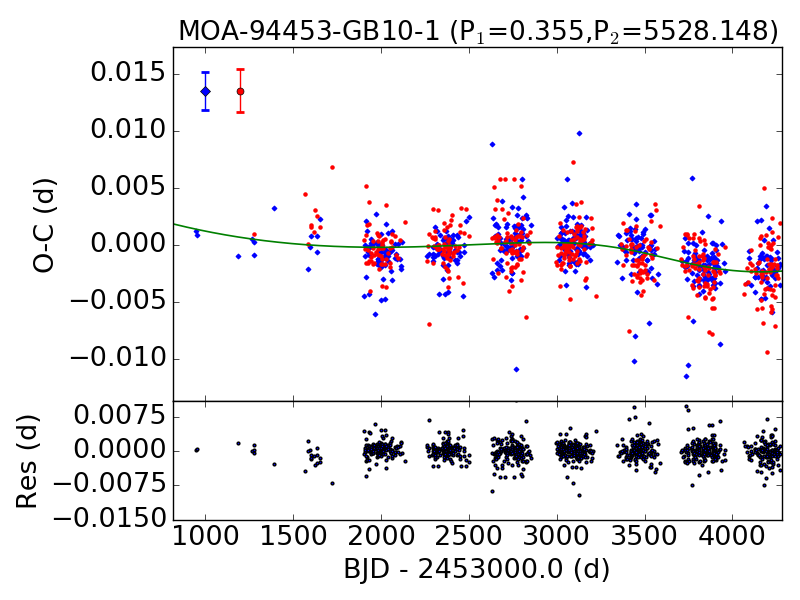}
\includegraphics[width=.32\textwidth]{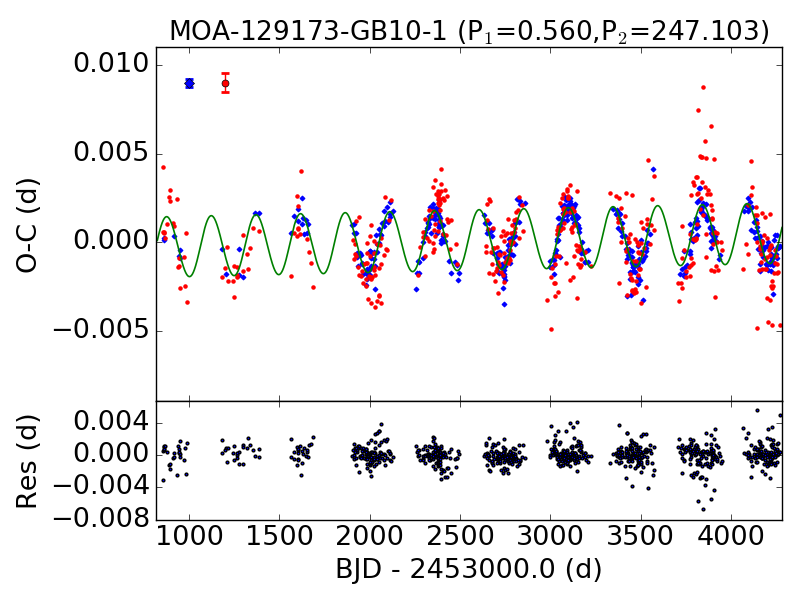}
\includegraphics[width=.32\textwidth]{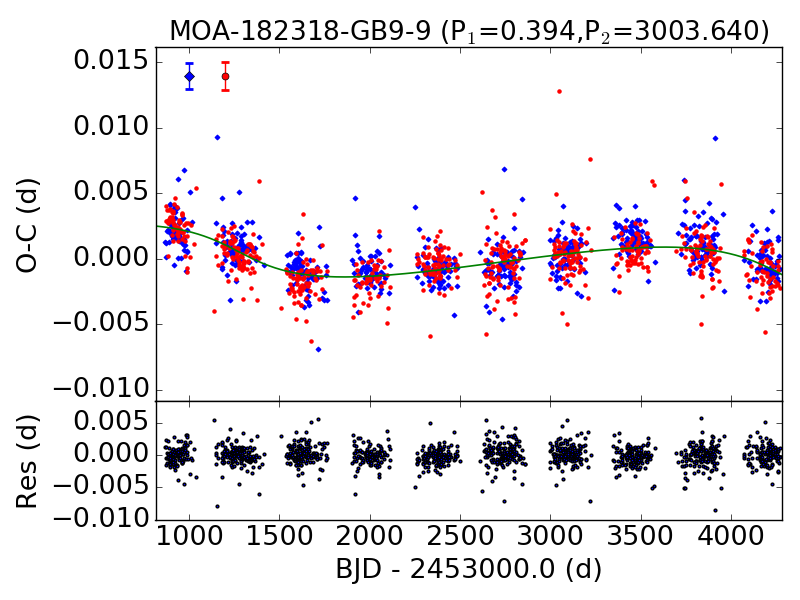}
\includegraphics[width=.32\textwidth]{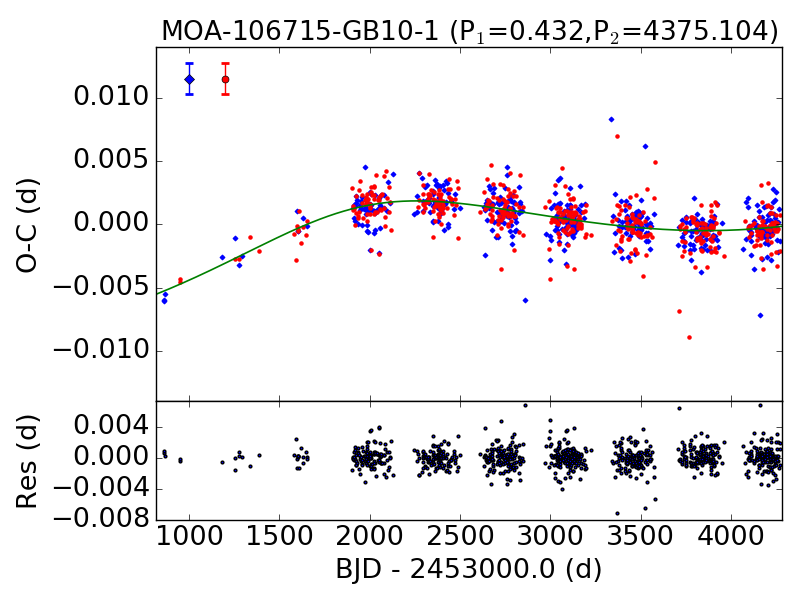}
\includegraphics[width=.32\textwidth]{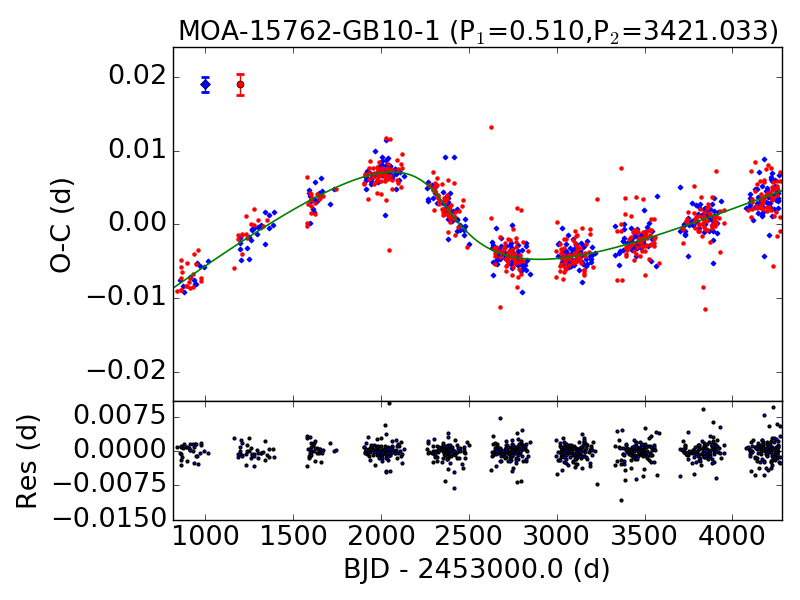}
\includegraphics[width=.32\textwidth]{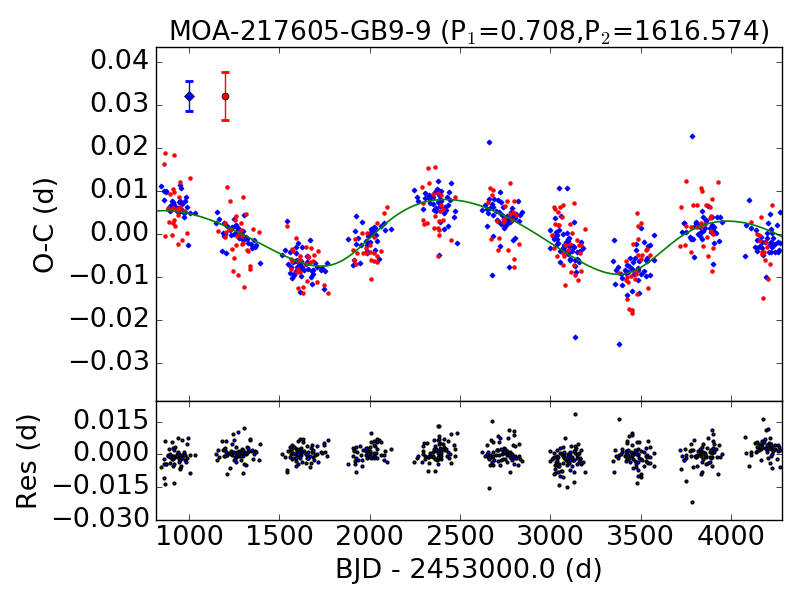}
\includegraphics[width=.32\textwidth]{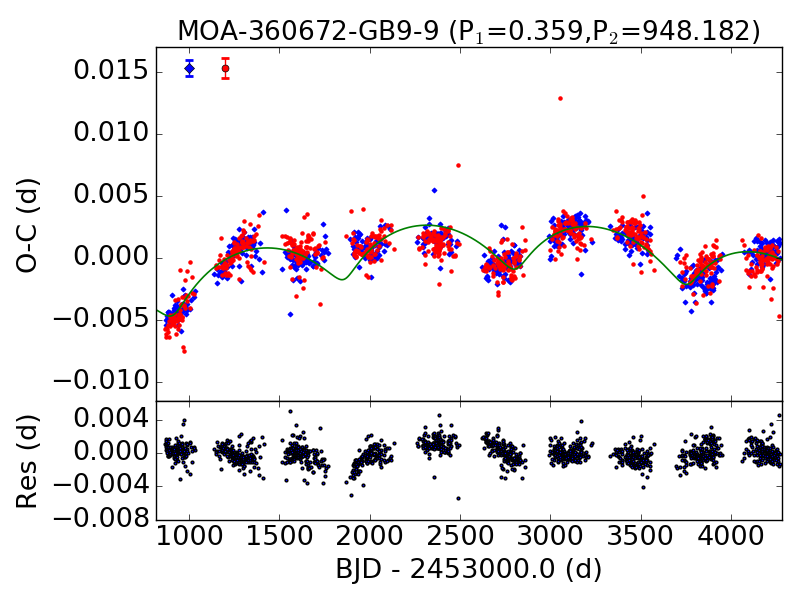}
\includegraphics[width=.32\textwidth]{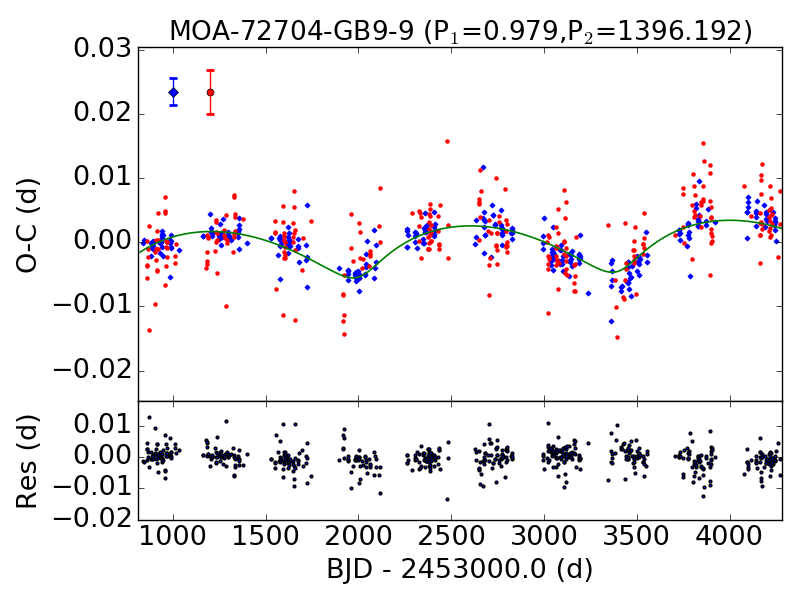}
\includegraphics[width=.32\textwidth]{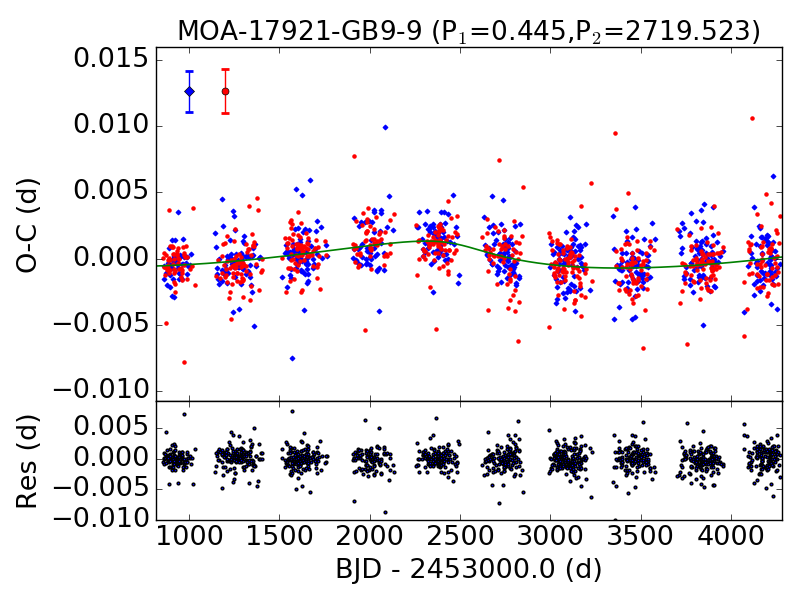}
\caption{(continued)}
\end{figure*}            

\begin{figure*}
\ContinuedFloat
\centering
\includegraphics[width=.32\textwidth]{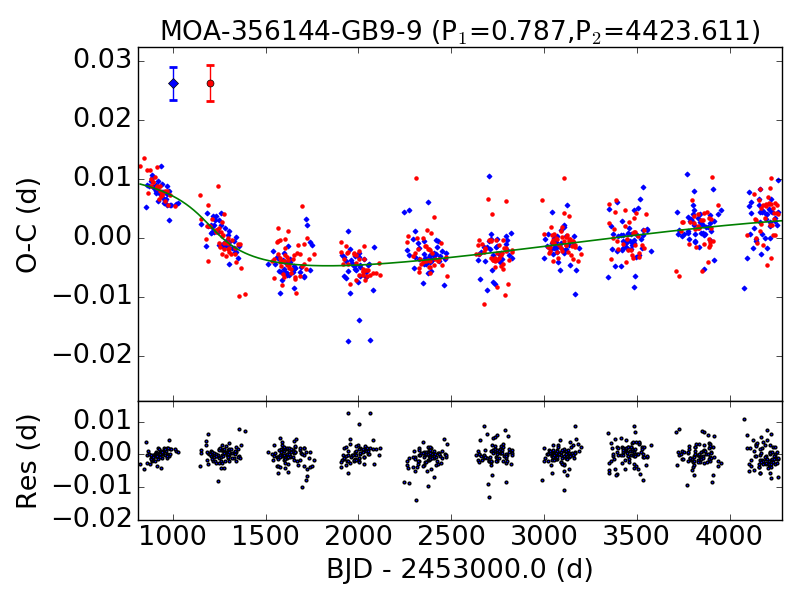}
\includegraphics[width=.32\textwidth]{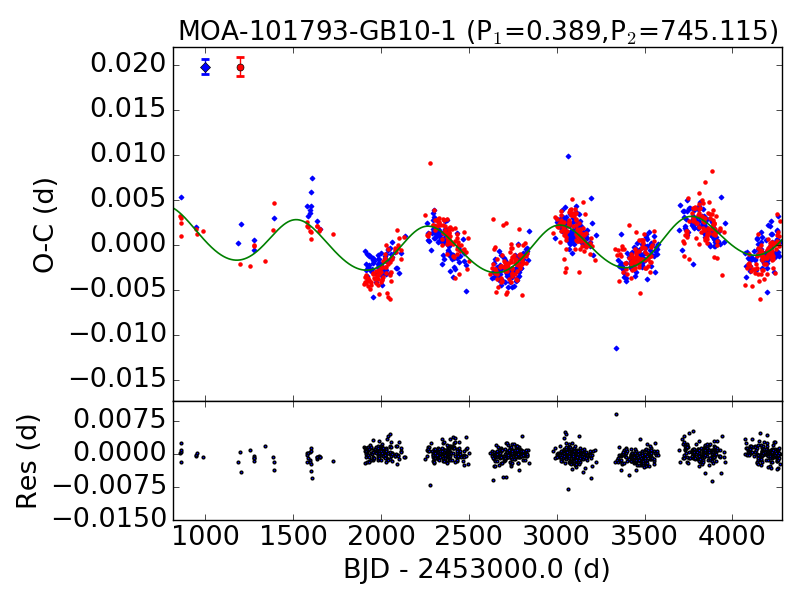}
\includegraphics[width=.32\textwidth]{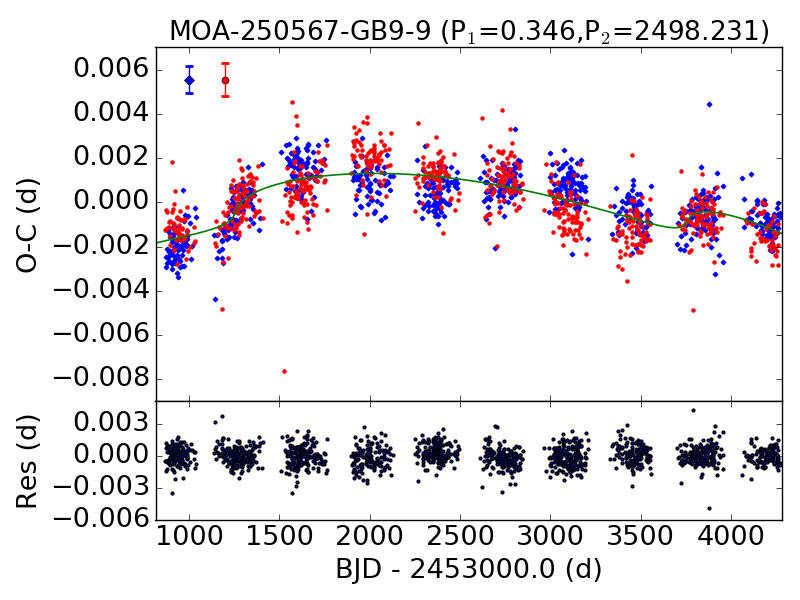}
\includegraphics[width=.32\textwidth]{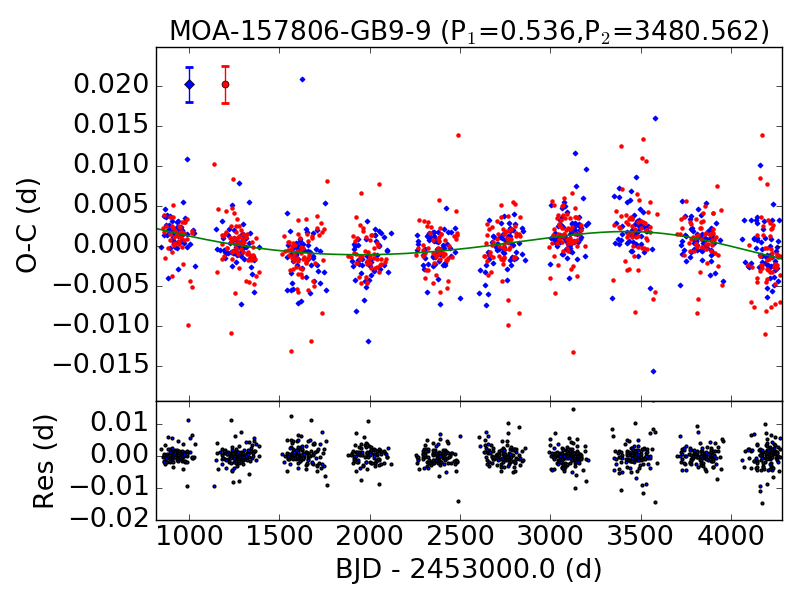}
\includegraphics[width=.32\textwidth]{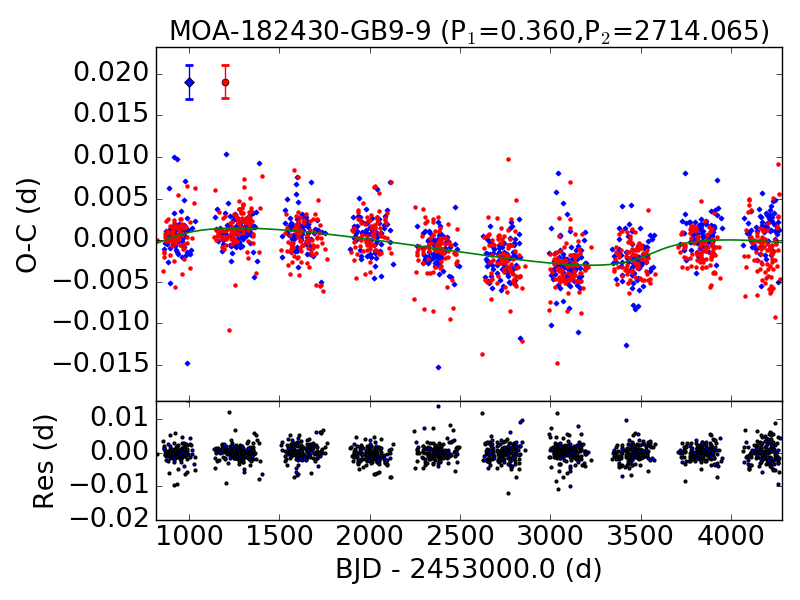}
\includegraphics[width=.32\textwidth]{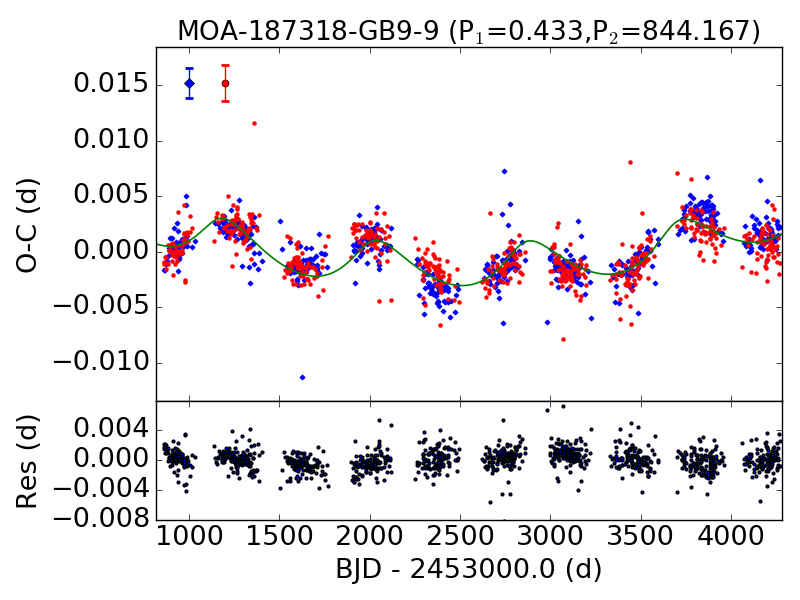}
\includegraphics[width=.32\textwidth]{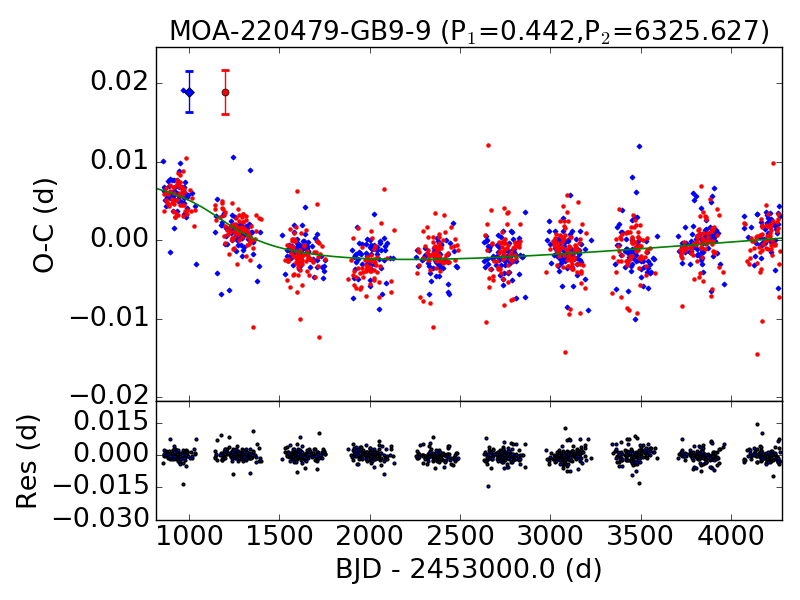}
\includegraphics[width=.32\textwidth]{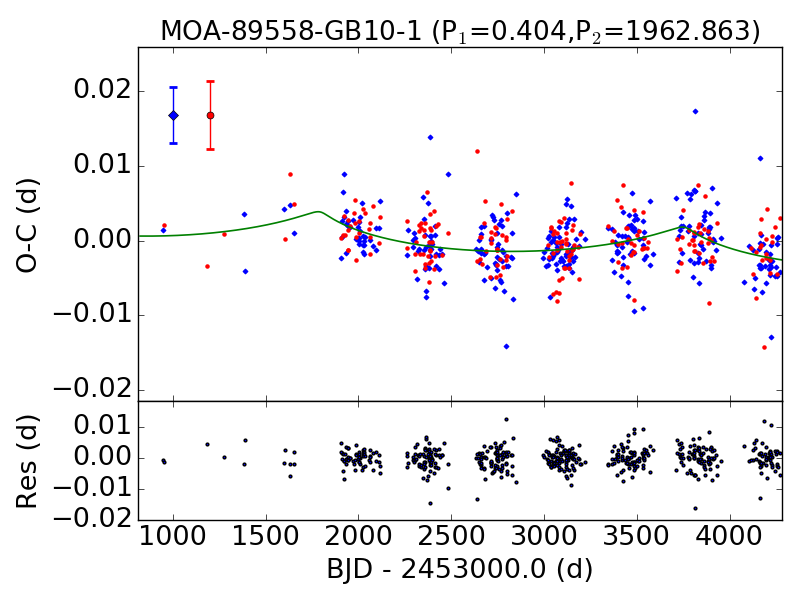}
\includegraphics[width=.32\textwidth]{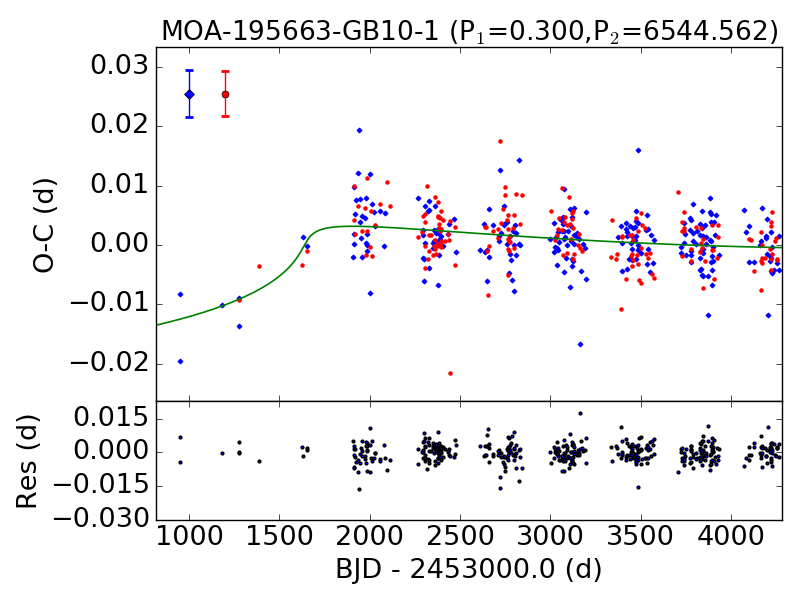}
\includegraphics[width=.32\textwidth]{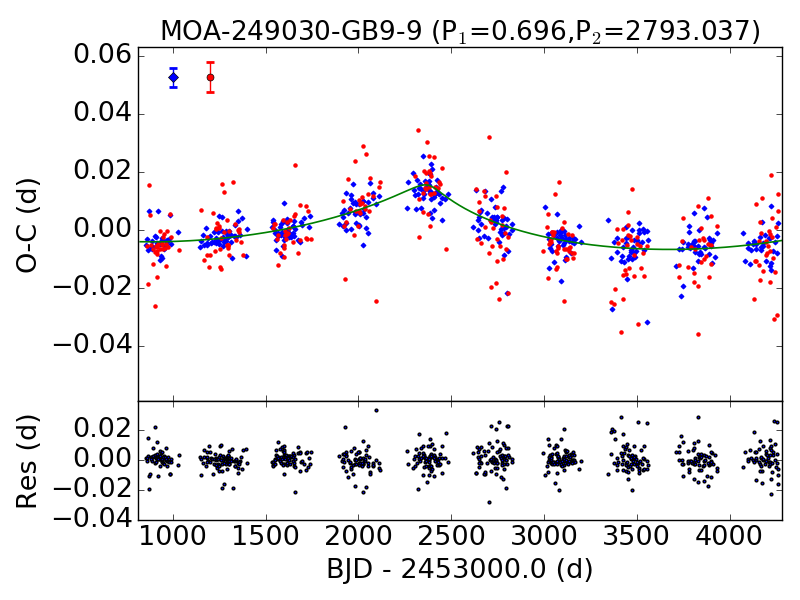}
\includegraphics[width=.32\textwidth]{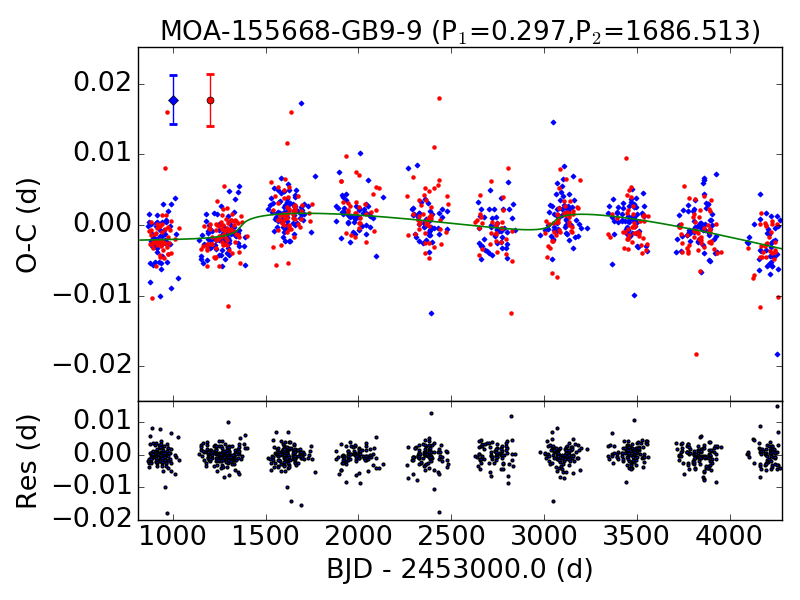}
\includegraphics[width=.32\textwidth]{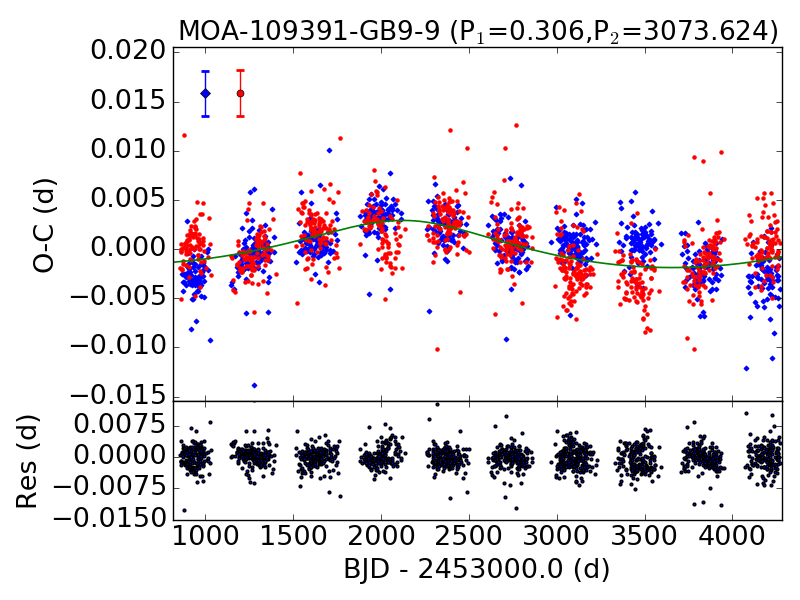}
\includegraphics[width=.32\textwidth]{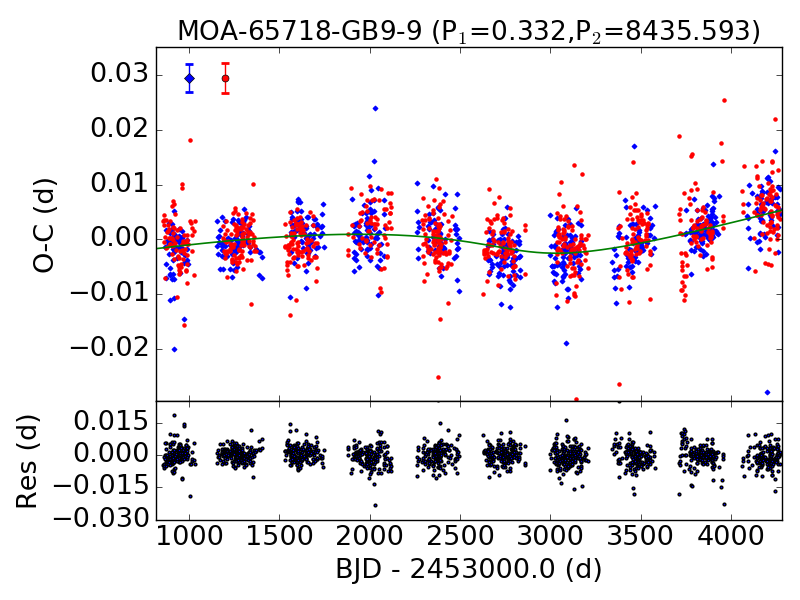}
\includegraphics[width=.32\textwidth]{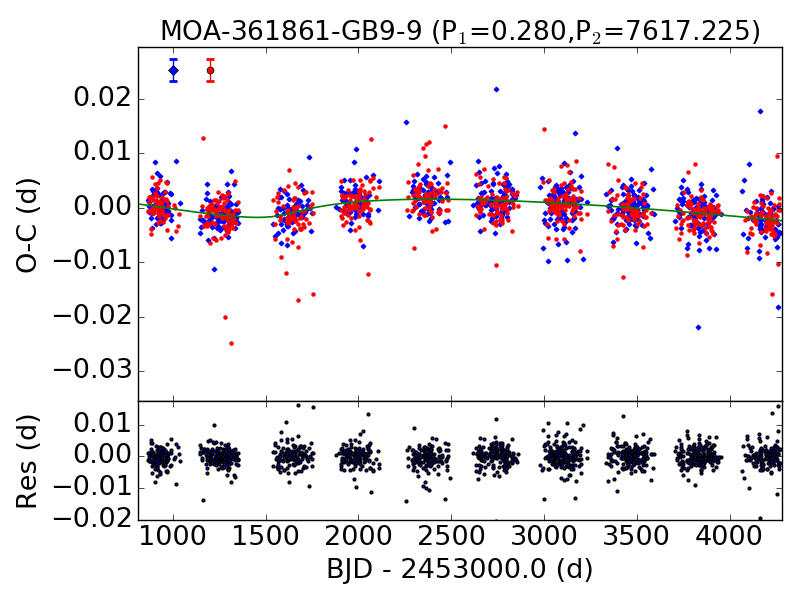}
\includegraphics[width=.32\textwidth]{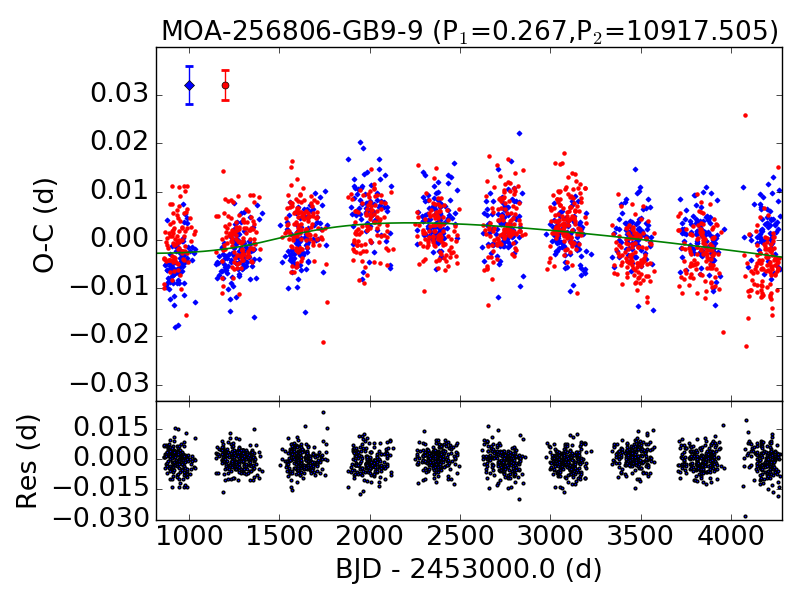}
\caption{(continued)}
\end{figure*} 

\begin{figure*}
\ContinuedFloat
\centering
\includegraphics[width=.32\textwidth]{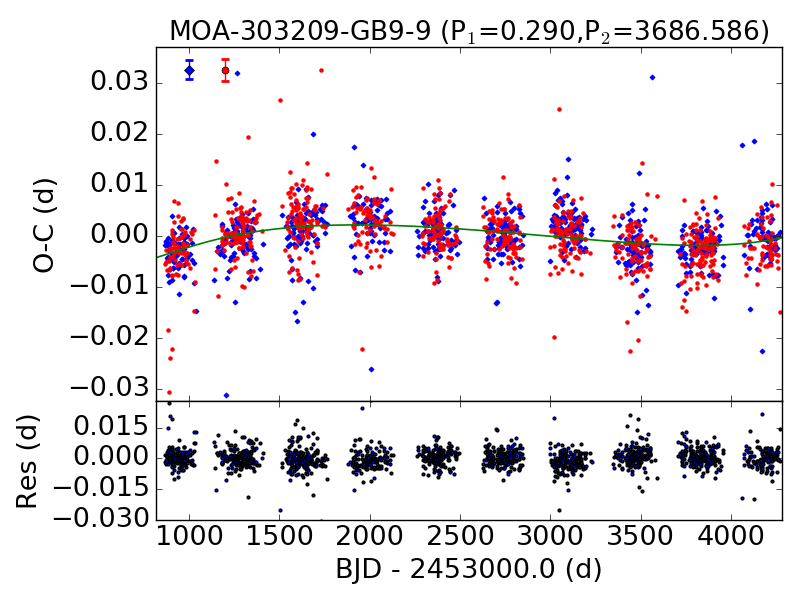}
\includegraphics[width=.32\textwidth]{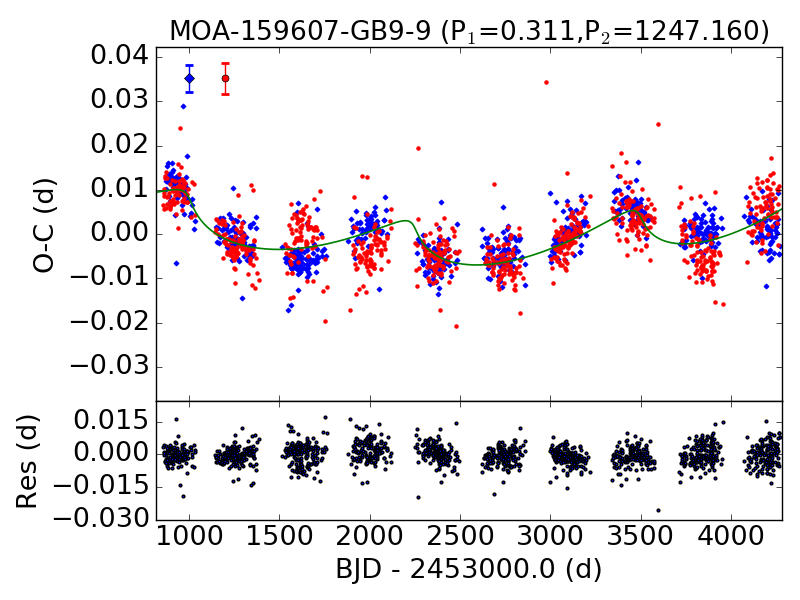}
\includegraphics[width=.32\textwidth]{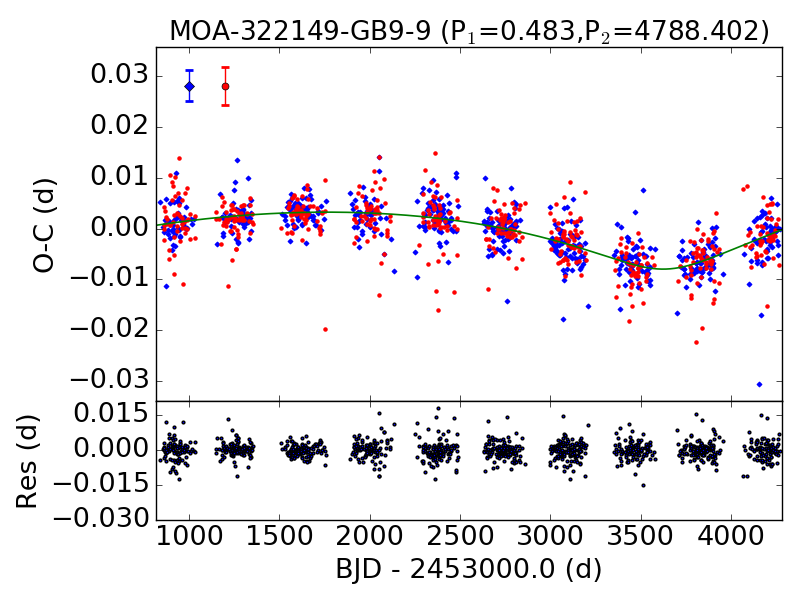}
\includegraphics[width=.32\textwidth]{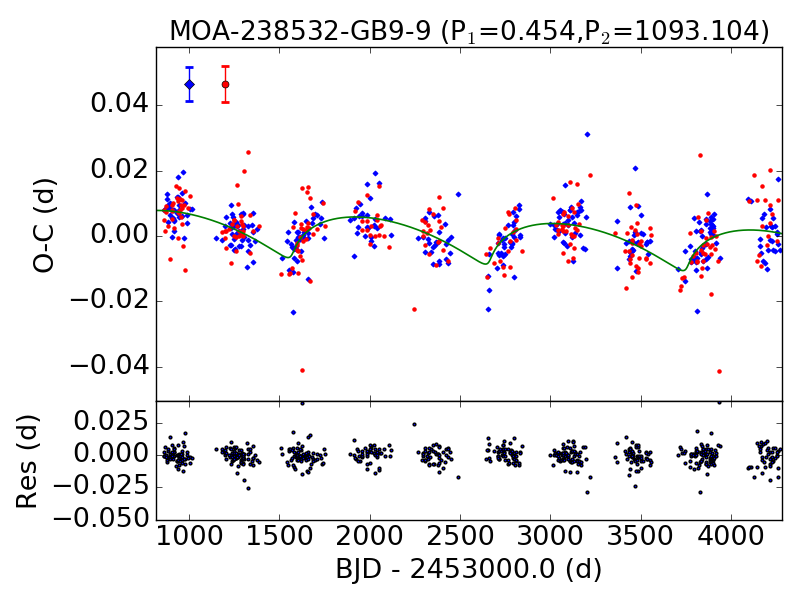}
\includegraphics[width=.32\textwidth]{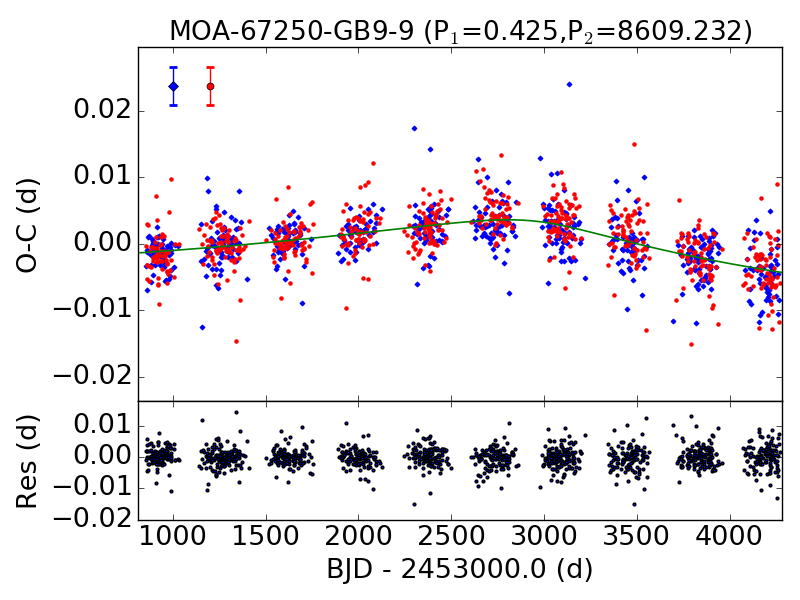}
\includegraphics[width=.32\textwidth]{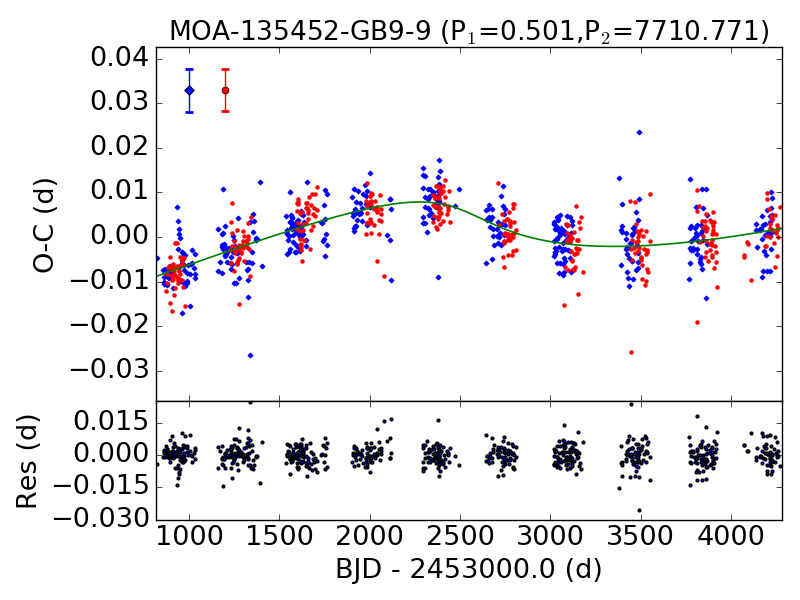}
\includegraphics[width=.32\textwidth]{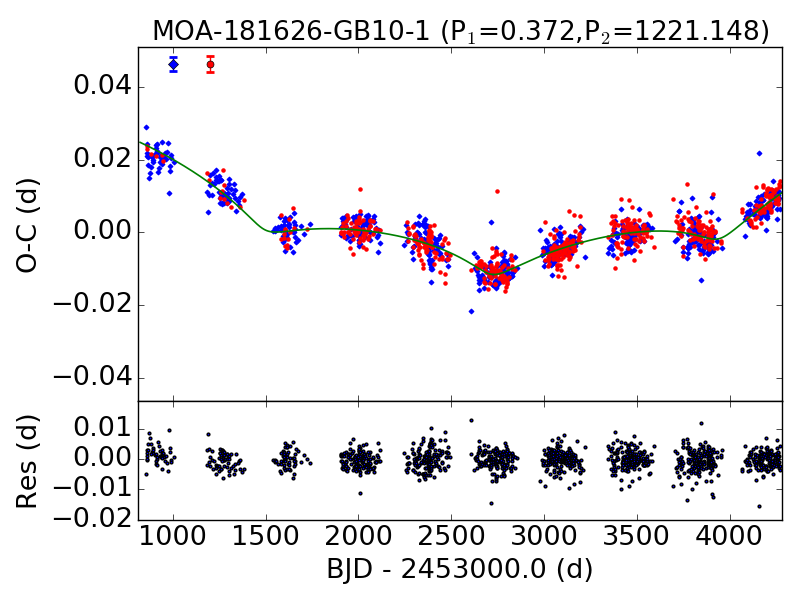}
\includegraphics[width=.32\textwidth]{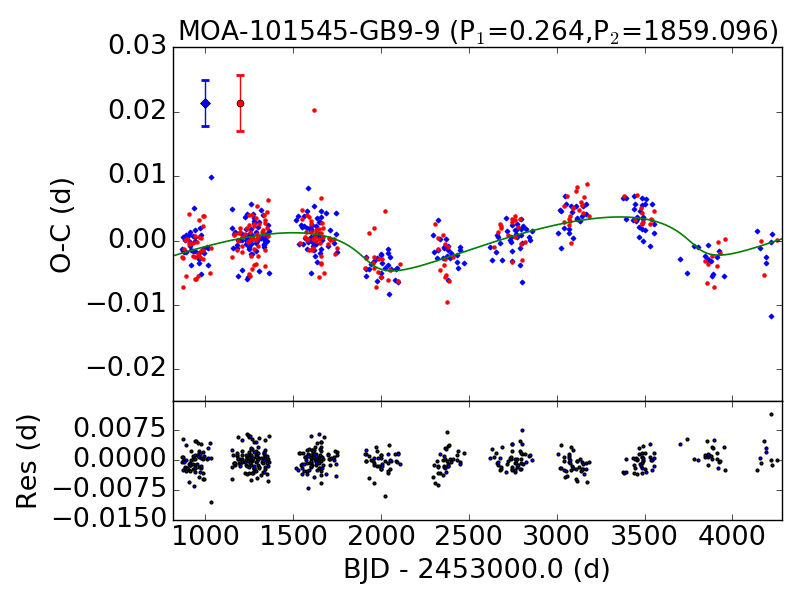}
\includegraphics[width=.32\textwidth]{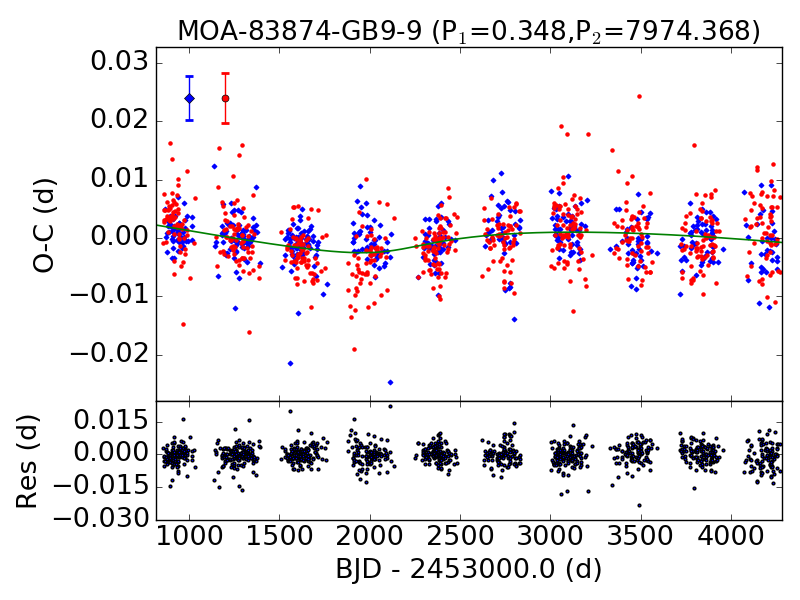}
\includegraphics[width=.32\textwidth]{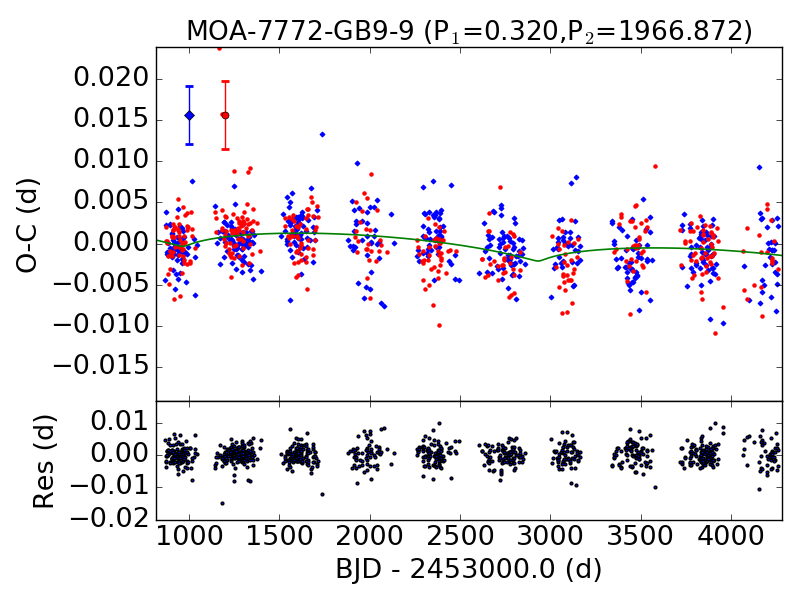}
\includegraphics[width=.32\textwidth]{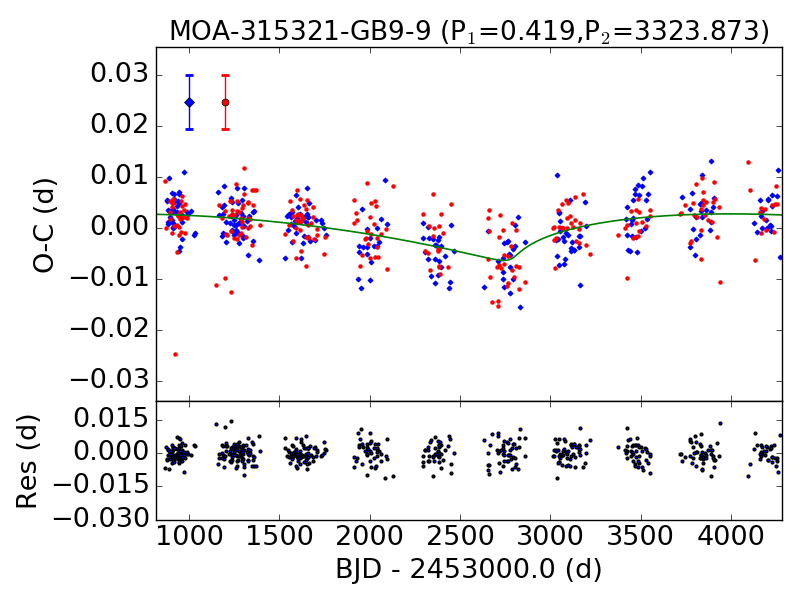}
\includegraphics[width=.32\textwidth]{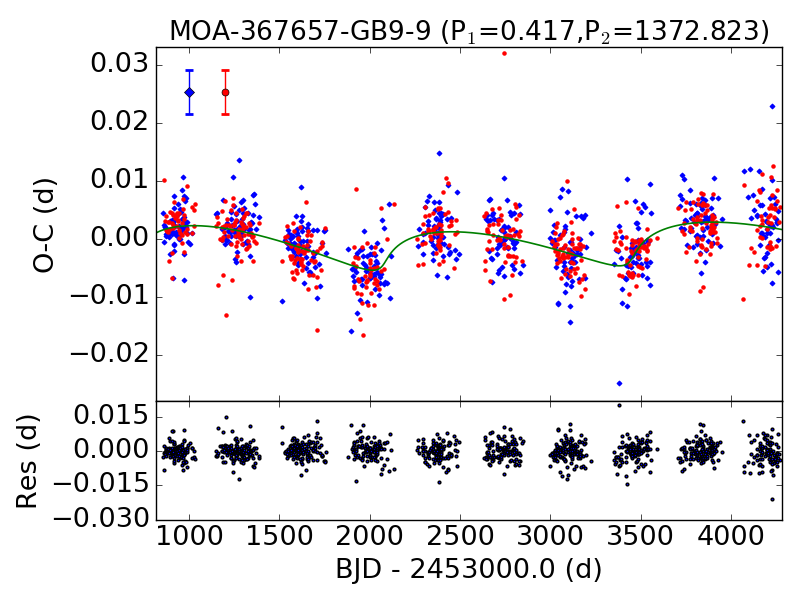}
\includegraphics[width=.32\textwidth]{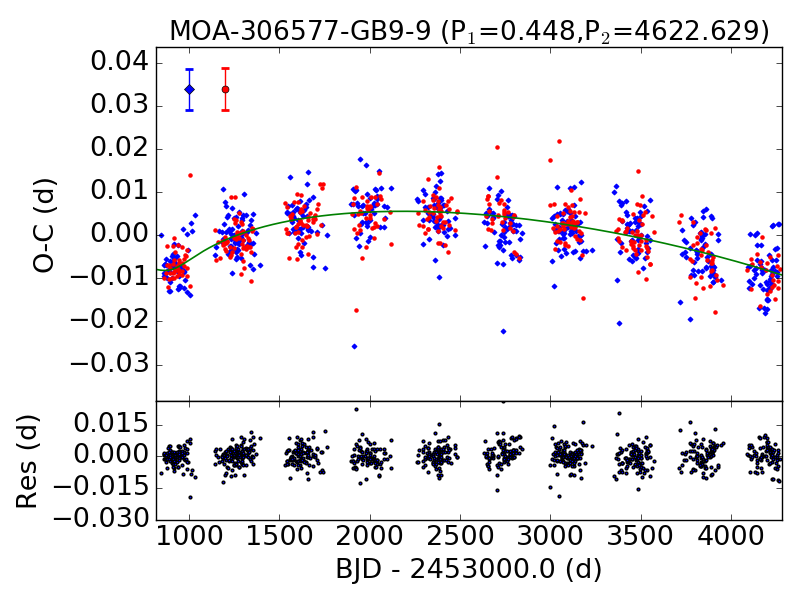}
\includegraphics[width=.32\textwidth]{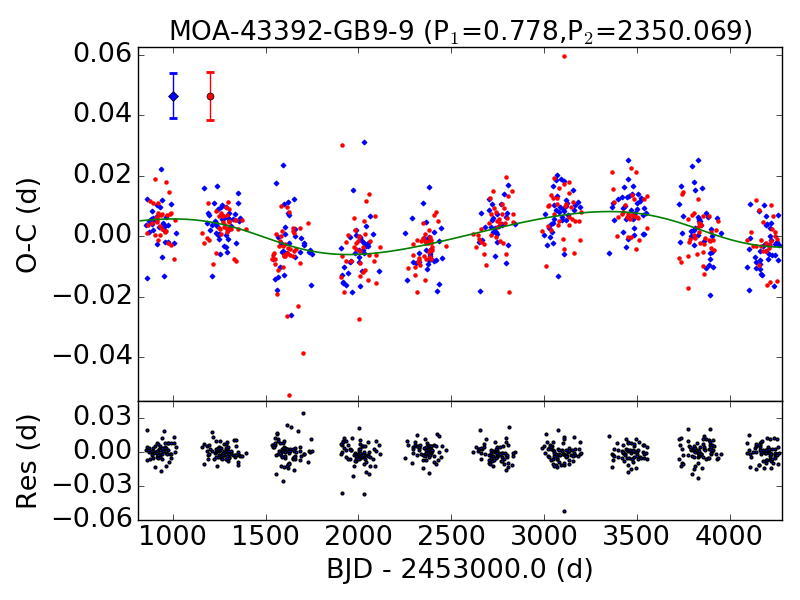}
\includegraphics[width=.32\textwidth]{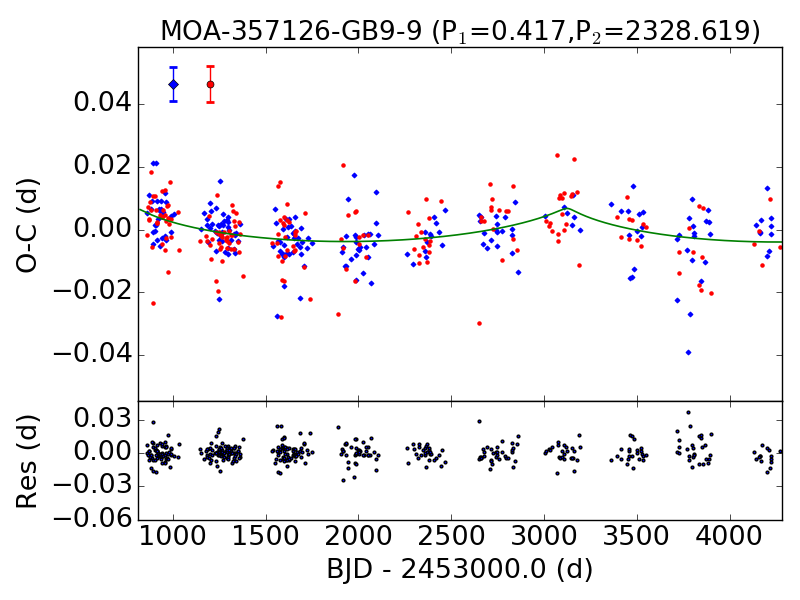}
\caption{(continued)}
\end{figure*} 

\begin{figure*}
\ContinuedFloat
\centering
\includegraphics[width=.32\textwidth]{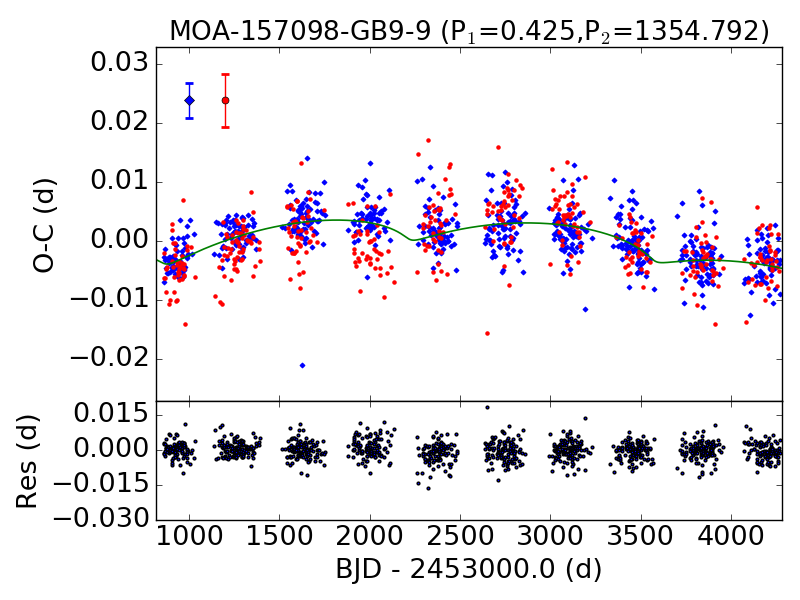}
\includegraphics[width=.32\textwidth]{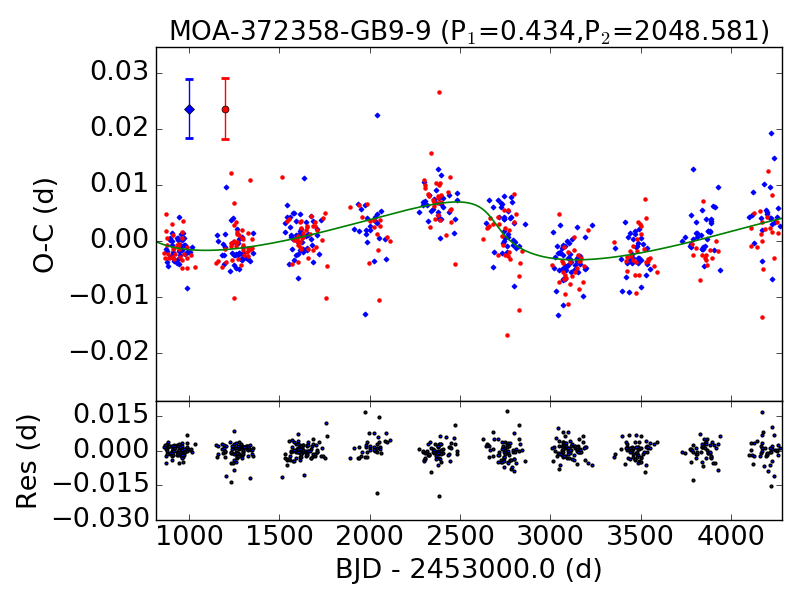}
\includegraphics[width=.32\textwidth]{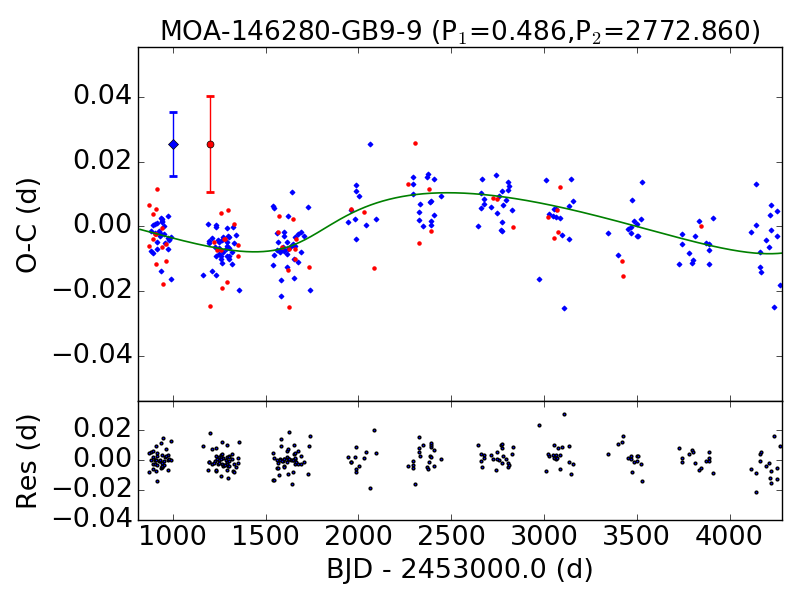}
\includegraphics[width=.32\textwidth]{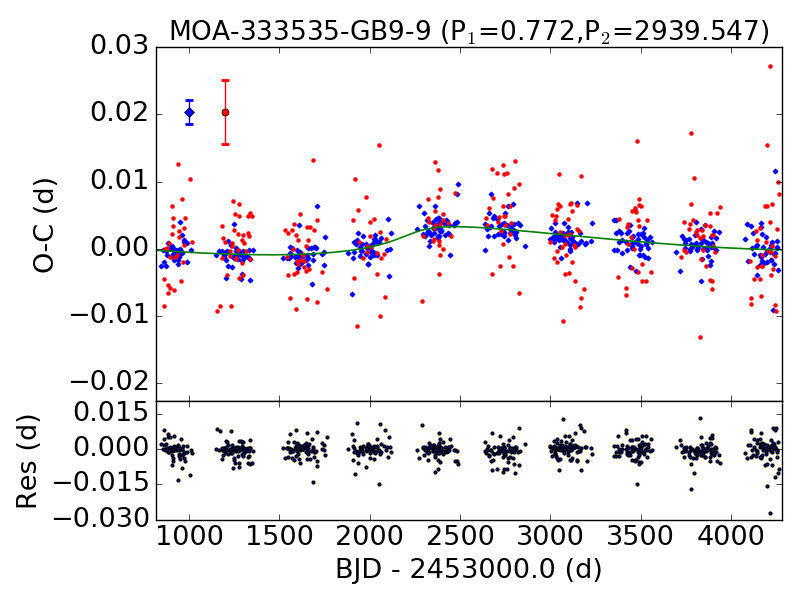}
\includegraphics[width=.32\textwidth]{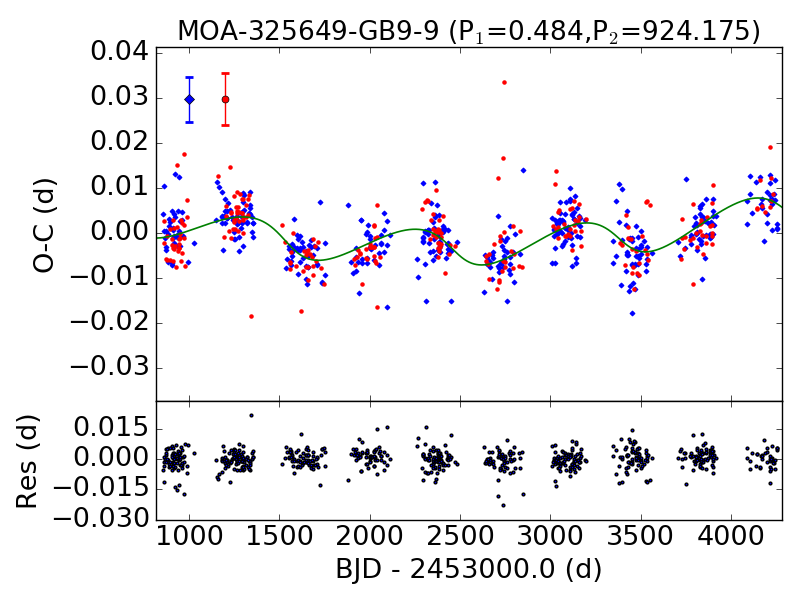}
\includegraphics[width=.32\textwidth]{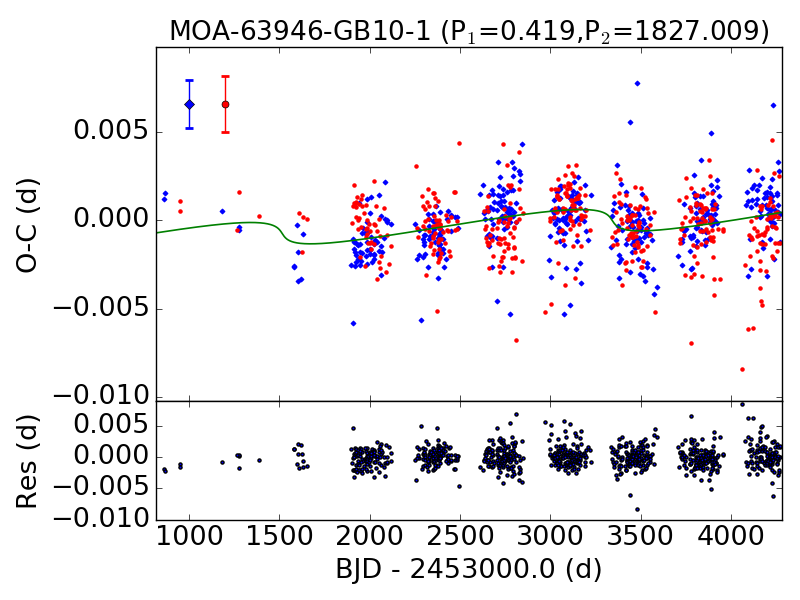}
\includegraphics[width=.32\textwidth]{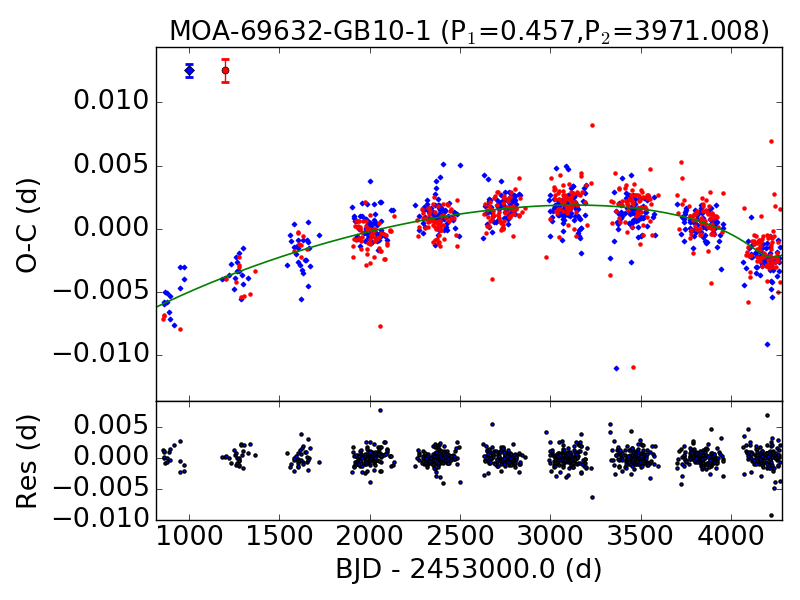}
\includegraphics[width=.32\textwidth]{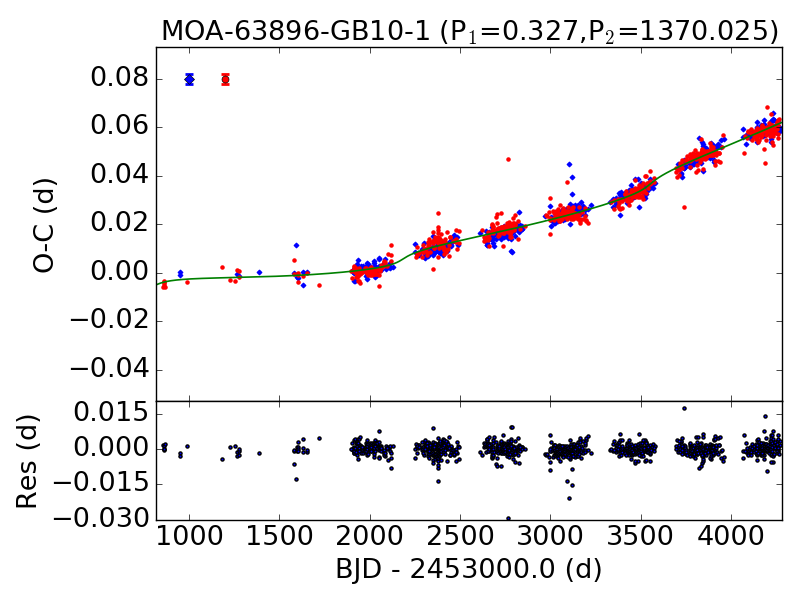}
\includegraphics[width=.32\textwidth]{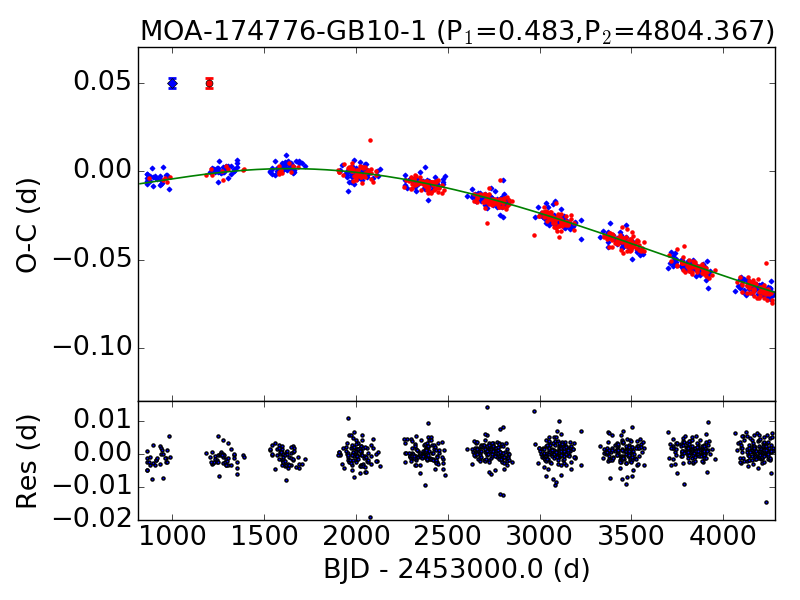}
\includegraphics[width=.32\textwidth]{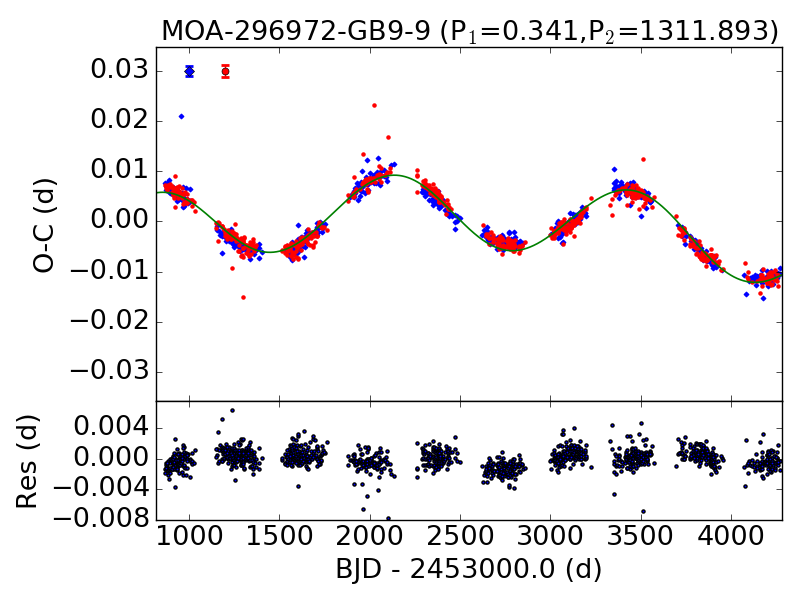}
\includegraphics[width=.32\textwidth]{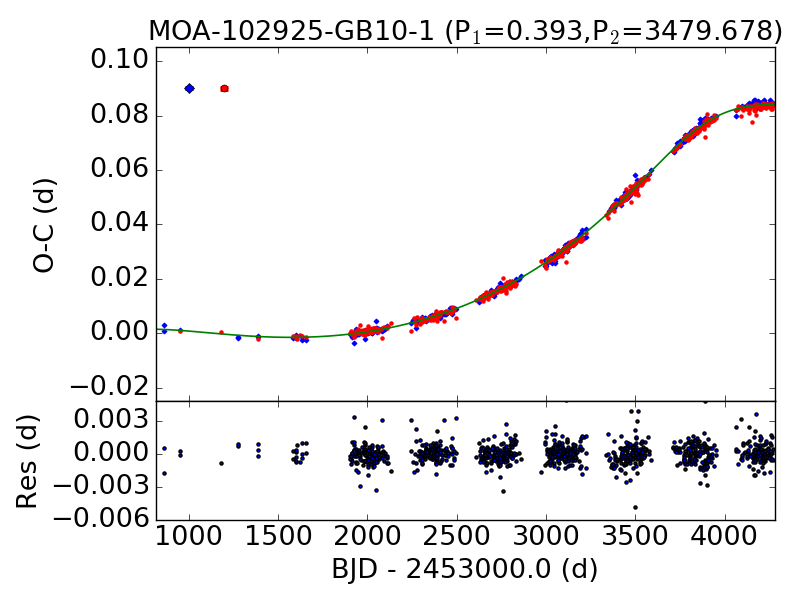}
\includegraphics[width=.32\textwidth]{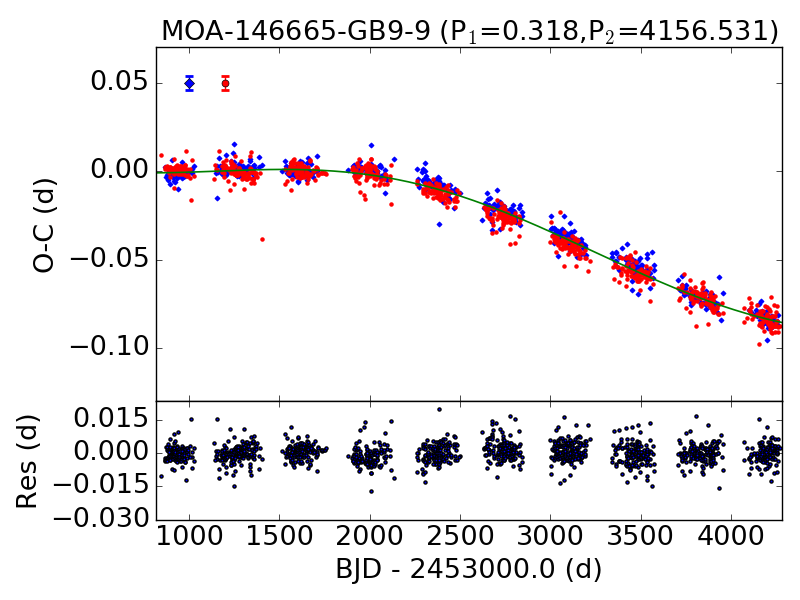}
\includegraphics[width=.32\textwidth]{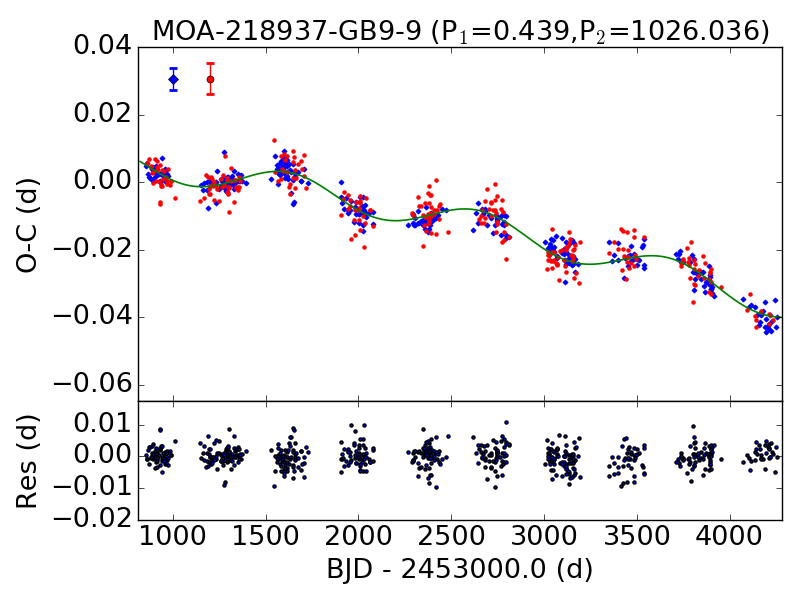}
\caption{(continued)}
\end{figure*} 

\subsubsection{Frequency of tertiary companions}
The period distribution of our EB sample is shown in Figure~\ref{fig:eb_p1_dis}. The peak occurs at around 0.5 day and the number of EBs declines rapidly when period is longer than 0.5 day. On the other side, there is a cut-off at $\sim0.2$ days. The lack of contact binaries below 0.2 days in the MOA EBs is consistent with the idea of the existence of a physical lower limit of the period of contact binaries \citep{1992AJ....103..960R}. Looking at the period distribution of the 91 EBs with detected LTTE signals, 69 of them (i.e.~$75\%$) are of periods $<0.5$ days, while none of them are of periods longer than 1.5 days. The overall frequency of EBs with detected LTTE signals is $91/542 = 0.168$. If we look at the distribution of the frequency of EBs with detected LTTE signals over the period as shown in Figure~\ref{fig:outer_freq}, it is interesting to note that the frequency basically increases as the period decreases, and the frequency reaches $13/20 = 0.65$ when the period is shorter than 0.3 days. When we further zoomed into the period range between 0.2 and 0.4 days, there are six EBs in our sample of periods shorter than 0.26 days and they all have the LTTE signals detected in their O-C diagrams, giving the frequency of having tertiary companions equal to 1. Figure~\ref{fig:etv_low_cb} shows the ETV curves of these six EBs. The periods of their tertiary companions range from $\sim$1500 days (or 4 years) to $\sim$8000 days (or 22 years). We have to emphasize that our estimation of tertiary companion frequency is very preliminary. To obtain robust estimation of the frequency of contact binaries with tertiary companions, the corrections which take all the selection effects and detection limitations into account have to be estimated through the population synthesis. It would require substantial follow-up work and is out of the scope of this paper.

\subsubsection{Outer eccentricity}
Another interesting property to look at is the distribution of the outer eccentricities. We plotted the outer eccentricity distributions in Figure~\ref{fig:outer_ecc_dis} with the number of bins of 10 and 20. In the case of outer eccentricity binned into 10 bins, the distribution was characterized by a peak at $e_{2} = 0.7$, while, interestingly, the second peak which contains 10 triple candidates was seen at $e_{2} > 0.9$. When we binned the outer eccentricity into 20 bins instead, an excess was even clearly noticed at $e_{2} > 0.95$. Taking uncertainties in the eccentricity into account, the outer eccentricities of these 10 triple candidates all still fell into the range of $e_{2} > 0.9$ except one which just fell into the range of $e_{2}$ from 0.8 to 0.9. Since the excess at $e_{2} > 0.9$ is still preserved for our triple candidates when the uncertainties are concerned, such an excess is not an artifact resulting from binning. 

Nonetheless, such high eccentric companions are expected to be so unstable that they would not survive owing to long-term instability or their eccentricities would not be still maintained to be so large if they formed with the inner binary systems roughly at the same time, given that contact binaries such as W UMa variables belong to old populations of ages about 4.4-4.6$\,$Gyr \citep{2014MNRAS.437..185Y}. Thus, whether the derived LTTE solutions were physical has to be examined carefully. We inspected the O-C diagrams of every EB with detected LTTE signals by eye. The LTTE solutions associated with $e_{2}>0.9$ turned out to have unique shapes with sharp turning points (see Figure~\ref{fig:etv_ecc_0.9}), indicating the possibility of sudden changes in their orbit periods. In particular, such sudden period changes are already noticeable in the O$-$C diagrams of MOA-284305-GB9-9, MOA-108463-GB9-9 and MOA-249394-GB9-9. Although the values of $\Delta$BIC of their LTTE fits are much larger than 10, it should be emphasized that the high value of $\Delta$BIC simply means that the LTTE model, eq.(\ref{eq:ltte}), which can be recognized as a mathematical model containing combination of sinusoidal terms, gives a better description than the pure parabolic model and does not guarantee that the LTTE fit is physically reliable. Since the LTTE solutions with extremely high outer eccentricities are probably unphysical, other reasons might be more appropriate to explain the observed ETVs of these ten MOA EBs. Abrupt changes in or sudden jumps of orbit periods are, in fact, not a rare phenomenon in close binaries. Dozens of close binaries, which belong to Algol- or W UMa-type, were reported to exhibit sudden jumps in their O$-$C diagrams (e.g. \citealt{2002PASP..114..650Q,1999A&A...341..799Q,1994AJ....107.1141N}). Mechanisms which might induce such sudden period jumps include sudden mass exchange \citep{1987A&A...172..155H} or mass loss \citep{2002A&A...390..555Y} via stellar flares, variations in the internal structures (i.e., convective envelopes) of binaries' active components \citep{2002PASP..114..650Q}, and the rapid accretion of binaries from the circumstellar matter \citep{2002A&A...390..555Y}. Also, the periodicity of the O$-$C diagrams might come from magnetic cycles arising from, e.g., the Applegate effect, which can produce quasi-cyclic ETVs, instead of LTTE from unseen tertiary companions. Despite the questionable reliability of the LTTE solutions, these ten MOA EBs show very interesting ETVs, which are worth taking notice of.

The cumulative distribution of the outer eccentricity of the MOA triple candidates was calculated (see Figure~\ref{fig:ecc_cpf}). If all 91 triple candidates are taken into account, the calculated distribution lies between the uniform distribution and the thermal distribution\footnote{Thermal eccentricity distribution refers to the distribution of eccentricities of a population of binary stars, where every member has interacted with each other and reached statistical equilibrium. The normalized distribution of such a population as a function of eccentricity is $f(e) = 2e\,de$, where $e$ is eccentricity, derived by \citeauthor{1919MNRAS..79..408J} (\citeyear{1919MNRAS..79..408J}).}. However, as the reliability of the LTTE solutions of extremely high outer eccentricities are quite questionable, inclusion of the triple candidates with $e_2>0.9$ might lead to an incorrect conclusion. We, therefore, excluded the triple candidates with $e_2>0.9$ and recalculated cumulative distribution. The recalculated cumulative distribution, in contrast to the case when all the triple candidates were included, resembles neither a linear nor a flat distribution, indicating that the issue of whether the detection of the triple candidates with very high outer eccentricities was real or not would lead to very different conclusions. 

The plot of outer eccentricity against tertiary period is shown in Figure~\ref{fig:e2_P2_plot}. The correlation coefficient was calculated to be 0.042, indicating no correlation between the outer eccentricity and tertiary period for our MOA sample.

\section{Discussion and Conclusions}
\label{sec:ltte_conclusion}
We carried out ETV analysis for the sample of MOA EBs of periods $<2$ days in two MOA subfields, GB9-9 and GB10-1, using the MOA-II data spanning 9.5 years. The sample contains 524 EBs, 436 and 106 in the GB9-9 and GB10-1 fields, respectively. The Bayesian information criterion was used as a measure for the model selection between ETV models with and without the LTTE term. In this way, we discovered 91 MOA EBs with detected LTTE signals, indicating the presence of tertiary orbiting companions (see Figure \ref{fig:etv_spot_eg}, \ref{fig:etv_low_cb}, \ref{fig:etv_ecc_0.9} and \ref{fig:etv_17}). The distribution of tertiary period for our 91 triple candidates peaked sharply at 2660 days (or 7.2 years), while there were no EB in the sample with any tertiary companion of orbiting period $P_{2}>30$ years. Given the fact that the data spanned only 9.5 years, it is obvious that the lack of detection of tertiary companions of $P_{2}>30$ years is a consequence of the data time span being not long enough. In addition, we suspect that the peak being at $2660$ days also resulted from a selection effect due to the data time span. Nonetheless, the significant decline in the distribution for $P_{2}<10^{3}$ days might be related to the formation of close and contact binaries although it might be also due to the presence of regular gaps in the ETV curves associated with the off-season periods. 

As our sample was homogeneous in terms of period, it would be interesting to see how the frequency of EBs with tertiary companions varies as a function of the inner binary period $P_{1}$. Particularly, the group of EBs of periods $<0.5$ days represented a homogeneous sample of contact binaries and the detection of LTTEs in the contact binaries in this period range should suffer from the least selection effect due to day-night cycles as indicated by the number of eclipse time measurement points we obtained. For our sample, there is an obvious tendency for short period contact binaries to be likely accompanied by tertiary companions. The frequency of our EBs with tertiary companions increases as $P_{1}$ decreases. For our 13 contact binaries of $P_{1}<0.3$ days, the frequency reaches a value of $0.65$. Looking into these 13 contact binaries, we further found that all six contact binaries of $P_{1}<0.26$ days are with tertiary binaries. Since all our detected tertiary companions are of orbiting periods $<10^{4}$, our results suggest that contact binaries of periods close to the 0.22-day contact binary limit are commonly accompanied by relatively close tertiary companions. Meanwhile, the outer eccentricity distribution for our 91 triple candidates behaved approximately as a linear function, but an excess at $e_{2}$ > 0.9 was observed. In addition, long-term flux variations were seen in the light curves of most of our triple candidates. In a few cases, the flux variations are seemingly correlated with the ETVs as inspected by eye. This kind of flux variation was also observed in OGLE EBs with cyclic ETVs \citep{2017AcA....67..115P}. The long-term flux variations might come from the third light from bright stars which orbit around the EBs (e.g. \citealt{2011Sci...332..216D}). Nonetheless, such variations might otherwise originate from changing luminosity of EBs' components associated with stellar magnetic activities or pulsations. In particular, the Applegate mechanism predicts cyclic variations in the luminosity and colours which are correlated with the orbital period variations (i.e., the O$-$C cycles) \citep{2002AN....323..424L,1998MNRAS.296..893L,1992ApJ...385..621A}. Since there is a possibility that the detected O$-$C cycles for those MOA samples were driven by the Applegate mechanism, it will be necessary to examine the correlation between the long-term flux variations and ETVs in order to have better judgment on the origins of their ETVs.

In addition to the actions we mentioned in the previous paragraph, there is also much follow-up work that can be carried out in the future. First of all, given the fact that the MOA fields we investigated overlap the OGLE fields and the OGLE observations began earlier than MOA, it is worth investigating the possibility of including the OGLE data to extend the time span for ETV analysis. On the other hand, since several mechanisms such as mass transfer and the Applegate mechanism which would be often present in contact binaries could induce long-term ETVs, the possibility of false positive detection of LTTE in our sample has to be a concern. In this sense, radial velocity measurements or direct imaging would be desirable to confirm our discoveries. Also, we investigated short period binaries only in the two MOA subfields, GB9-9 and GB-10-1. We did not exploit the entire MOA EB catalogue that was established by \citeauthor{2017MNRAS.470..539L}$\,$(\citeyear{2017MNRAS.470..539L}). Therefore, the study of the multiplicity of contact binaries using a larger sample from the current MOA EB catalogue should be a task that can be carried out in the near future.

\section*{Acknowledgements}
M.C.A.~Li acknowledges the contribution of NeSI high-performance computing facilities to the results of this research. NZ's national facilities are provided by the NZ eScience Infrastructure and funded jointly by NeSI's collaborator institutions and through the Ministry of Business, Innovation \& Employment's Research Infrastructure programme. URL https://www.nesi.org.nz. NJR is a Royal Society of New Zealand Rutherford Discovery Fellow. AS is a University of Auckland Doctoral Scholar. TS acknowledges financial support from the Japan Society for the Promotion of Science (JSPS) under grant numbers JSPS23103002, JSPS24253004 and JSPS26247023. NK is supported by Grant-in-Aid for JSPS Fellows. The MOA project is supported by JSPS grants JSPS25103508 and JSPS23340064 and by the Royal Society of New Zealand Marsden Grant MAU1104.





\bibliographystyle{latest}
\bibliography{microlensing,eclipsingBinaries}





\label{lastpage}
\end{document}